\documentclass[journal]{IEEEtran}

\usepackage{amsmath, amsfonts, amssymb, amsthm, amsbsy, amscd, bm} 
\usepackage{mathrsfs}
\usepackage{cite}
\usepackage{array}
\usepackage{rotating}
\usepackage{hyperref}
\usepackage{subfigure}
\usepackage{comment,array}
\usepackage{graphicx}
\usepackage[table]{xcolor}

\usepackage{cite}
\usepackage{mwe}
\usepackage[percent]{overpic}

\makeatletter
\def\hlinewd#1{%
  \noalign{\ifnum0=`}\fi\hrule \@height #1 \futurelet
   \reserved@a\@xhline}
\makeatother

\definecolor{mygreen1}{rgb}{0, .5, 0}
\definecolor{myred1}{rgb}{.5, 0, 0}
\definecolor{mygray1}{rgb}{0.2,.2,.2}
\definecolor{mygray2}{rgb}{.5,.5,.5}
\definecolor{category1}{rgb}{1, 0.9, 0.8}
\definecolor{category2}{rgb}{0.9,1,1}
\definecolor{category3}{rgb}{1, 1, 0.9}
\definecolor{category4}{rgb}{0.9, 0.9, 1}
\definecolor{category5}{rgb}{1, 0.95, 0.9}
\definecolor{category6}{rgb}{0.95, 0.85, 0.9}
\definecolor{category7}{rgb}{0.9, 1, 0.9}
\definecolor{category8}{rgb}{0.85, 0.9, 0.85}
\definecolor{category9}{rgb}{0.75, 0.8, 0.7}

\usepackage{amsmath,graphicx}
\usepackage{amsfonts, amssymb, amsbsy, amscd, bm} 
\usepackage{mathrsfs}
\usepackage{cite}
\usepackage{array}
\usepackage{booktabs}
\usepackage{tikz}
\usepackage{subfigure}
\usepackage{multicol}
\usepackage{multirow}
\usepackage{rotating}
\usepackage{arydshln}
\usepackage{pstricks}
\usepackage{enumerate}
\usepackage{color}
\usepackage{hyperref}
\usepackage{algorithmic}
\usepackage[lined,ruled,linesnumbered]{algorithm2e}
\usepackage{microtype}
\usepackage{booktabs}
\usepackage{float}
\usepackage{siunitx}
\usepackage{cite}

\makeatletter
\def\hlinewd#1{%
  \noalign{\ifnum0=`}\fi\hrule \@height #1 \futurelet
   \reserved@a\@xhline}
\makeatother

\newcommand{\mcrot}[4]{\multicolumn{#1}{#2}{\rlap{\rotatebox{#3}{#4}~}}} 
\newcommand*{\twoelementtable}[3][l]%
{%
    \renewcommand{\arraystretch}{0.8}%
    \begin{tabular}[t]{@{}#1@{}}%
        #2\tabularnewline
        #3%
    \end{tabular}%
}

\definecolor{mygreen1}{rgb}{0, .4, 0}
\definecolor{mygreen2}{rgb}{0, .8, 0}

\graphicspath{{figs/}}

\ifCLASSINFOpdf
\else 
\fi

\begin{document}
%

\title{Encoding Visual Sensitivity by MaxPol Convolution Filters for Image Sharpness Assessment}

%
%
%

\author{Mahdi~S.~Hosseini,~\IEEEmembership{Member,~IEEE,},
        Yueyang~Zhang,~\IEEEmembership{Student member,~IEEE,}
        and~Konstantinos~N.~Plataniotis,~\IEEEmembership{Fellow,~IEEE}
\thanks{Authors are with The Edward S. Rogers Sr. Department of Electrical and Computer Engineering, University of Toronto, Toronto, ON M5S 3G4, Canada e-mail: \url{mahdi.hosseini@mail.utoronto.ca}.

This paper has supplementary downloadable material available at \url{https://github.com/mahdihosseini/HVS-MaxPol} and \url{https://sites.google.com/view/focuspathuoft}, provided by the author. The materials include (a) MATLAB package for No-Reference (NR) Image Sharpness Assessment (ISA) of Natural Images, and (b) FocusPath a benchmark database in digital pathology for validation of image quality assessment of whole slide imaging systems. Contact \url{mahdi.hosseini@mail.utoronto.ca} for further questions about this work.}} 

%
%

\markboth{Accepted for Publication in IEEE Transaction on Image Processing}%
{Shell \MakeLowercase{\textit{et al.}}: Bare Demo of IEEEtran.cls for IEEE Journals}
%



\maketitle

\begin{abstract}
In this paper, we propose a novel design of Human Visual System (HVS) response in a convolutional filter form to decompose meaningful features that are closely tied with image sharpness level. No-reference (NR) Image sharpness assessment (ISA) techniques have emerged as the standard of image quality assessment in diverse imaging applications. Despite their high correlation with subjective scoring, they are challenging for practical considerations due to high computational cost and lack of scalability across different image blurs. We bridge this gap by synthesizing the HVS response as a linear combination of Finite Impulse Response (FIR) derivative filters to boost the falloff of high band frequency magnitudes in natural imaging paradigm. The numerical implementation of the HVS filter is carried out with MaxPol filter library that can be arbitrarily set for any differential orders and cutoff frequencies to balance out the estimation of informative features and noise sensitivities. Utilized by HVS filter, we then design an innovative NR-ISA metric called ``HVS-MaxPol'' that (a) requires minimal computational cost, (b) produce high correlation accuracy with image sharpness level, and (c) scales to assess synthetic and natural image blur. Specifically, the synthetic blur images are constructed by blurring the raw images using Gaussian filter, while natural blur is observed from real-life application such as motion, out-of-focus, luminance contrast, etc. Furthermore, we create a natural benchmark database in digital pathology for validation of image focus quality in whole slide imaging systems called ``FocusPath'' consisting of $864$ blurred images. Thorough experiments are designed to test and validate the efficiency of HVS-MaxPol across different blur databases and state-of-the-art NR-ISA metrics. The experiment result indicates that our metric has the best overall performance with respect to speed, accuracy and scalability.
\end{abstract}
%

%

\section{Introduction}\label{sec:intro}
\IEEEPARstart{I}{mages} of natural scenery in the context of image processing follow an inverse magnitude frequency response proportional to $1/\omega$, with $\omega$ being the spatial frequency. Such decay implies that the magnitude response of low frequencies is more powerful than that of its high band. Therefore, without processing the raw signals, humans may visually lose fine image edges and perceive only coarse information that can lead to blurry observations. However, in the human visual system (HVS), the distribution of frequency information is perceived equally across the whole spectrum. In fact, the HVS introduces a sensitivity response to the input visual contents to compensate the energy loss of high frequency information and make them as important as the low frequencies \cite{field1987relations, brady1995s, field1997visual}. Biologically speaking, the neurons in the humans’ visual cortex are able to automatically tune the amplitude frequencies, balancing out the existing falloff of the high frequency spectrum. After such tuning, the average response remains constant (balanced) across a wide range of frequencies, which enables human beings to visualize any natural objects clearly, regardless of their size and distance.

The HVS could be modeled as a linear convolution process, with the natural scene modeled as an input power function that decays in the frequency domain, the processed signal in human’s visual cortex as another output power function that remains constant in the frequency domain, and HVS as a convolution filter. Under such assumptions, when a human visualizes a natural scene, the input power function of the scene is convolved with a pre-designed convolution filter inspired by this human’s HVS to generate the output power function of the processed image signal
\begin{align*}
I_{\text{O}} \approx I * h_{\text{HVS}},
\end{align*}

where $I_{\text{O}}$ is the output image signal perceived by the human visual cortex, $I$ is the natural image signal and $h_{HVS}$ is the convolution filter simulating the HVS response. To understand how HVS works, this is no more than synthesizing the associated convolution filter, which boosts the power amplitude of the higher frequency to the same level as that of the lower frequency. Under such a model, the sharpness level of an image is obtained by ``deconvolving'' the visual content with HVS filtering process. In discrete computational modeling, many existing non-reference (NR) image sharpness assessment (ISA) approaches make use of this fact to grade the sharpness level of images. Examples include but are not limited to the gradient map approaches which approximate HVS response by finite impulse response (FIR) filters \cite{bahrami2014fast, li2016image, li2016no, li2017no, liu2017quality, li2016no, xue2014gradient}, variational methods \cite{bahrami2014fast, bahrami2016efficient, vu2012bf} and the contrast map techniques \cite{gu2018learning, gvozden2018blind, guan2015no, liu2012image}. However, these existing methods are suboptimal in the sense that they cannot fully manifest the reality of the HVS response. 

In this paper, we aim to synthesize a convolution filter that mimics human visual sensitivity response which can correct back the falloff of natural scenery's amplitude frequency for meaningful feature extraction. These features are directly tied with natural image blur that can be efficiently encoded for sharpness assessment. In particular, we model the falloff of natural image amplitude frequency using the generalized Gaussian (GG) distribution \cite{subbotin1923law}. The shape and scale of the distribution conforms with the decay response of natural images in the frequency domain. To compensate for such falloff, the inverse frequency response of GG is considered to be the representative of the synthesized HVS. For numerical approximation, we fit this model with a linear combination of FIR derivative filters, where its numerical solution is provided by the MaxPol filter library introduced in \cite{HosseiniPltaniotis_MaxPol_TIP_2017, HosseiniPltaniotis_MaxPol_SIAM_2017}. This library is capable of generating various FIR kernels with different order of derivatives and cutoff frequencies. The multiple derivatives can be used to compensate the falloff of the blur images in the frequency domain and make a balanced average response for observation. The cutoff frequency of the filters also introduces a balanced way of keeping informative features in image decomposition and canceling the high band frequencies that are mostly related to noise artifacts. This newly synthesized HVS-like FIR filter provides a unique framework for image decomposition where the blur features are well detected across different image frequencies. Employed by this HVS filter, we then propose our unique framework for image sharpness scoring for NR-ISA development.

\begin{table}[htp]
\renewcommand{\arraystretch}{1.3}
\caption{List of existing NR-ISA metrics categorized into seven approaches: learning-based, gradient map, contrast map, wavelet, phase coherency, luminance map, total variation, and singular value decomposition (SVD).}
\label{table_deblurring_methods}
\centering
\scriptsize
\begin{tabular}{lp{.35cm}p{.5cm}p{.05cm}p{.05cm}p{.05cm}p{.05cm}p{.05cm}p{.05cm}p{.05cm}p{.05cm}p{.05cm}p{.05cm}p{.05cm}}
\hlinewd{1.5pt}
Author & Year &
\mcrot{1}{c}{45}{\textcolor{mygray1}{Learning Algorithm}} &
\mcrot{1}{c}{45}{\textcolor{mygray1}{Gradient Map}} &
\mcrot{1}{c}{45}{\textcolor{mygray1}{Fourier Transform}} &
\mcrot{1}{c}{45}{\textcolor{mygray1}{Contrast Map}} &
\mcrot{1}{c}{45}{\textcolor{mygray1}{Wavelet Domain}} &
\mcrot{1}{c}{45}{\textcolor{mygray1}{Phase Coherence}} &
\mcrot{1}{c}{45}{\textcolor{mygray1}{Luminance Map}} &
\mcrot{1}{c}{45}{\textcolor{mygray1}{Color Variation}} &
\mcrot{1}{c}{45}{\textcolor{mygray1}{Local/Total Variation}} &
\mcrot{1}{c}{45}{\textcolor{mygray1}{Spectrum Map}} &
\mcrot{1}{c}{45}{\textcolor{mygray1}{SVD}} &
\mcrot{1}{c}{45}{\textcolor{mygray1}{Edge}}
\\ \hlinewd{1.5pt}

\hlinewd{.75pt}
\rowcolor{category1}
Yang \cite{yang2018assessing} & $2018$ & DNN & & & & & & & & & & & \\

\rowcolor{category1}
Gu \cite{gu2018learning} & $2018$ & Regr. & & & $\textcolor{mygray1}{\bullet}$ & & & $\textcolor{mygray1}{\bullet}$ & $\textcolor{mygray1}{\bullet}$ & & & &
\\ 

\rowcolor{category1}
Yu \cite{yu2017shallow} & $2017$ & CNN & & & & & & & & & & &
\\

\rowcolor{category1}
Yu \cite{yu2016cnn} & $2017$ & CNN & & & & & & & & & & &
\\ 

\rowcolor{category1}
Li \cite{li2017no} & $2017$ & SVR & $\textcolor{mygray1}{\bullet}$ & & & & & & & & $\textcolor{mygray1}{\bullet}$ & $\textcolor{mygray1}{\bullet}$ &
\\

\rowcolor{category1}
Li \cite{li2016image} & $2016$ & Dict. & $\textcolor{mygray1}{\bullet}$ & & & & & & & & & &
\\

\rowcolor{category1}
Gu \cite{gu2015no} & $2015$ & Regr. & & & & & & & & & & &
\\ 


\rowcolor{category1}
Mavridaki \cite{mavridaki2014no} & $2014$ & SVM & & $\textcolor{mygray1}{\bullet}$ & & & & & & & & &
\\

\rowcolor{category1}
Ye \cite{ye2012unsupervised} & $2012$ & KMean & & & & & & & & & & &
\\ \hlinewd{.75pt}

\rowcolor{category2}
Hosseini \cite{mahdi2018image} & $2018$ & & $\textcolor{mygray1}{\bullet}$ & & & & & & & & & &
\\

\rowcolor{category2}
Liu \cite{liu2017quality} & $2017$ & & $\textcolor{mygray1}{\bullet}$ & & & & $\textcolor{mygray1}{\bullet}$ & & & & $\textcolor{mygray1}{\bullet}$ & &
\\

\rowcolor{category2}
Li \cite{li2016no} & $2016$ & & $\textcolor{mygray1}{\bullet}$ & & & & & & & & & &
\\

\rowcolor{category2}
Bahrami \cite{bahrami2014fast} & $2014$ & & $\textcolor{mygray1}{\bullet}$ & & & & & & & $\textcolor{mygray1}{\bullet}$ & & &
\\ \hlinewd{.75pt}

\rowcolor{category3}
Gvozden \cite{gvozden2018blind} & $2017$ & & & & $\textcolor{mygray1}{\bullet}$ & $\textcolor{mygray1}{\bullet}$ & & & & & & &
\\

\rowcolor{category3}
Guan \cite{guan2015no} & $2015$ & & & & $\textcolor{mygray1}{\bullet}$ & & & & & & & &
\\

\rowcolor{category3}
Liu \cite{liu2012image} & $2012$ & & & & $\textcolor{mygray1}{\bullet}$ & & & $\textcolor{mygray1}{\bullet}$ & & & & &
\\ \hlinewd{.75pt}

\rowcolor{category4}
Vu \cite{vu2012fast} & $2012$ & & & & & $\textcolor{mygray1}{\bullet}$ & & & & & & &
\\
\hlinewd{.75pt}
Ninassi \cite{ninassi2008performance} & $2008$ & & & & & $\textcolor{mygray1}{\bullet}$ & & & & & & &
\\ 
Ferzli \cite{ferzli2005no} & $2005$ & & & & & $\textcolor{mygray1}{\bullet}$ & & & & & & &
\\
\hlinewd{.75pt}

\rowcolor{category5}
Leclaire \cite{leclaire2015no} & $2015$ & & & & & & $\textcolor{mygray1}{\bullet}$ & & & & & &
\\ 

\rowcolor{category5}
Hassen \cite{hassen2013image} & $2013$ & & & & & & $\textcolor{mygray1}{\bullet}$ & & & & & &
\\ \hlinewd{.75pt}

\rowcolor{category6}
Lee \cite{lee2016toward} & $2016$ & & & & & & &  $\textcolor{mygray1}{\bullet}$& $\textcolor{mygray1}{\bullet}$ & & & &
\\ \hlinewd{.75pt}

\rowcolor{category7}
Bahrami \cite{bahrami2016efficient} & $2016$ & & & & & & & & & $\textcolor{mygray1}{\bullet}$ & & &
\\

\rowcolor{category7}
Vu \cite{vu2012bf} & $2012$ & & & & & & & & & $\textcolor{mygray1}{\bullet}$ & $\textcolor{mygray1}{\bullet}$ & &
\\ \hlinewd{.75pt}

\rowcolor{category8}
Sang \cite{sang2014no} & $2014$ & & & & & & & & & & & $\textcolor{mygray1}{\bullet}$ &
\\ \hlinewd{.75pt}

\rowcolor{category9}
Ferzli \cite{ferzli2009no} & $2009$ & & & & & & & & & & & & $\textcolor{mygray1}{\bullet}$ \\
\rowcolor{category9}
Sadaka \cite{sadaka2008no} & $2008$ & & & & & & & & & & & & $\textcolor{mygray1}{\bullet}$ \\
\rowcolor{category9}
Ferzli \cite{ferzli2007no} & $2007$ & & & & & & & & & & & & $\textcolor{mygray1}{\bullet}$ \\
\rowcolor{category9}
Ferzli \cite{ferzli2006human} & $2006$ & & & & & & & & & & & & $\textcolor{mygray1}{\bullet}$ \\
\rowcolor{category9}
Marziliano \cite{marziliano2002no} & $2002$ & & & & & & & & & & & & $\textcolor{mygray1}{\bullet}$ \\
\hlinewd{.75pt}
\end{tabular}
\end{table}

\subsection{Related Work in Blind Image Sharpness Assessment}\label{sec:relate}
NR-ISA has recently emerged as a special branch of image quality assessment (IQA) for blindly grading the blur or sharpness level of images. An overview of these methods is listed in Table I, where we divide the methods into nine different categories. Metrics based on gradient domain maps include maximum local variation \cite{bahrami2014fast}, energy concentration in image patches \cite{li2016no}, and local saliency map \cite{liu2017quality}. Contrast map is widely used to develop a NR-ISA metric. Other methods combine brightness and color information \cite{gu2018learning}, transform contrast values of pixel into the wavelet domain \cite{gvozden2018blind}, apply the just noticeable blur metric (JNBM) to the pre-calculated contrast map \cite{guan2015no}, and merge the contrast map with structural change and luminance distortion \cite{liu2012image}. Developments based on the wavelet domain have also been studied to decompose the image into several wavelet subbands to compute the log-energy of the decomposed features at pixel level for assisting sharpness evaluation \cite{vu2012fast}, or to compute the absolute values of vertical and horizontal bands \cite{ferzli2005no}. The phase domain is also commonly used to develop NR-ISA. The theoretical development and relationship between phase domain and sharpness index is studied by Leclaire et al \cite{leclaire2015no}. Furthermore, Hassen et al \cite{hassen2013image} compute the phase at each location of the image, called Local Phase Coherence (LPC), which is further defined as a sharpness score. Variational maps are also reported as a useful feature. The map has been used in several methods of NR metric development in conjunction with other feature methods \cite{lee2016toward, gu2018learning}. The transfer image domain into total variation and local variation for image analysis is another example to develop NR-ISA metric \cite{bahrami2016efficient, bahrami2014fast}. The spectral and spatial domains is used in \cite{vu2012bf} based on the local maximum total variation domain to construct a third map for sharpness scoring. Metrics based on singular value decomposition to evaluate the image sharpness is also used in \cite{sang2014no}. The dictionary based method to learn sparse representation is also used in \cite{li2016image} to obtain sparse coefficients that are closely tied with image sharpness. Finally the k-mean algorithm is also used in \cite{ye2012unsupervised} to extracted sharpness features. Similar studies to encode HVS for blur image assessment have been also made. Marziliano et al \cite{marziliano2002no} proposed a perception based metric by detecting and analyzing the edge response of a given image. In \cite{ferzli2006human}, a HVS based Mean Just-Noticeable Blur (MJNB) model is defined to weight the detected edge response for sharpness scoring. Furthermore, a probability summation model is introduced in \cite{ferzli2007no,ferzli2009no} to assist JNB analysis as a NR-ISA metric. Based on the fact that humans’ judgment of blur is influence by the image distortion in salient regions, Sadaka et al \cite{sadaka2008no} propose to construct a salient map, upon which the probability summation and JNB model are applied. Also, Discrete Wavelet Transform can be a useful tool to simulate HVS by decomposing images into perceptual bands and using masking functions to remove the masking effect \cite{ninassi2008performance}. In recent years, various machine learning algorithms have been applied to the sharpness analysis of images. One common tool is the regression model in \cite{gu2018learning, gu2015no} which performs regression on previously extracted features to approximate sharpness score. Another strong learning algorithm is support vector regression (SVR) in \cite{li2017no, yu2017shallow} that passes the previously obtained features into convolution neural network (CNN) models for sharpness learning. The CNN model is usually trained to detect the strength of image edges. The application of NR-ISA metric development has been also studied recently in digital pathology application \cite{yang2018assessing}.

\subsection{Contributions}
The NR-ISA metric has high potential for different applications of the image acquisition pipeline, such as quality check control of high-throughput scanning solutions in digital pathology, astronomy, consumer imaging devices such as camera, smart-phones, computing-pads, etc. Owing to the limitations of computational speed, the majority of existing methods lag behind the real-time analysis (known as on-the-fly processing) which is mandated by aforementioned imaging platforms. After experimenting several recent NR-ISA methods on various databases, we observed that majority of the methods such as in \cite{vu2012bf, gu2015no, li2016image, li2017no} produce acceptable correlation accuracy with respect to subjective quality scoring. However, they are time consuming in terms of computational speed. In contrast, methods such as in \cite{bahrami2014fast, leclaire2015no, mahdi2018image} consume much less time for computation, but they are prone to errors dealing with diverse natural bluring types. In this context, there is a high demand for a NR-ISA metric that is fast and performs with high accuracy. The following lists the contributions made in this paper

\begin{itemize}
\item Propose an effective approach for modeling a HVS-like FIR filter that mimics the visual sensitivity response, where the numerical framework is accommodated by the MaxPol library filter solution.
\item Implement a novel NR-ISA metric called ``HVS-MaxPol'' which uses the newly synthesized HVS-filter to extract meaningful features that are closely tied with human visual blur perception.
\item Construct a medical imaging database of digital pathology for natural image sharpness assessment, called ``FocusPath''.
\item Perform comprehensive comparisons over diverse databases and state-of-the-art NR-ISA metrics in terms of statistical performance accuracy, computational complexity and scalability.
\end{itemize}

Our early related work \cite{mahdi2018image} proposes a NR-ISA metric called Synthetic-MaxPol. This metric directly replaces HVS by the first and the third order derivatives for rough approximation of HVS. In this paper we take a quite different approach by mimicing the boosting energy of HVS on the band-pass frequency. This is done by considering the HVS filter as a series of frequency polynomials that are equivalent to the superposition of multiple derivative operators in the spatial domain. The coefficients of each derivative operator is defined by fitting this superposition model into the inverse falloff of the amplitude frequency of the natural images. Here, we model such falloff using the General Gaussian (GG) distribution as a closed-form solution to approximate the unkown coefficients of high order derivatives.

The remainder of the paper is as follows. We design HVS convolution filter in Section \ref{sec_HVS_MaxPol} and utilize the filter in Section \ref{IQA_measure} for NR-ISA metric development. Section \ref{section_digital_pathology} introduces a new digital pathology database for NR-ISA benchmarking in natural medical imaging. The experiments are provided in Section \ref{sec_experiment} and the paper is concluded in Section \ref{conclusion}.


\section{Convolutional Filter Design of Visual Sensitivity Model}\label{sec_HVS_MaxPol}
\begin{figure}[htp]
\centerline{
{\includegraphics[width=0.3\textwidth]{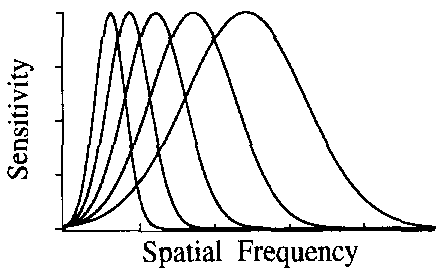}}
}\vspace{-.05in}
\centering
\caption{Model of spatial frequency sensitivity with maximum frequency range of is $32$ cycles/deg, curtsey of Field et al. \cite{field1987relations, field1997visual}.}
\label{spatial_frequency_sensitivity_model}
\end{figure}

In this section, we design a new convolutional filter that numerically synthesizes the visual sensitivity model for blur perception. The target applications considered in this paper cover divers blurring types such as point-spread-function (PSF), out-of-focus, motion, haze, luminance, and atmospheric turbulence, that all manifest themselves in natural imaging paradigm \cite{ciancio2011no, virtanen2015cid2013}. Here, we are keen in first place to study how human visual system would respond into blur effect in general? This interest stems from the fact that it is us as ``human beings'' whom grade the image sharpness for subjective evaluations. Therefore, it is critical to understand that how HVS perceives blur? In other words, how HVS respond to visual blur for quality judgment. In fact, this problem has been well studied in psychology to model visual sensitivity  \cite{field1987relations, brady1995s, field1997visual}. In Figure \ref{spatial_frequency_sensitivity_model} we show the exact replica of this model adopted from \cite{field1987relations, field1997visual} which demonstrates the model of spatial frequency response of the human visual sensitivity.

As it shown, the frequency sensitivity behaves similar to a pass-band filter which dramatically boosts mid-frequencies and then attenuates the higher-band (aka stop-band). It worth noting that this sensitivity function also conforms with the contrast sensitivity function proposed in \cite{daly1992visible} which yields a band-pass filter to encode visual sensitivity. Here, we propose that such dramatic rising can be fitted by a series of frequency polynomials of even orders
\begin{align}
\hat{h}_{\text{HVS}}(\omega)\equiv\sum\limits^{N}_{n=1} c_n \hat{d}_{2n}(\omega) =  \sum\limits^{N}_{n=1} (-1)^n c_n \omega^{2n}.\label{eq_HVS_3}
\end{align} 

The frequency polynomials $(i\omega)^{2n}$ are equivalent to the derivative operators in the spatial domain. The superposition of multiple frequency polynomials which fits the boosting effect, is equivalent to superposition of the high order derivative operators in the spatial domain
\begin{align}
h_{\text{HVS}}(x)\equiv c_1d_2(x) + c_2d_4(x) + \hdots + c_N d_{2N}(x). \label{eq_HVS_2}
\end{align} 
where $d_{2n}(x)$ is the $2n$th derivative operator. Note that the Fourier transform of the even derivative operator is $\hat{d}_{2n}(\omega)=(i\omega)^{2n}=(-1)^n\omega^{2n}$.

\begin{figure}[htp]
\centerline{
\subfigure[$h_{GG}(x)$]{\includegraphics[height=0.16\textwidth]{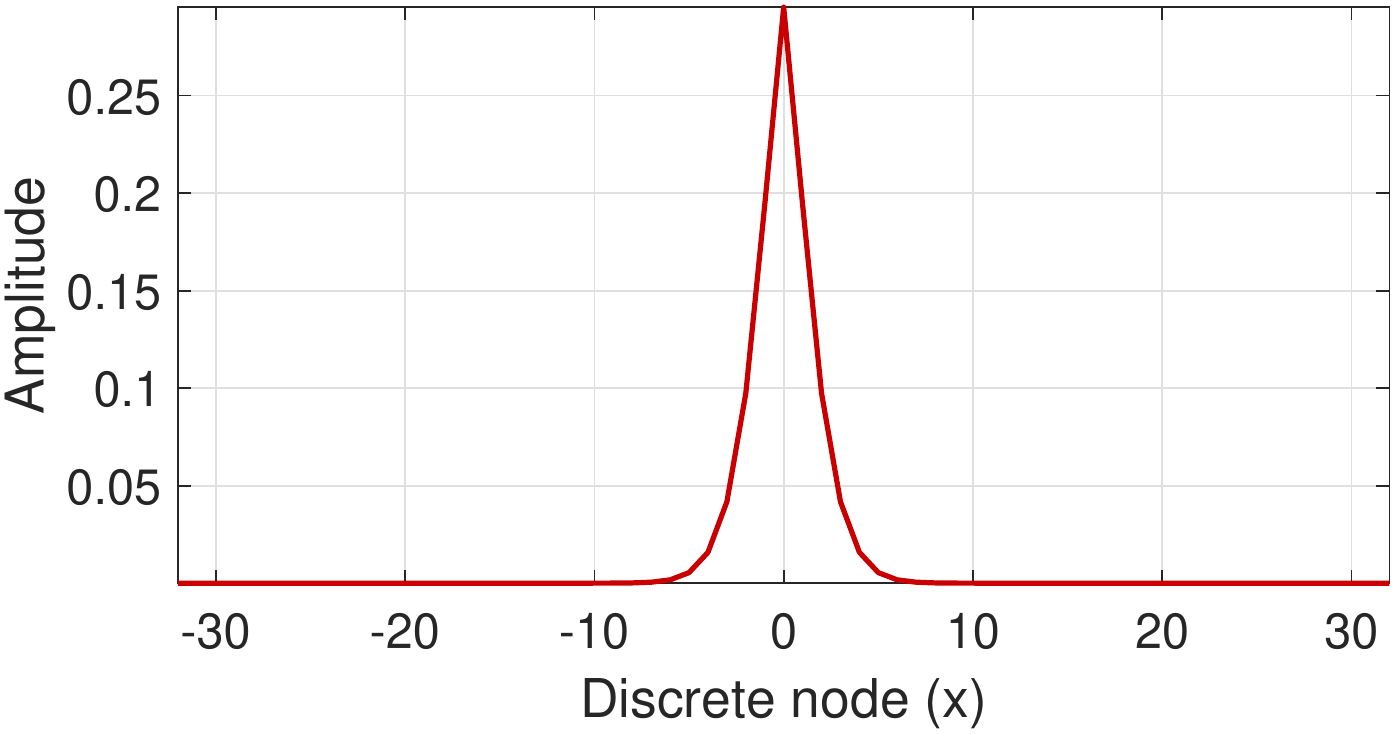}}
}
\centerline{
\subfigure[$\hat{h}_{GG}(\omega)$]{\includegraphics[height=0.16\textwidth]{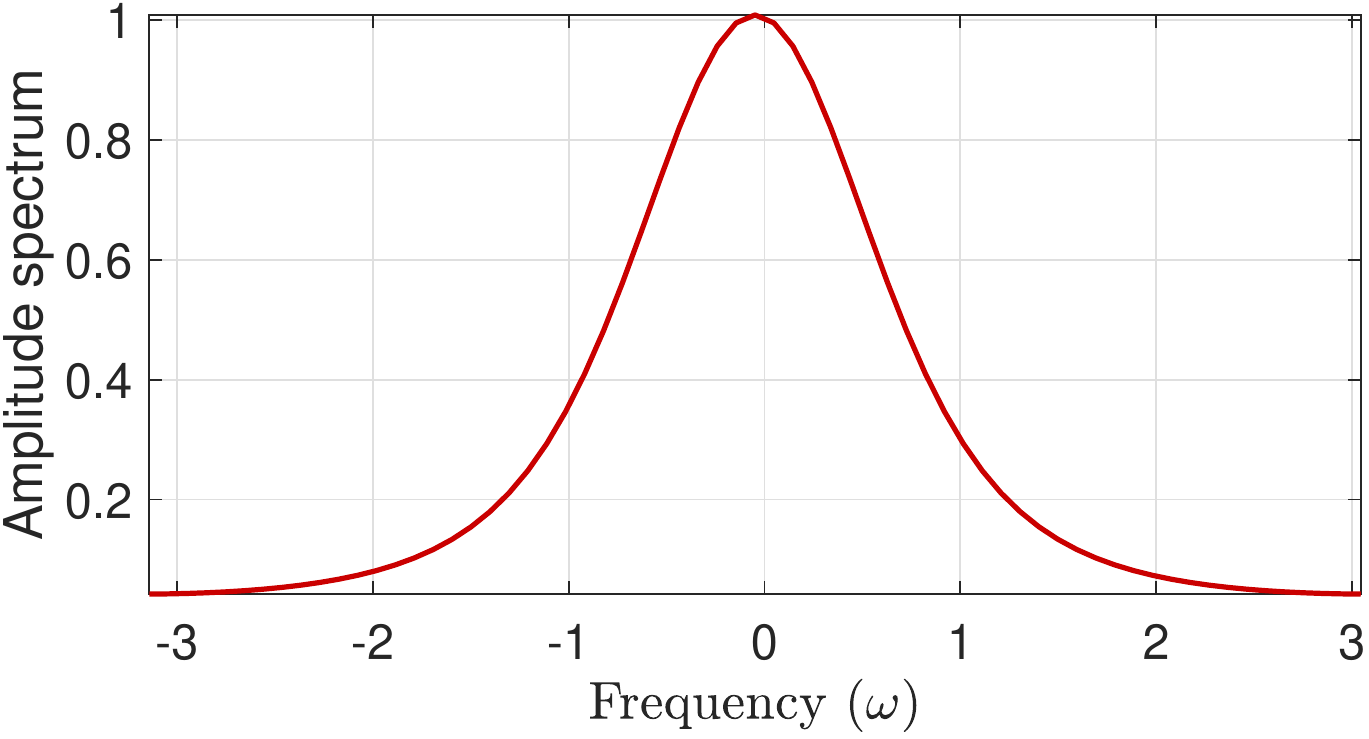}}
}
\centerline{
\subfigure[$h_{\text{HVS}}(x)$]{\includegraphics[height=0.16\textwidth]{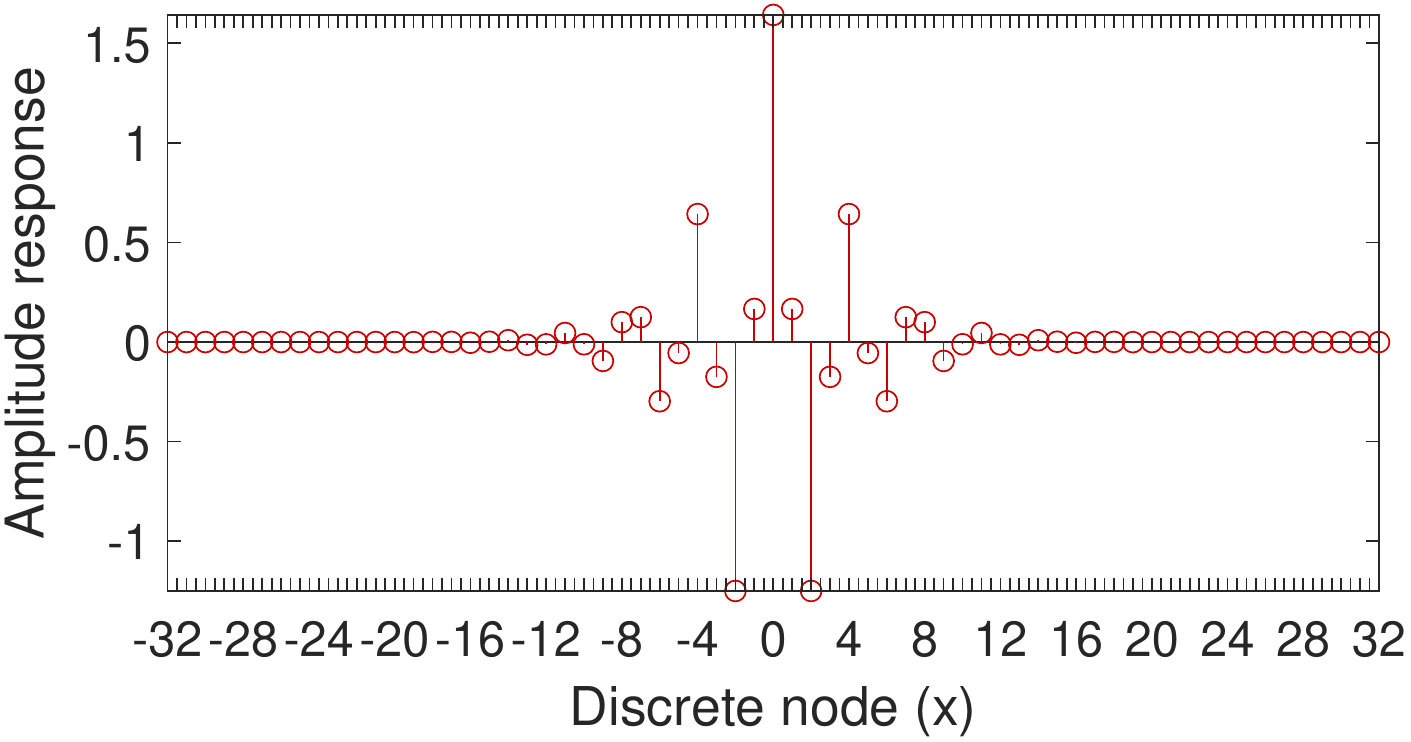}}
}
\centerline{
\subfigure[$\hat{h}_{\text{HVS}}(\omega)$]{\includegraphics[height=0.16\textwidth]{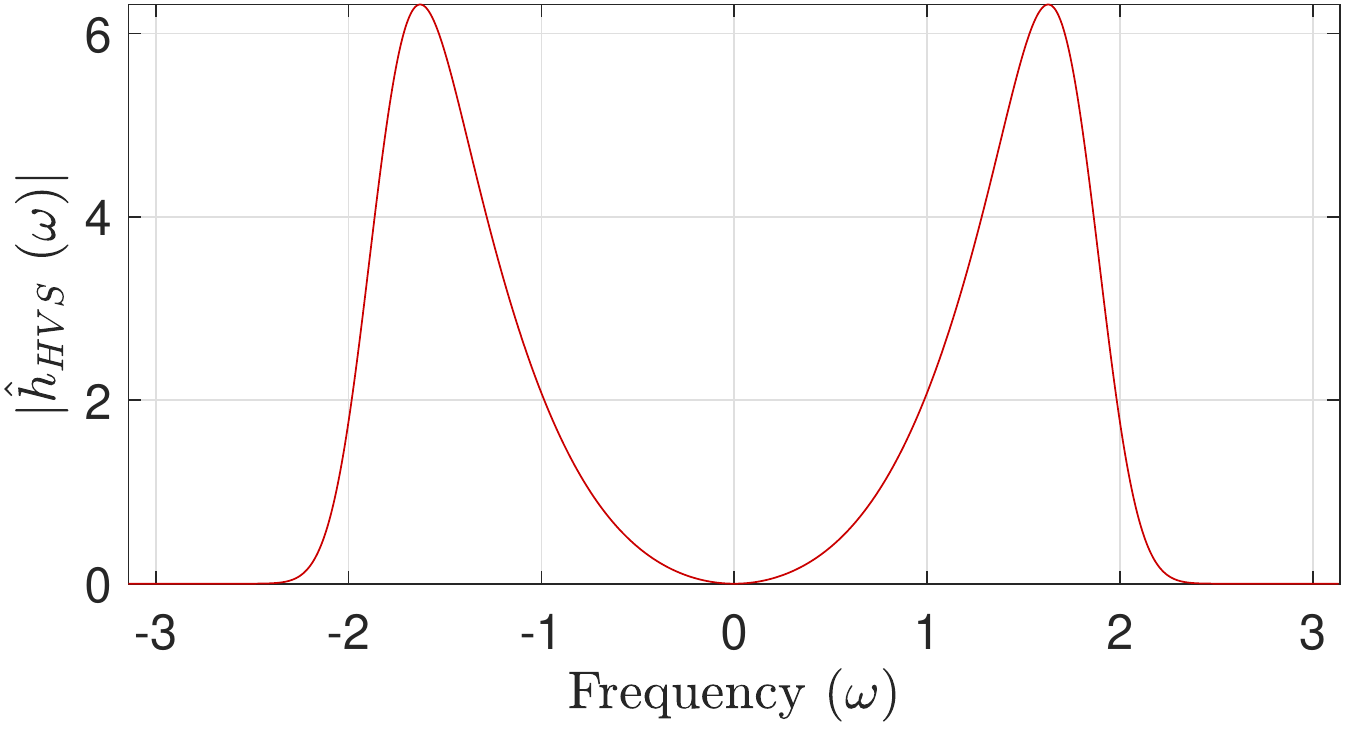}}
}
\centering
\caption{Example plot of a generalized Gaussian distribution in the (a) spatial and (b) frequency domains. Here $\alpha=1.7$ and $\beta=1.4$. The approximated FIR filter of the HVS is shown by its (c) impulse response, and (d) frequency response. The cutoff design is $\omega_c\approx 0.6\pi$}
\label{generalized_Gaussian_plots}
\end{figure}

The objective here is to design a finite impulse response (FIR) kernel $h_{\text{HVS}}$, where its convolution with natural images yields equal average response in amplitude frequency domain
\begin{align}
h_{\text{GG}}(x) \ast h_{\text{HVS}}(x) = \delta(x). \label{eq_HVS_1}
\end{align}
where, $h_{\text{GG}}(x)$ is a lowpass-kind filter that mimics the falloff of the natural image amplitude frequencies. As mentioned earlier, the close approximation such falloff is $\omega^{-\gamma}$, where $\gamma\approx 1$ \cite{field1987relations, brady1995s, field1997visual}. We propose a closed form solution to such falloff approximation in natural images using the generalized Gaussian distribution \cite{subbotin1923law}
\begin{align}\label{blur_model_eq_1}
h_{GG}(x) = \frac{1}{2\Gamma(1+1/\beta)A(\beta,\alpha)}\exp{-\Big{\vert}\frac{x}{A(\beta,\alpha)}\Big{\vert}^\beta}
\end{align}
where $\beta$ defines the \textit{shape} of the distribution function, $A(\beta,\alpha) = \left(\alpha^2\Gamma(1/\beta)/\Gamma(3/\beta)\right)^{1/2}$ is the scaling parameter, and $\Gamma(\cdot)$ is the Gamma function $\Gamma(z) = \int^{\infty}_{0}e^{-t}t^{z-1}dt, \forall z>0$. For instance, the standard Gaussian distribution, i.e. second order GG, is a variant of this model, where $\beta=2$ and $A(2, \alpha)$ \cite{subbotin1923law, kotz2012laplace}. An example of the GG filter kernel is shown in Figure \ref{generalized_Gaussian_plots} for  preset scale and shape parameters. The frequency spectrum of the filter (in the same figure) demonstrates the falloff of the frequency on the high frequency band, a property considered for natural images.

An equivalent representation in (\ref{eq_HVS_1}) is that the frequency response of HVS filter has an inverse relation to the natural image frequency model, i.e. $\hat{h}_{\text{HVS}}(\omega)=\hat{h}^{-1}_{\text{GG}}(\omega)$. Therefore, the unknown coefficients $c_n$ are approximated by fitting the model design in (\ref{eq_HVS_3}) to the inverse response $\hat{h}^{-1}_{\text{GG}}(\omega)$ to minimize the fitting error
\begin{align}
c_n\leftarrow\arg\min_{c_n}\|\hat{h}_{\text{HVS}}(\omega) - \hat{h}^{-1}_{\text{GG}}(\omega)\|^2_2
\label{eq_HVS_5}
\end{align}
which can be solved using the non-linear least square in \cite{kelley1999iterative}. In particular, we consider the lowpass design
\begin{align}
\hat{h}_{\text{HVS}}(\omega)=
\left\{\begin{array}{ll}
\sum\limits^{N}_{n=1} (-1)^n c_n \omega^{2n}, & 0\leq\omega\leq\omega_c \\
0, & \omega \geq \omega_c
\end{array}\right.
\label{eq_HVS_4}
\end{align}
where $\omega_c$ is the cutoff frequency. Once the unknown coefficients are approximate in (\ref{eq_HVS_5}), the HVS filter is constructed by the superposition of derivatives in (\ref{eq_HVS_2}). For numerical approximation of the the lowpass derivative filters in (\ref{eq_HVS_4}), we used MaxPol\footnote{\url{https://github.com/mahdihosseini/MaxPol}}, a package to solve numerical differentiation \cite{HosseiniPltaniotis_MaxPol_TIP_2017, HosseiniPltaniotis_MaxPol_SIAM_2017}. Each derivative operator $d_{2n}(x)$ can be approximated using different orders of polynomials i.e. filter tap length and cutoff frequency to meet the lowpass criterion defined in (\ref{eq_HVS_4}). Figure \ref{generalized_Gaussian_plots}.(c)-(d) demonstrates an example of the impulse and frequency responses of the HVS filter design.
\section{No-Reference Image Sharpness Assessment} \label{IQA_measure}
In this section, we elaborate on the different steps used to build the image sharpness metric from a digital image. Note that we avoid using color information by converting all color images into grayscale image for processing. The proposed algorithms consists of four main operations discussed in subsequent sections.

\subsection{Background Check}
We design a background check condition to exclude the image pixels with dark values. We set a simple hard threshold to exclude all pixels with normalized grayscale values smaller than $0.05$. This is because the HVS can hardly extract useful information from an extremely dark environment for interpretation. Note that this step only excludes the extreme dark backgrounds and does not have any influence in most cases.

\subsection{MaxPol Variational Decomposition}
We decompose the grayscale image $I\in\mathbb{R}^{N_1\times N_2}$ using the HVS filter $h_{\text{HVS}}(x)$ designed in Section \ref{sec_HVS_MaxPol} along the horizontal and vertical axes 
\begin{align}
[\nabla_{\text{HVS}}{I}(x), \nabla_{\text{HVS}}{I}(y)] = [{I}\ast{h_{\text{HVS}}}^T, {I}\ast{h_{\text{HVS}}}]
\label{equation_2}
\end{align}

It is worth noting that the selection of the cutoff band $\omega_c$ in designing HVS filter is usually set such that to avoid noise amplification on the decomposed images. Majority of the natural images studied in our experiments contain high signal-to-noise-ratio (SNR). Therefore, we empirically set the cutoff frequency to extract meaningful features from wide image frequency band to mitigate the information loss. Figure \ref{procedure_demo}.(b)-(c) demonstrates the horizontal and vertical decomposition of an image shown in Figure \ref{procedure_demo}.a (we used only grayscale for decomposition). While the kernel is highly sensitive on sharp edges, it avoids noise amplification on smooth areas. The distribution of the decomposed features using the HVS filter is shown in Figure \ref{generalized_Gaussian_and_sigmoid_plots} for both horizontal and vertical features.

\subsection{Rectified Linear Unit}
\begin{figure}[htp]
\centerline{
\subfigure[Image ${I}$]{\includegraphics[height=0.18\textwidth]{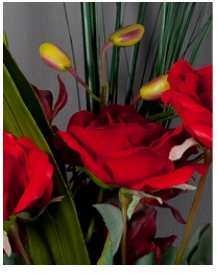}}\hspace{.01in}
\subfigure[${I}\ast{h_{\text{HVS}}}$]{\includegraphics[height=0.18\textwidth]{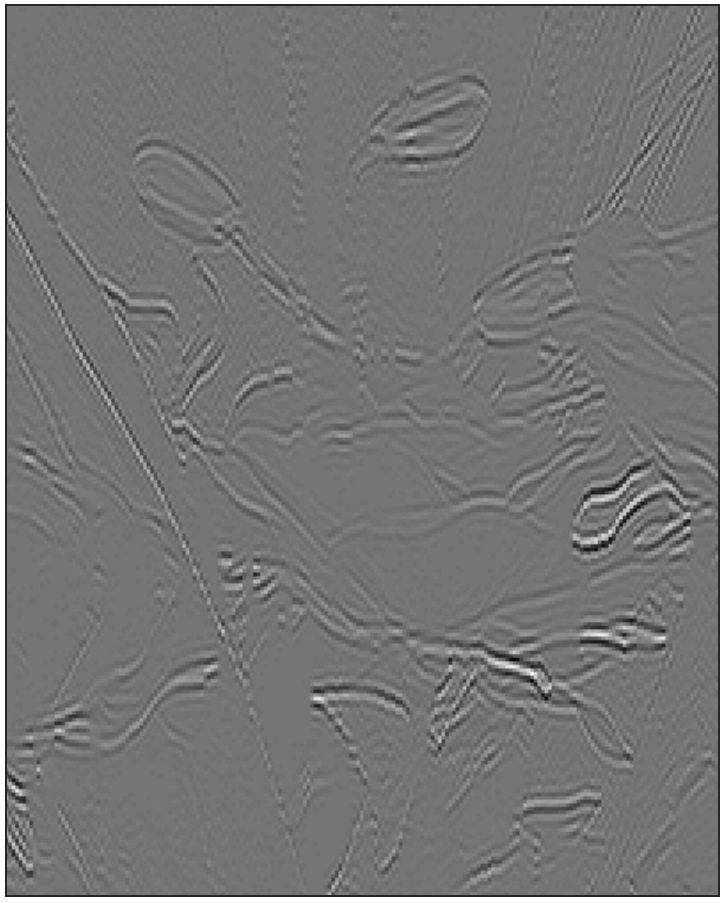}}\hspace{.01in}
\subfigure[${I}\ast{h_{\text{HVS}}}^T$]{\includegraphics[height=0.18\textwidth]{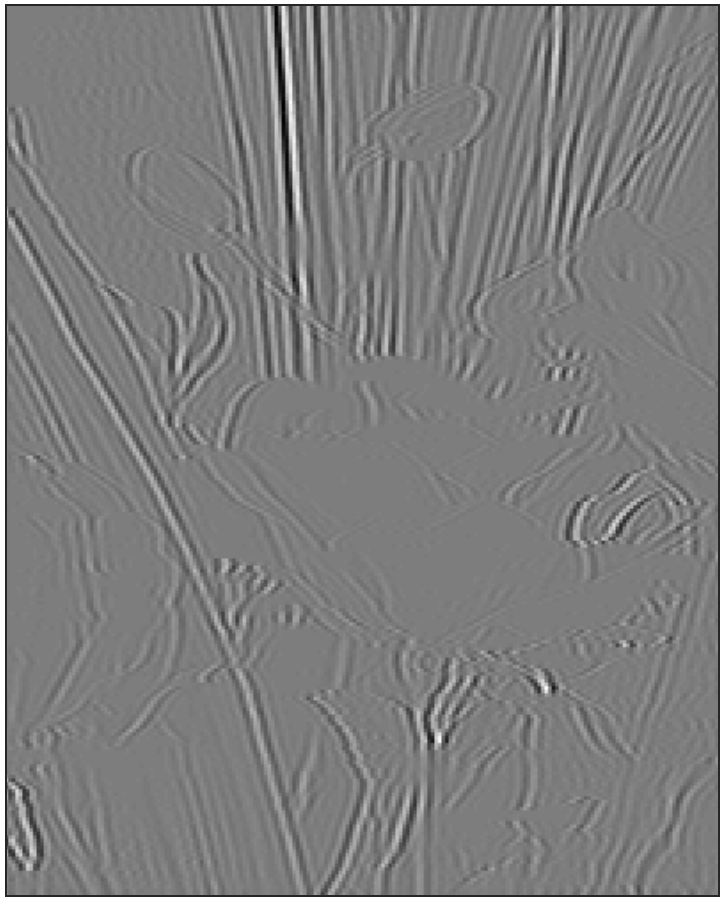}}
}\vspace{.1in}
\centerline{
\subfigure[$R(\nabla_{\text{HVS}}{I}(x))$]{\includegraphics[height=0.18\textwidth]{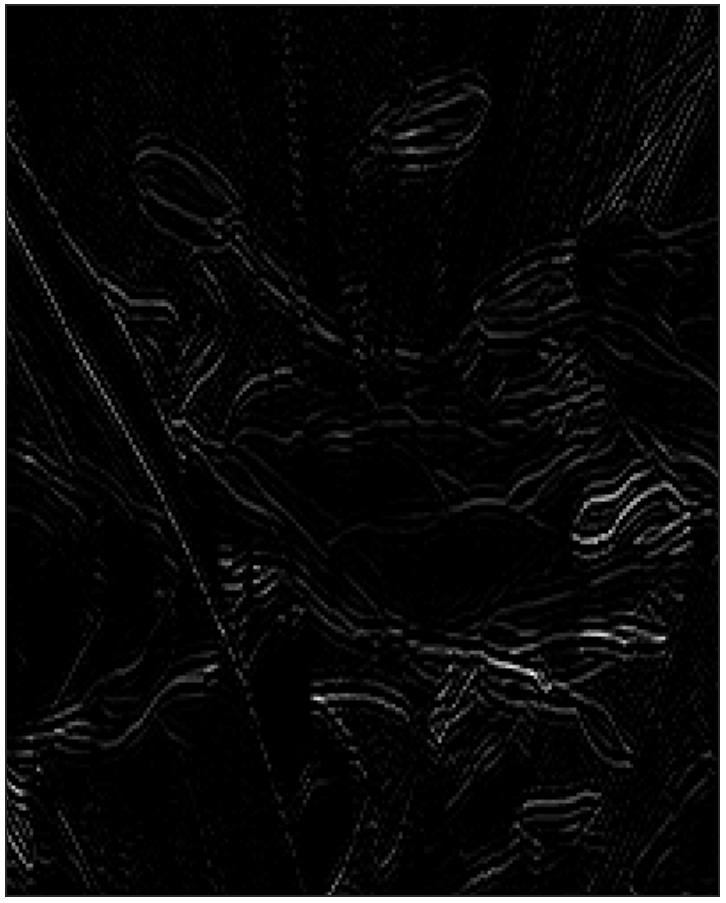}}\hspace{.01in}
\subfigure[$R(\nabla_{\text{HVS}}{I}(y))$]{\includegraphics[height=0.18\textwidth]{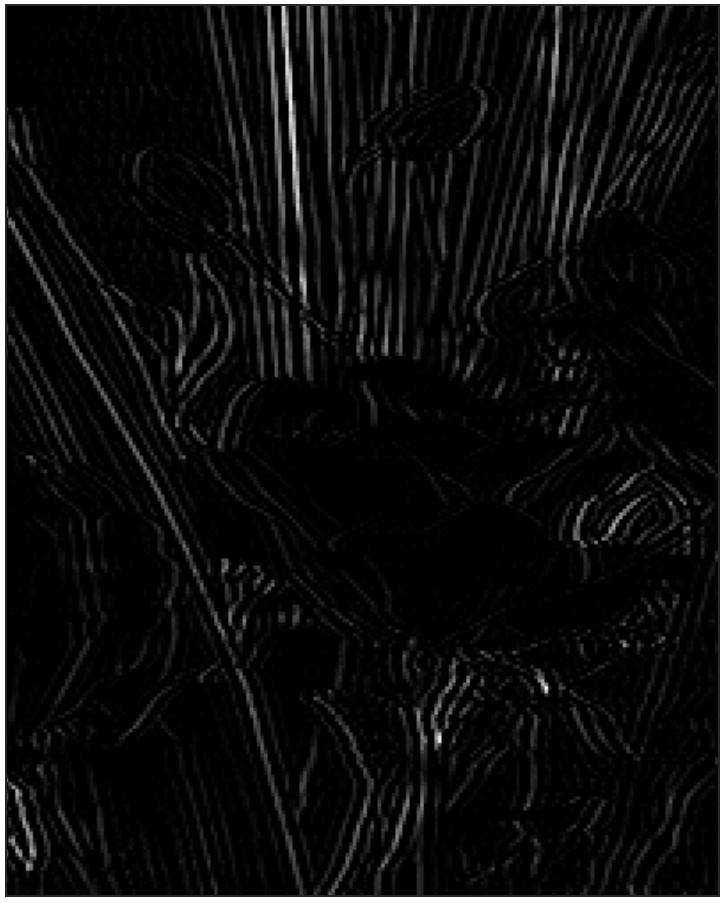}}\hspace{.01in}
\subfigure[$M_{\text{HVS}}$]{\includegraphics[height=0.18\textwidth]{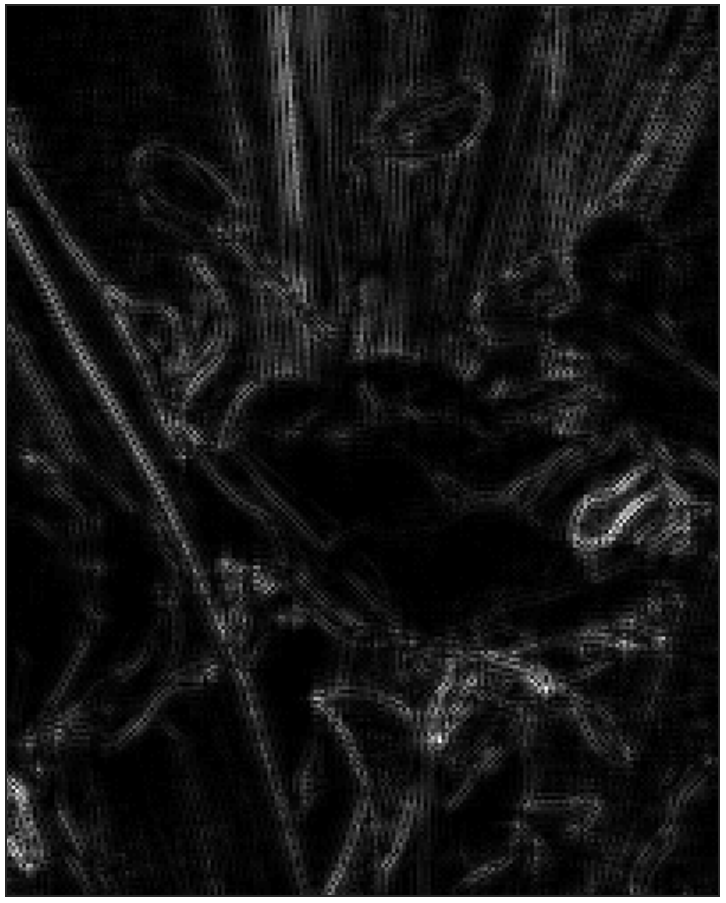}}
}\vspace{.1in}
\centering
\caption{Demo images after each single operation from the HVS-MaxPol NR-ISA metric. (a) is the original demo image. (b) is the horizontal decomposition of the image using the HVS-MaxPol kernel. (c) is the vertical decomposition of the image using the HVS-MaxPol kernel. (d) is the activated horizontal decomposition using ReLU. (e) is the activated vertical decomposition using ReLU. (f) is the feature map.}
\label{procedure_demo}
\end{figure}

The HVS filter is a symmetric FIR kernel and provides mostly redundant features on one side of the histogram distribution. To this end, we select features activating beyond zero similar to rectified linear unit (ReLu) function after the convolution operation
\begin{align}
R(x) = \max(x, 0).
\end{align}
Figure \ref{procedure_demo}.(e)-(f) demonstrates the activated features beyond zero after decomposition. Figure \ref{procedure_demo}.(e)-(f) demonstrates the activated profiles using the decomposed features in Figure \ref{procedure_demo}.(b)-(c). After obtaining the feature vector, we define two operations in parallel in order to select meaningful pixels for blur analysis.

\subsection{Operation 1: Feature Map}
We construct the feature map using the vector decomposition $\nabla_{\text{HVS}}{I}(x)$ and $\nabla_{\text{HVS}}{I}(y)$ defined by (\ref{equation_2}) in the $\ell_{\frac{1}{2}}$--norm space as follows
\begin{align}
M_{\text{HVS}} = \left(|R(\nabla_{\text{HVS}}{I}(x))|^{\frac{1}{2}}+|R(\nabla_{\text{HVS}}{I}(y))|^{\frac{1}{2}}\right)^{2}.
\label{equation_3}
\end{align}
The feature map $M_{\text{HVS}}$ encodes the edge significance in $\ell_{\frac{1}{2}}$--norm space. This promotes the sparsity of decomposed coefficients by weakening minor perturbations while preserving significant coefficients related to image edges. Figure \ref{procedure_demo}.d demonstrates the feature map using the activated profiles in Figure \ref{procedure_demo}.(e)-(f). The histogram distribution of the decomposed features along horizontal and vertical axes are also shown in Figure \ref{generalized_Gaussian_and_sigmoid_plots}.a.

\subsection{Operation 2: Pixel Selection}
\begin{figure}[htp]
\centerline{
\subfigure[Histogram bin plot]{\includegraphics[height=0.17\textwidth]{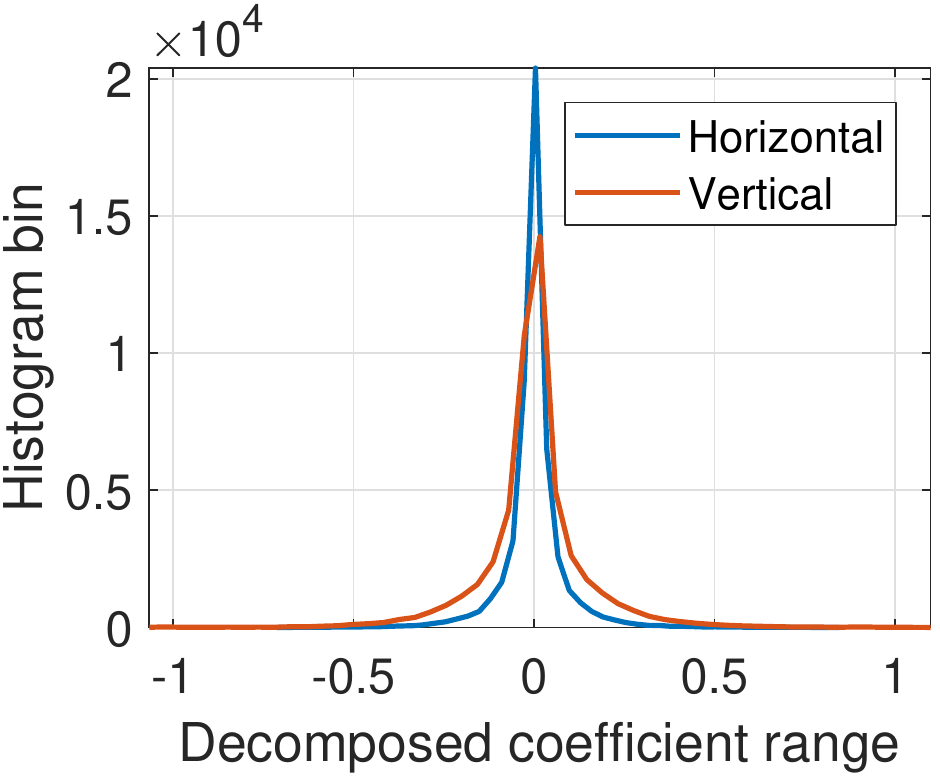}}\hspace{.1in}
\subfigure[Nonlinear mapping]{\includegraphics[height=0.15\textwidth]{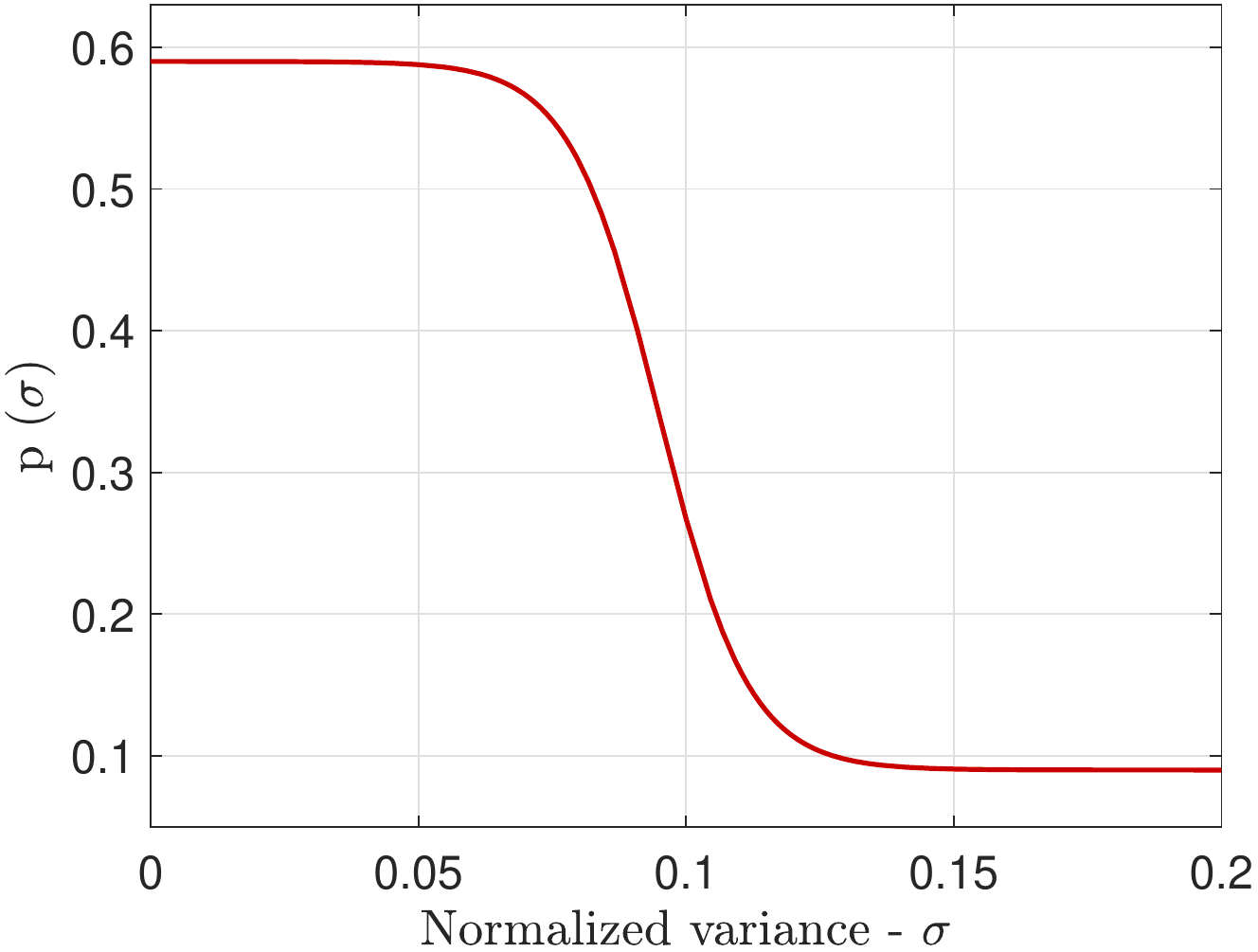}}
}
\centering
\caption{(a) is the plot of histogram bins of horizontal and vertical decomposition maps using the HVS-MaxPol kernel. (b) is the nonlinear mapping (sigmoid-like) function on normalized variance.}
\label{generalized_Gaussian_and_sigmoid_plots}
\end{figure}

To determine the number of pixels that will be reserved in the feature map $M_{\text{HVS}}$, we design a non-linear projection function in \ref{equation_5}. The input variable $\sigma$ is the 95th percentile of cumulative distribution function of the feature vector $[R(\nabla_{\text{HVS}}{I}(x)), R(\nabla_{\text{HVS}}{I}(y))]$ (i.e. $\sigma = CDF(V, 95\%)$, where $V = [R(\nabla_{\text{HVS}}{I}(x)), R(\nabla_{\text{HVS}}{I}(y))]$). The output $p(\sigma)$ refers to the ratio of retained pixels. The image plot of this projection is shown in Figure \ref{generalized_Gaussian_and_sigmoid_plots}.b. A lower value of $\sigma$ is an indication of highly sparse image in the decomposed HVS domain $[\nabla_{\text{HVS}}{I}(x), \nabla_{\text{HVS}}{I}(y)]$ where more pixel coefficients are kept to exploit pertinent sharpness information. High $\sigma$ is related to high blur images where we keep less coefficients to avoid over fitting.
\begin{align}
p(\sigma) = \frac{1}{4}(1-\tanh(60(\sigma-0.095)))+0.09.
\label{equation_5}
\end{align}

\subsection{Adaptive Hard-Thresholding}
After building the feature map in Operation 1 and determining the number of retained pixels in Operation 2, similar to \cite{vu2012bf}, we keep a subset of pixels $\overline{M}_{\text{HVS}}$ from feature map pixels $M_{\text{HVS}}$ to eliminate shallow coefficients that are less related to blur features
\begin{align}
\overline{M}_{\text{HVS}} = \textit{sort}_d(M_{\text{HVS}})_k,~k\in\{1,2,\hdots,p(\sigma)N_1N_2\},
\label{equation_4}
\end{align}
where $N_1$ and $N_2$ refers to the number of pixel along vertical and horizontal axes of the image feature, $\textit{sort}_d$ is the sorting operator in descending form and the notation $|\overline{M}_{\text{HVS}}|$ refers to the cardinality of subset $\overline{M}_{\text{HVS}}$. 

\subsection{Central Moments Information}
The HVS map obtained from (\ref{equation_3}) contains high order moment features that are decomposed by the superposition of different order of derivatives defined in Section \ref{sec_HVS_MaxPol}. These features are related mostly to the image edges and contain high frequency components. To extract meaningful features from such a map, we measure the $m$th central moment of $\overline{M}_{\text{HVS}}$ for feature extraction
\begin{align}
\mu_m = \mathbb{E}\left[(\overline{M}_{\text{HVS}} - \mu_0)^m\right]
\label{equation_6}
\end{align}
where $\mathbb{E}[\cdot]$ is the expectation and $\mu_0 = \frac{1}{|\Omega|}\sum_{k\in\Omega}\overline{M}_{\text{HVS}}(k)$ is the average value of the remaining features. The $m$th moment $\mu_m$ encodes the $m$th power of deviation of variables $\overline{M}_{\text{HVS}}$ from its mean $\mu_0$. Here the negative logarithm of the moment is considered as the final score 
\begin{align}
\text{C} = - \log{\mu_{m}}
\label{equation_8}
\end{align}
which represents the sharpness score related to an individual HVS filter.

\subsection{Combination of Multiple Kernels}
Often it is meaningful to deploy more than one HVS filters for feature extraction due to the complexity of blur type in an image of interest. Here we propose a method using two HVS filters, called HVS-2. To simulate HVS using two kernels, we linearly combine the sharpness scores from the two different HVS filters. We find the weights of the linear combination by modifying the standard Image Quality Assessment (IQA) measurement equation shown(\ref{equation_9}) \cite{sheikh2006statistical, dixon2014pathology}. 
The IQA measurement takes two sets of objective scores $X$ and subjective scores $Y$ as input and scores the correlation. Specifically, the algorithm first performs a non-linear projection defined in (\ref{equation_9}). Next, the best projection of objective scores, $\hat{Y}$, is compared with the subjective scores $Y$ to show the correlation between $X$ and $Y$ in terms of the metrics such as PLCC, SRCC, KRCC, and RMSE \cite{video2003final, sheikh2006statistical}. The nonlinear fitting is realized by tuning five parameters to minimize the RMSE between $\hat{Y}$ and $Y$, that is $\|Y-\hat{Y}\|_{2}^{2}$. The equation of $\hat{Y}$ is:
\begin{align}
\hat{Y} = Q(X) = k_{1}\Big(\frac{1}{2}+\frac{1}{1+e^{k_{2}(X-k_{3})}}\Big) + k_{4}X + k_{5}
\label{equation_9}
\end{align}
where $k_i$ are the tuning parameters to be determined. Here, we revise this fitting problem for multiple objective scoring by substituting $X$ with a linear combination shown in (\ref{equation_12}). In this case, we will have two more weight parameters $w_1, w_2$ for tuning.
\begin{align}
X = M \times W,
\label{equation_12}
\end{align}
where
\begin{align}
M = 
\begin{bmatrix} 
C^{(1)}_{1} & C^{(1)}_{2} \\
C^{(2)}_{1} & C^{(2)}_{2} \\
\vdots & \vdots \\
C^{(N)}_{1} & C^{(N)}_{2}
\end{bmatrix}~\text{and} ~W = \begin{bmatrix} w_{1} \\ w_{2}\end{bmatrix},
\label{equation_10}
\end{align}
where $C^{(i)}_{j}$ refers to the objective scoring of the $i$th image using the $j$th HVS filter (In our case, we only have two HVS filters). By substituting (\ref{equation_10}) into (\ref{equation_9}), the revised IQA metric for multiple feature measurement gives 
\begin{align}
\hat{Y} = Q(M W) = k_{1}\Big(\frac{1}{2}+\frac{1}{1+e^{k_{2}(MW-k_{3})}}\Big) + k_{4}MW + k_{5}.
\label{equation_11}
\end{align}
The parameters are identified by fitting the revised IQA $\hat{Y}$ in (\ref{equation_11}) to the subjective scores accordingly. Note that the parameters $k_1, k_2, k_3, k_4, k_5, w_1, w_2$ are optimzed at the same time and we only need $w_1, w_2$. The detailed parameter tuning procedures are explained in Section V.A.
After determining the weights $w_1$ nad $w_2$, the final NR-ISA score for HVS-2 is
\begin{align}
\text{C} = w_1 * C_1 + w_2 * C_2
\end{align}
where $C_i$ is the NR-ISA score generated by a single HVS filter.
\section{FocusPath: A Digital Pathology Archive}\label{section_digital_pathology}
The primary purpose of proposing this pathology database is to evaluate the performance of different non-reference sharpness assessment metrics on pathological images. In whole slide imaging (WSI) system, tissue slides are scanned automatically where out-of-focus scans are a common problem. The common practice in digital pathology (DP) is to install the tissue slide on a slide-holder and feed the slide holder into the scanner. The level of the tissue slide is automatically moved up/down in quarter micron ($0.25\mu$) resolution to adjust the focus level with respect to the focal length of the optical lens installed in the scanner. If the position of the stage is not adjusted properly, the tissue depth level will be miss-aligned with focal length, and hence the image will be perceived in blur. Such irregularities need to be detected and addressed accordingly by sending a feedback to the system to retake the scan or adding a quality check (QC) control on the scanning process.

\begin{figure}[htp]
\centerline{
\subfigure[$Z_{1}$]{\includegraphics[height=0.055\textwidth]{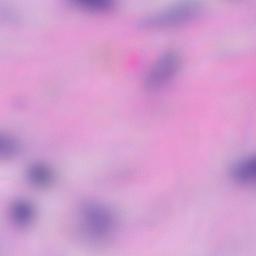}}\hspace{-.025in}
\subfigure[$Z_{2}$]{\includegraphics[height=0.055\textwidth]{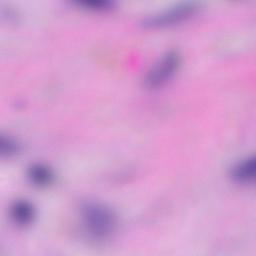}}\hspace{-.025in}
\subfigure[$Z_{3}$]{\includegraphics[height=0.055\textwidth]{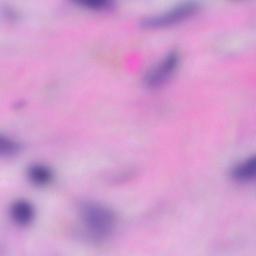}}\hspace{-.025in}
\subfigure[$Z_{4}$]{\includegraphics[height=0.055\textwidth]{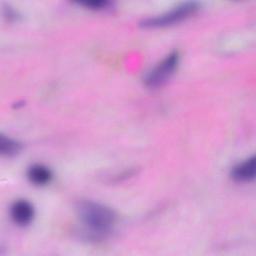}}\hspace{-.025in}
\subfigure[$Z_{5}$]{\includegraphics[height=0.055\textwidth]{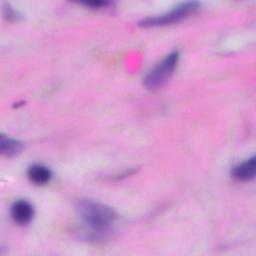}}\hspace{-.025in}
\subfigure[$Z_{6}$]{\includegraphics[height=0.055\textwidth]{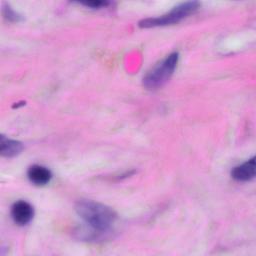}}\hspace{-.025in}
\subfigure[$Z_{7}$]{\includegraphics[height=0.055\textwidth]{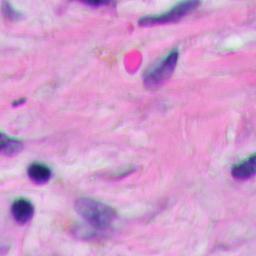}}\hspace{-.025in}
\subfigure[$Z_{8}$]{\includegraphics[height=0.055\textwidth]{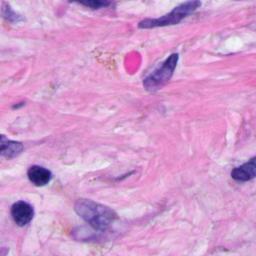}}
}\vspace{-.05in}
\centerline{
\subfigure[$Z_{9}$]{\includegraphics[height=0.055\textwidth]{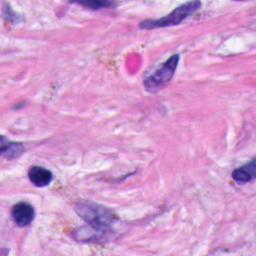}}\hspace{-.025in}
\subfigure[$Z_{10}$]{\includegraphics[height=0.055\textwidth]{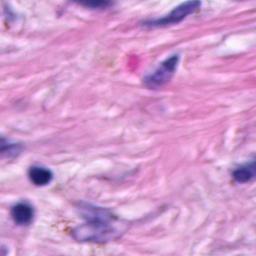}}\hspace{-.025in}
\subfigure[$Z_{11}$]{\includegraphics[height=0.055\textwidth]{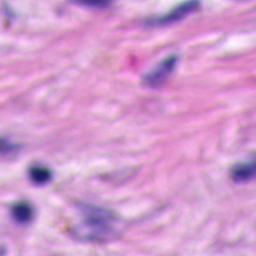}}\hspace{-.025in}
\subfigure[$Z_{12}$]{\includegraphics[height=0.055\textwidth]{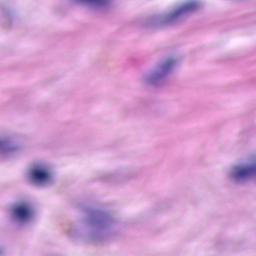}}\hspace{-.025in}
\subfigure[$Z_{13}$]{\includegraphics[height=0.055\textwidth]{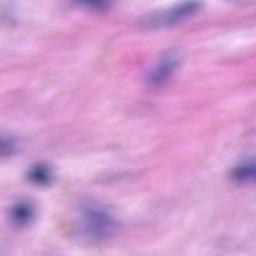}}\hspace{-.025in}
\subfigure[$Z_{14}$]{\includegraphics[height=0.055\textwidth]{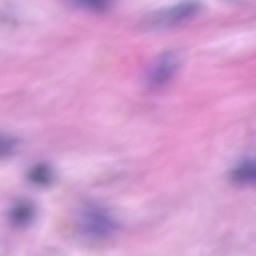}}\hspace{-.025in}
\subfigure[$Z_{15}$]{\includegraphics[height=0.055\textwidth]{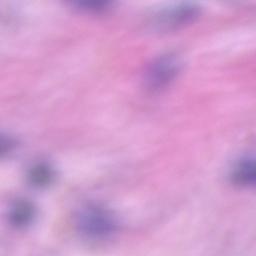}}\hspace{-.025in}
\subfigure[$Z_{16}$]{\includegraphics[height=0.055\textwidth]{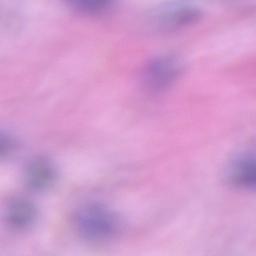}}
}\vspace{-.05in}
\caption{The sample images of Slice $1-16$. Sample images of different Slice index vary in blur levels, with the clearest one in the middle. All the samples come from Slide 1, Strip 0 and Position 1.}
\label{Slice_Example}
\end{figure}

All the digital pathological images included are captured as WSIs using the Huron TissueScope LE1.2 \cite{dixon2014pathology}. The scanner automatically digitizes, processes and stores the images of tissue slides in .tif format. To create a database with high diversity, the tissue slides used are cut from 9 distinct types of organs (i.e. 9 \textit{Slides}). When scanning through a whole organ tissue slide, the scanner captures multiple image patches all in one, which are called \textit{Strips}. We include 2 strips for each tissue slide to ensure we take different morphological structure across a wide tissue area. The WSI scanner generates several in-depth \textit{Z-stacks} for each strip, where each z-stack is referred as a \textit{Slice}. Different z-stacks represent different vertical positions of the WSI camera, which is highly related to the image blur. We include 16 z-stacks for every image strip of a certain tissue slide in the pathology database. In addition, we randomly choose 3 different positions and accordingly crop 3 image patches of $1024 \times 1024\ pixel\ size$ from the raw image captured by the whole slide imaging system. To summarize, the total number of images included is $864$, cropped from $9\ Slides \times 2\ Strips \times 3\ Positions \times 16\ Slices$. The database is called \texttt{FocusPath} and is publicly published in \footnote{FocusPath database \url{https://sites.google.com/view/focuspathuoft}}. A sample of 16 different focus levels is shown in Figure \ref{Slice_Example}. Note that we have an extended version of FocusPath containing $8640$ images extracted from $30$ positions instead of $3$ for ISA analysis. Since the database is too big, we avoid uploading it online, but it can be requested from the authors on demand.

\begin{table*}[htp]
\scriptsize
\centering
\caption{Sample images from each slide in FocusPath with different stain types.}
\label{slide_samples}
\begin{tabular}{c|c|c|c|c|c|c|c|c|c}
\hlinewd{1.5pt}
Slide \# & Side 1 & Slide 2 & Slide 3 & Slide 4 & Slide 5 & Slide 6 & Slide 7 & Slide 8 & Slide 9\\ \hlinewd{1.5pt}
Stain Type & Trichrome & H\&E & Mucicarmine & IRON (FE) & AFB & Congo Red (CR) & PAS & Grocott & H\&E\\
{\hspace{-.05in} Example \hspace{-.05in}} 
& {\hspace{-.05in}\includegraphics[width=0.07\textwidth]{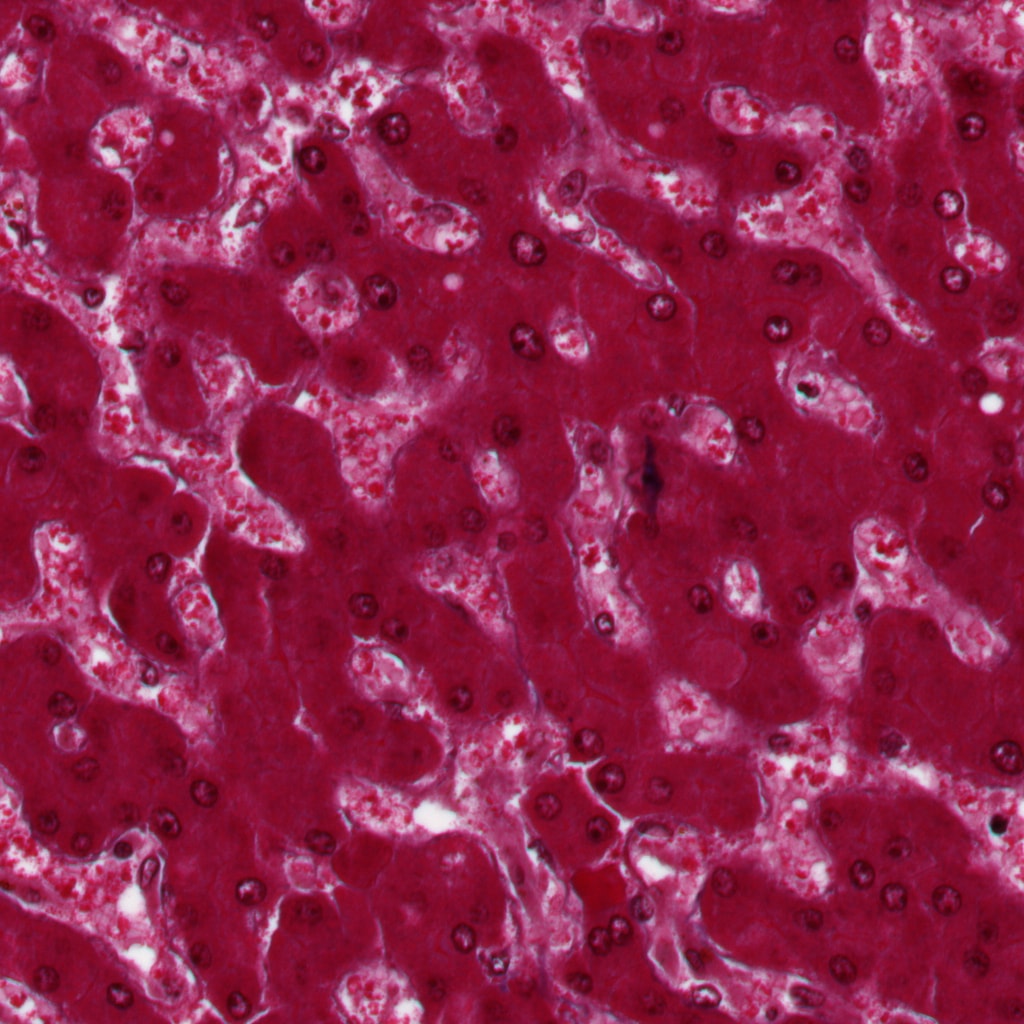}\hspace{-.05in}} 
& {\hspace{-.05in}\includegraphics[width=0.07\textwidth]{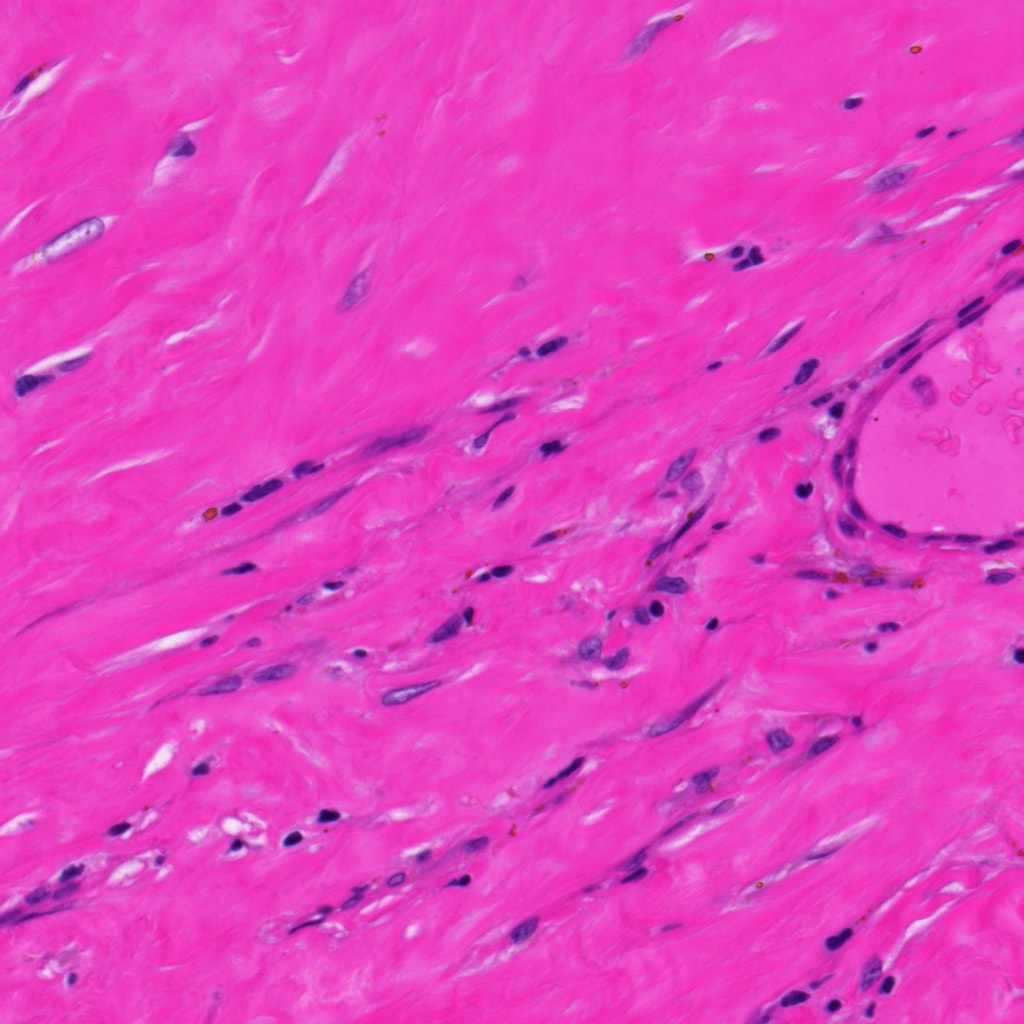}\hspace{-.05in}}
& {\hspace{-.05in}\includegraphics[width=0.07\textwidth]{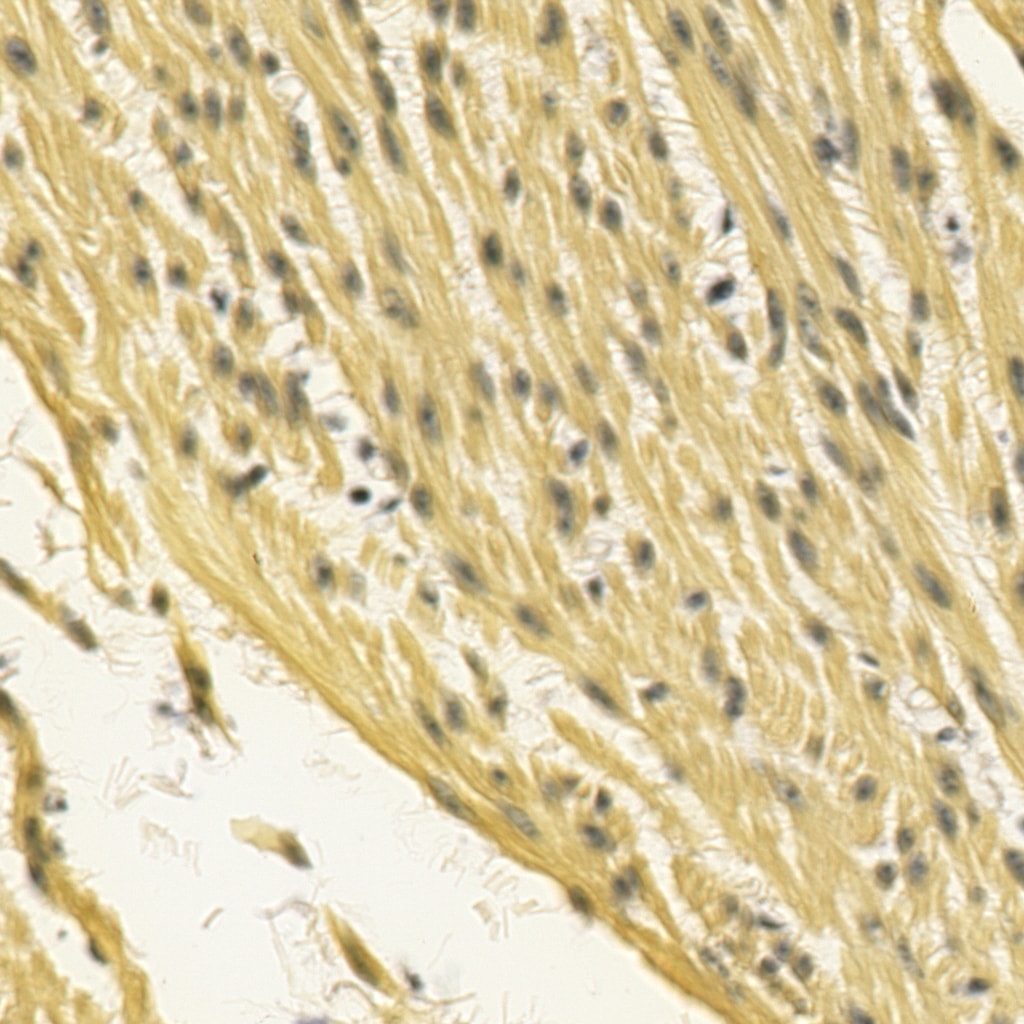}\hspace{-.05in}}
& {\hspace{-.05in}\includegraphics[width=0.07\textwidth]{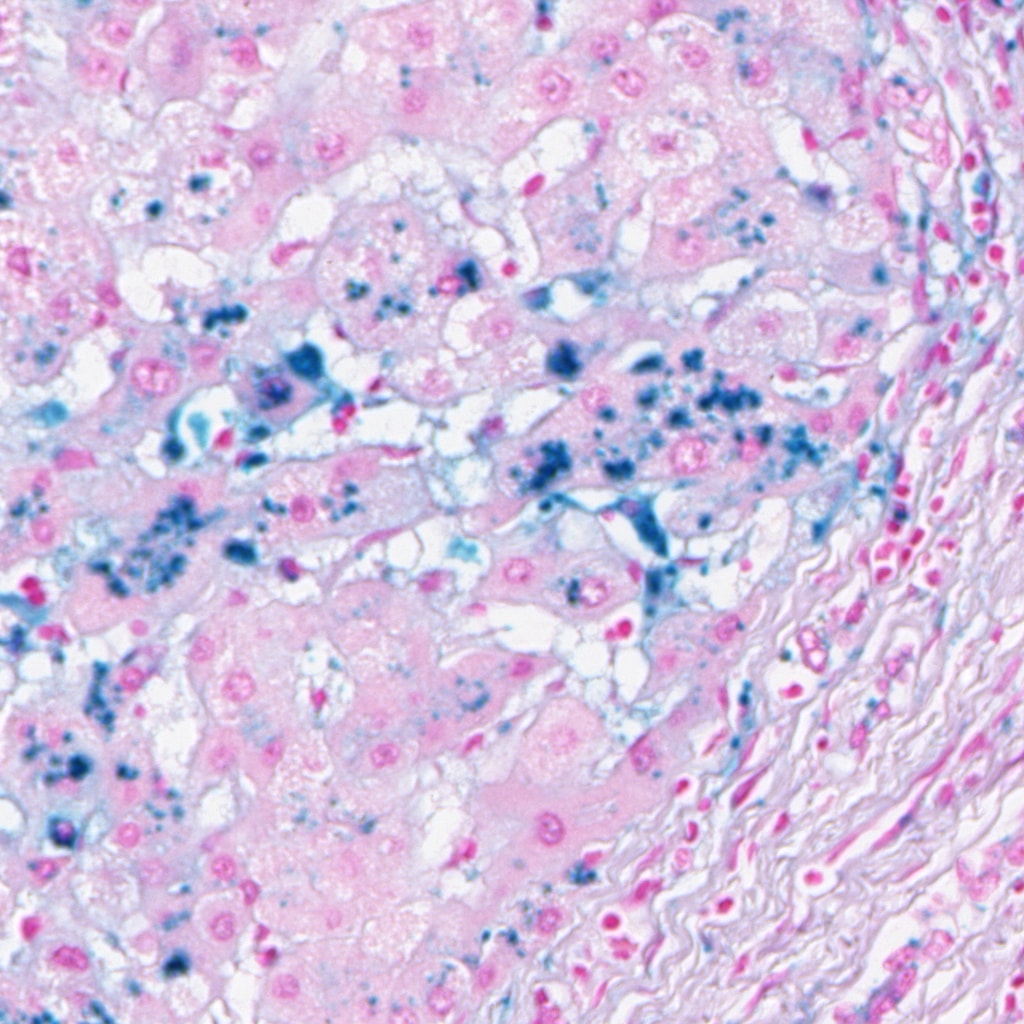}\hspace{-.05in}}
& {\hspace{-.05in}\includegraphics[width=0.07\textwidth]{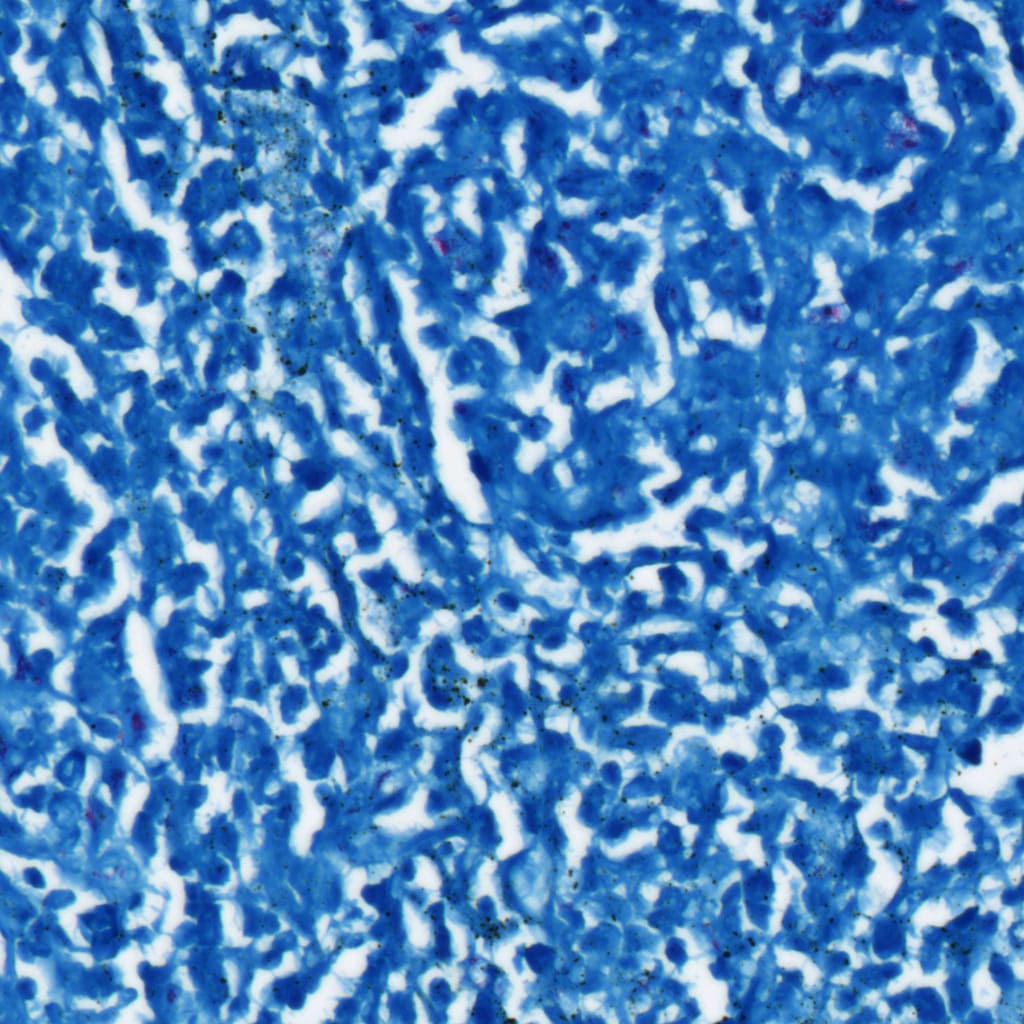}\hspace{-.05in}}
& {\hspace{-.05in}\includegraphics[width=0.07\textwidth]{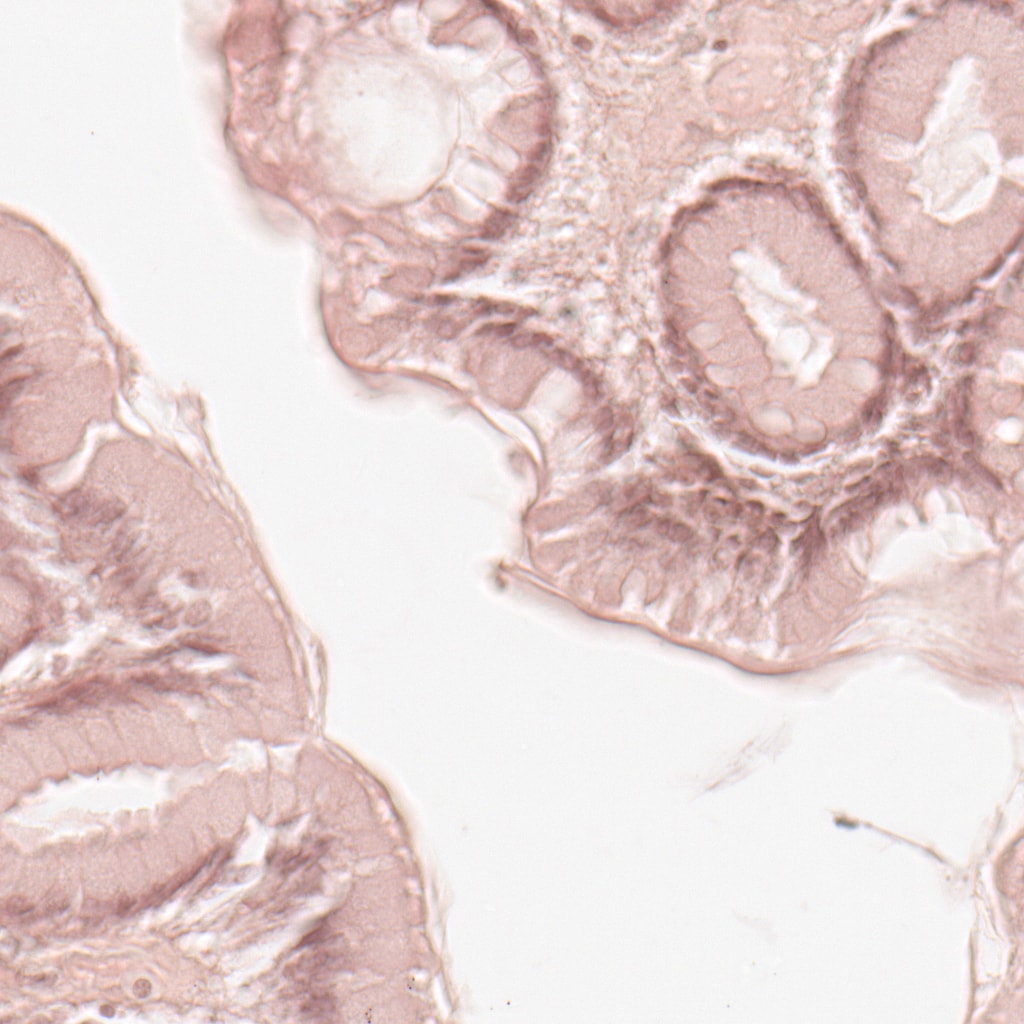}\hspace{-.05in}}
& {\hspace{-.05in}\includegraphics[width=0.07\textwidth]{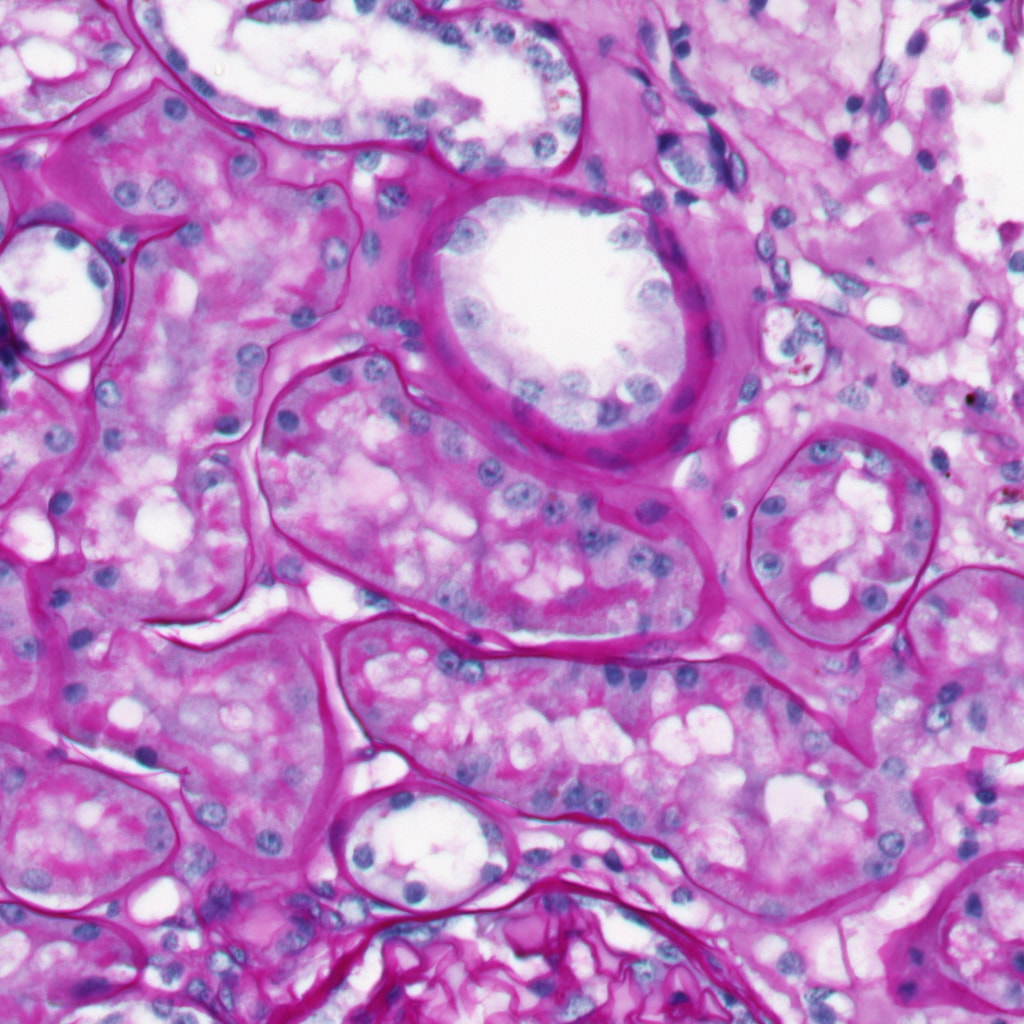}\hspace{-.05in}}
& {\hspace{-.05in}\includegraphics[width=0.07\textwidth]{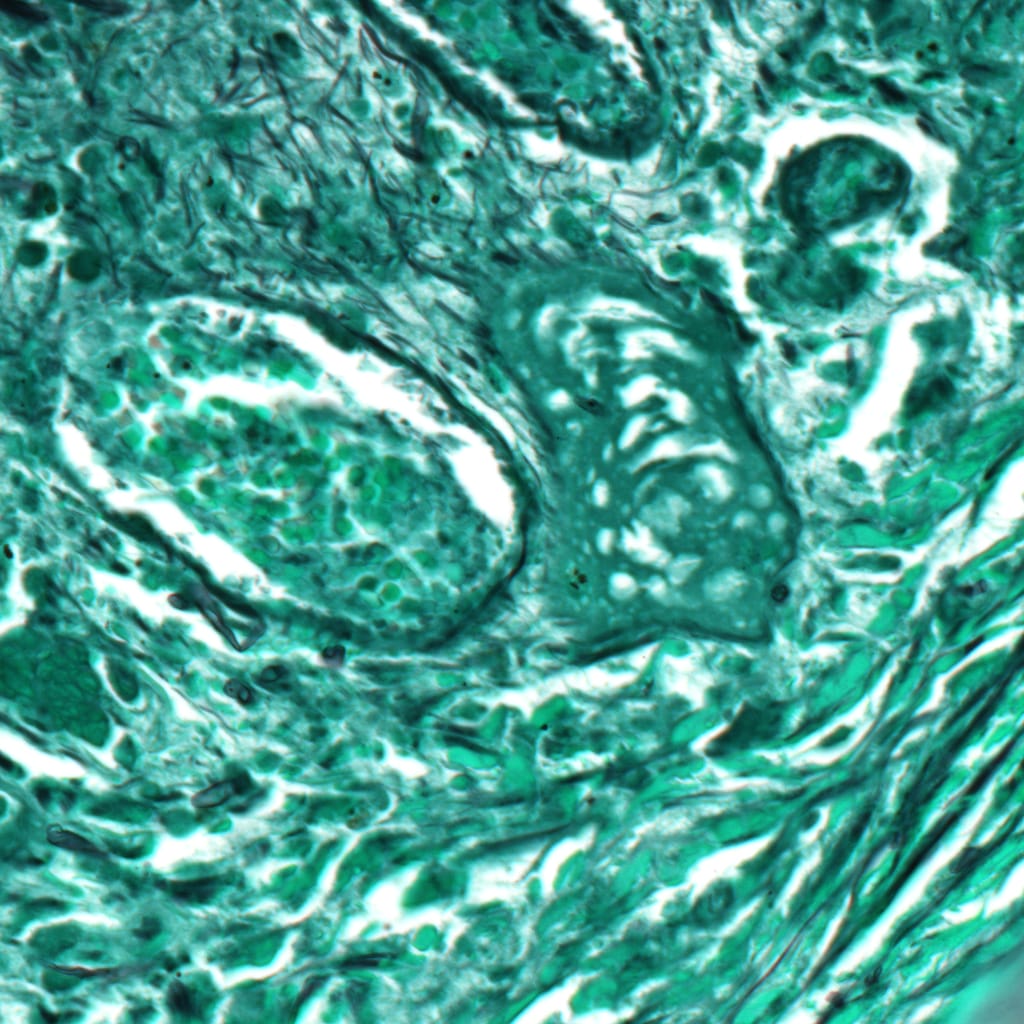}\hspace{-.05in}}
& {\hspace{-.05in}\includegraphics[width=0.07\textwidth]{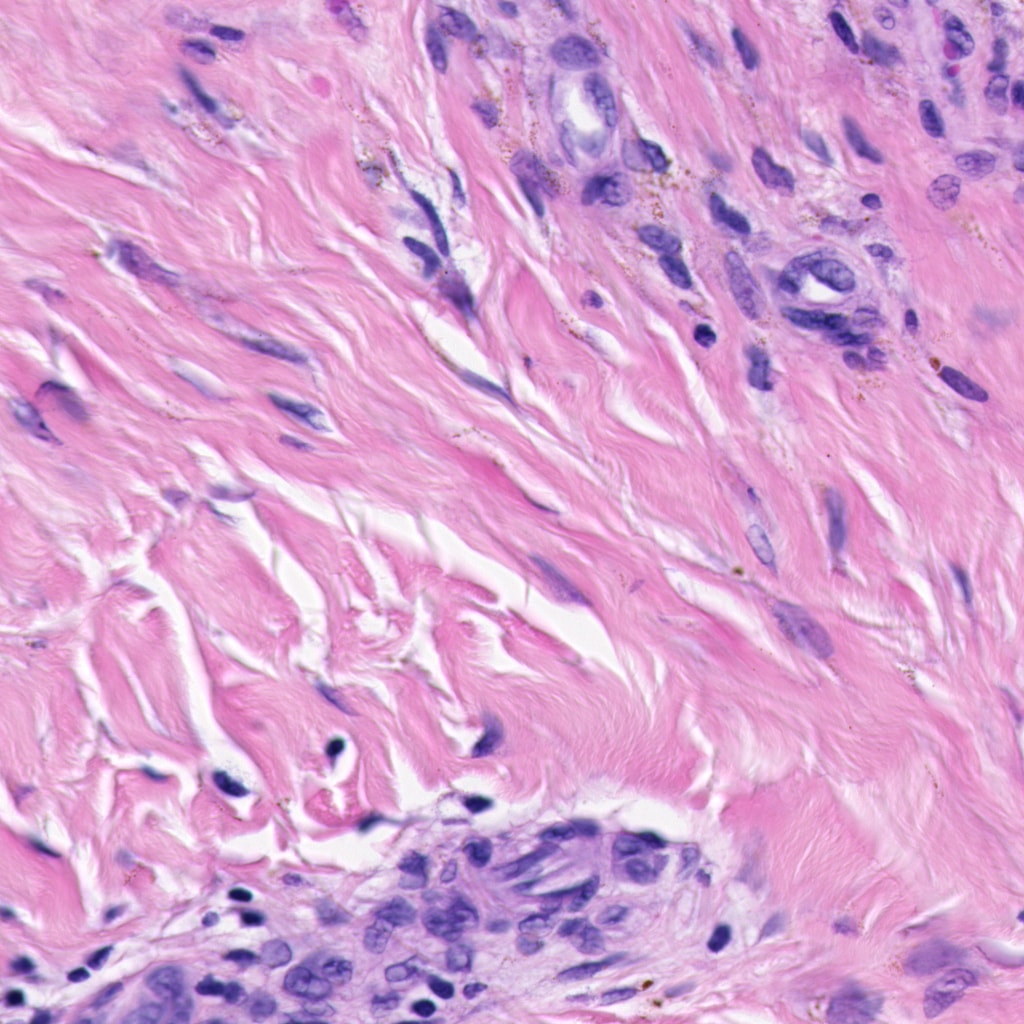}\hspace{-.05in}}\\
\hlinewd{1pt}
\end{tabular}
\end{table*}

\begin{figure}[htp]
\centerline{
\subfigure[focus score]{\includegraphics[height=0.2\textwidth]{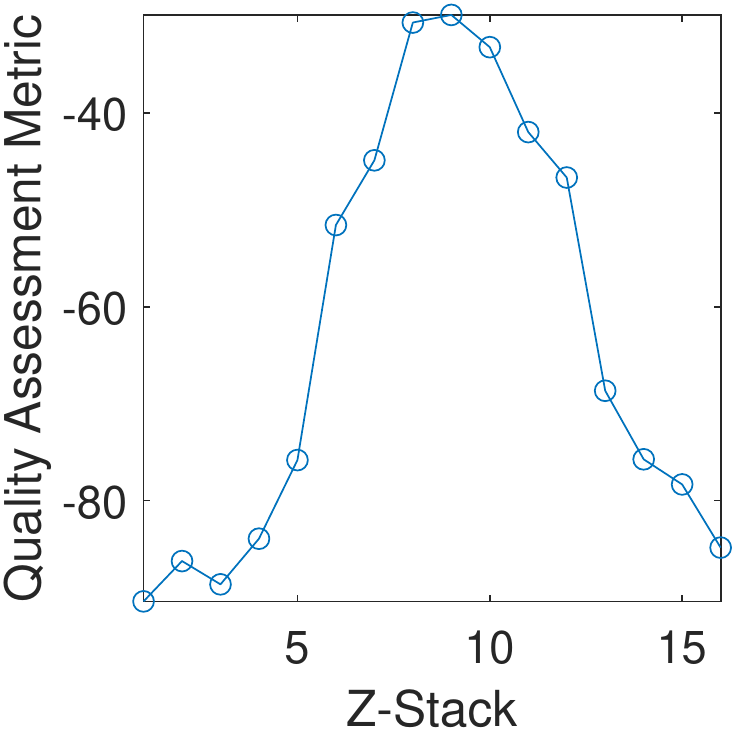}}\hspace{.3in}
\subfigure[z-level]{\includegraphics[height=0.2\textwidth]{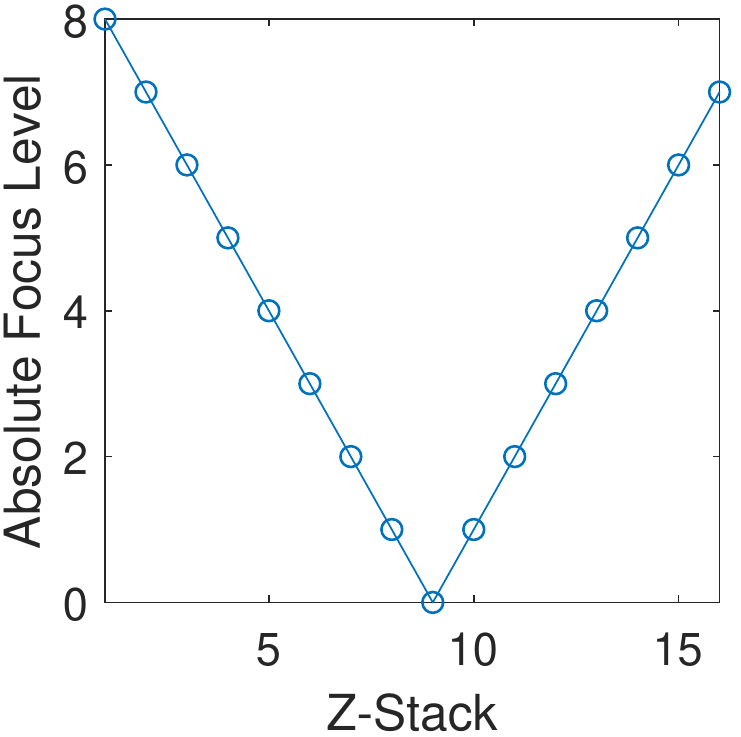}}
}\vspace{-.05in}
\centering
\caption{The focus scores generated using the proposed sharpness assessment metric and the z-level scores. The score peak indicates the clearest image among the set of sample images. In this sense, we pick the corresponding slice index of the peak as $Z^{*}$ for this set of images. The focus level scores are assigned in term of the selected $Z^{*}$. All the samples are selected from Slice 1-16, Slide 1, Strip 0 and Position 1.}
\label{score_curve}
\end{figure}

To generate a ``subjective-like'' score for FocusPath database, we assigned a ``focus level'' for each image that is determined by the difference between the best camera position $Z^{*}$ and the position of interest $Z_{i}$. In other words, the score represents how far the camera is away from its best focus level when scanning a certain image slide. Specifically, every 16 images of the same content but different blur levels have a common $Z^{*}$ and every image has a corresponding $Z_{i}$. A sample set of 16 images is shown in Figure \ref{Slice_Example}. The $Z^{*}$ value among every 16 images is determined by the proposed non-reference sharpness assessment metric in this paper. Since $Z^{*}$ is found among 16 different slices of the same tissue content, the scores will guarantee to obtain the best focus level. The plot of focus level scores across 16 images is shown in Figure \ref{score_curve}. The mathematical representation of the focus level for sharpness scoring is defined by
\begin{align}
\text{Focus Level at } Z_{i} \leftarrow Z_{i} - Z^{*}, i = {1,2,3 ... 16}
\end{align}

To make it clearer, we list the relevant information about the FocusPath in Table \ref{database_info}. In addition, as mentioned before, we include 9 tissue slides in FocusPath. They are from different types of organs with different staining types, which guarantees diversity and effectiveness of the database. Further information about the tissue slides and sample images are available in Table \ref{slide_samples}.

\begin{table}[htp]
\centering
\caption{Detailed information about the FocusPath database.}
\label{database_info}
\begin{tabular}{l|c}
\hlinewd{1.5pt}
Features              & \multicolumn{1}{l}{Description} \\ \hlinewd{1.5pt}
\# of Slides          & 9                                \\ 
\# of Strips          & 2                                \\ 
\# of Positions       & 3                                \\ 
\# of Slices(Z-Stack) & 16                               \\ 
Image Format          & .tiff                            \\ 
Image Size            & $1024 \times 1024$               \\ 
Pixel Resolution      & $0.25\mu$                        \\ 
Optical Zoom          & $40X$                            \\ 
Color Variation       & Diverse Gamut                    \\ 
Focus Resolution      & $1\mu$                           \\ 
Background Ratio      & \textless 50\%                   \\ \hlinewd{1pt}
\end{tabular}
\end{table}

\section{Experiments}\label{sec_experiment}
To evaluate the proposed HVS-MaxPol\footnote{\url{https://github.com/mahdihosseini/HVS-MaxPol}}, we conduct experiments in terms of statistical correlation accuracies, computational complexity and scalability. All of these evaluation criteria are highly related to the practical application of a NR-ISA metric. For comparison, we select 9 state-of-the-art non-reference sharpness assessment techniques, including $\text{S}_{3}$ \cite{vu2012bf}, MLV \cite{bahrami2014fast}, Kang's CNN \cite{kang2014convolutional}, $\text{ARISM}_{\text{C}}$ \cite{gu2015no}, GPC \cite{leclaire2015no}, SPARISH \cite{li2016image}, RISE \cite{li2017no}, Yu's CNN \cite{yu2017shallow} and Synthetic-MaxPol \cite{mahdi2018image}.

\subsection{Parameter Tuning}
A single HVS-MaxPol kernel contains three parameters to tune, where two parameters are the scale $\alpha$ and shape $\beta$ of the associated GG blur kernel used for the construction of the HVS response, and the third parameter is the cutoff frequency $\omega_{c}$ set for the stop band design. For more information on these parameters, please refer to Section \ref{sec_HVS_MaxPol}. An additional parameter is also defined in Section \ref{IQA_measure} to extract different order of moment $\mu$ from decomposed image features. To determine the single kernel with proper parameter adjustment yielding the highest accuracy, we perform a grid search on $\alpha$, $\beta$, $\mu$ and $\omega_{c}$ using a partial set of synthetic and natural dataset. After determining the best single kernel, we further perform another grid search to find the second best kernel that achieves the best overall accuracy. The second kernel is combined with the previously-determined best kernel using the combination weight selection defined in Section \ref{IQA_measure}. Specifically, we combine 2-dimensional $W\ =\ [w_1\ w_2]$ and 5-dimensional $K\ =\ [k_1\ k_2\ k_3\ k_4\ k_5]$ into a single 7-dimensional vector $\bar{K}$ to solve a non-linear fitting optimization problem, where the optimal value for $\bar{K}$ will be returned. We denote the optimal value of $\bar{K}$ as $[\hat{K}\ \hat{W}]$ and thus $M\hat{W}$ is the optimized objective scores. We select the optimal weight matrix $\hat{W}$ in terms of the PLCC performances tuned on FocusPath and CID2013 separately. In this case, HVS-MaxPol-1 uses the best single kernel and HVS-MaxPol-2 uses the combination of the best two kernels. All the parameters are independently tuned based for synthetic and natural dataset and the final values for each parameter are listed in Table \ref{best setting}.
	
\begin{table}[htp]
\centering
\caption{The best settings for HVS-MaxPol-1 and HVS-MaxPol-2 over natural and synthetic dataset from grid search. The search range of $\alpha$ is $0.7-3$, $0.8-2$ for $\beta$, $3-23$ for $\omega_{c}$ and $2-14$ for $\mu$.}
\label{best setting}
\begin{tabular}{l|cc|cc}
\hlinewd{1.5pt}
            & \multicolumn{2}{c|}{Synthetic} & \multicolumn{2}{c}{Natural} \\ 
            & HVS-1  & HVS-2  & HVS-1 & HVS-2 \\ \hlinewd{1.5pt}
$\alpha$       & 0.7           & 0.7, 0.7      & 1.7          & 1.7, 0.7     \\
$\beta$        & 0.8           & 0.8, 0.9      & 1.4          & 1.4, 0.8     \\
$\text{cutoff} (\omega_{c})$  & 19            & 19, 20        & 13           & 13, 26       \\
$\text{moment} (\mu)$ & 20            & 20, 12        & 12           & 12, 4       \\ \hlinewd{1pt}
\end{tabular}
\end{table}

\subsection{Blur Dataset Selection}
The statistical correlation analysis are performed on seven distinct image blur dataset. Specifically, we include four synthetic blur image dataset, LIVE \cite{sheikh2006statistical}, CSIQ \cite{larson2010most}, TID2008 \cite{ponomarenko2009tid2008}, and TID2013 \cite{ponomarenko2015image}, including 145, 150, 100 and 125 Gaussian blurred images, respectively. We also include three natural blur image dataset, BID \cite{ciancio2011no}, CID2013 \cite{virtanen2015cid2013} and our newly introduced FocusPath \footnote{FocusPath dataset \url{
https://sites.google.com/view/focuspathuoft}}, with 586, 474 and 864 images respectively. The image content of BID and CID2013 varies from human beings to natural landscapes while the image content of FocusPath is related to pathological organs.

\begin{figure}[htp]
\scriptsize
\centerline{
\subfigure[BID]{\includegraphics[height=0.1\textwidth]{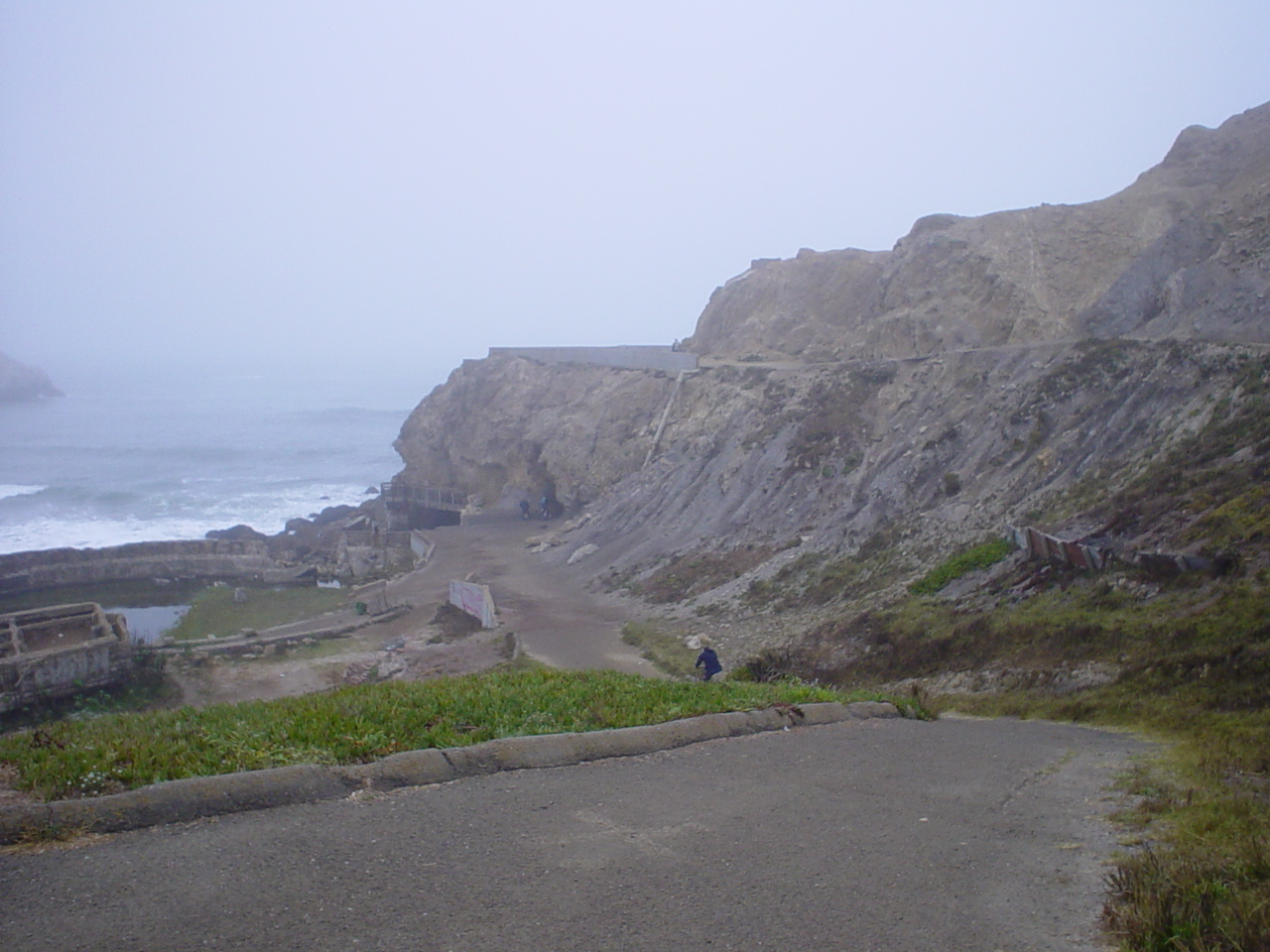}}\hspace{.01in}
\subfigure[CID2013]{\includegraphics[height=0.1\textwidth]{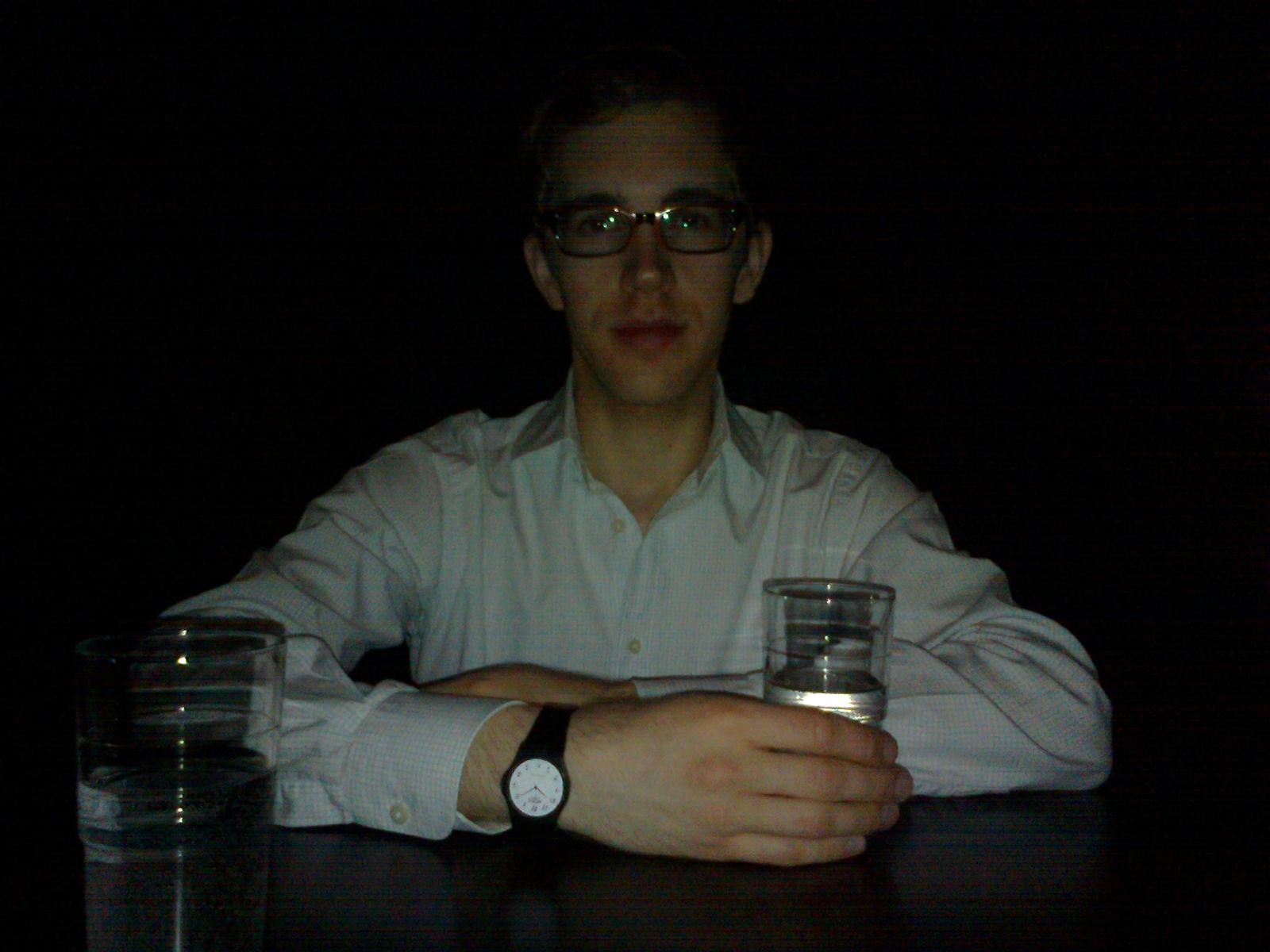}}\hspace{.01in}
\subfigure[FocusPath]{\includegraphics[height=0.1\textwidth]{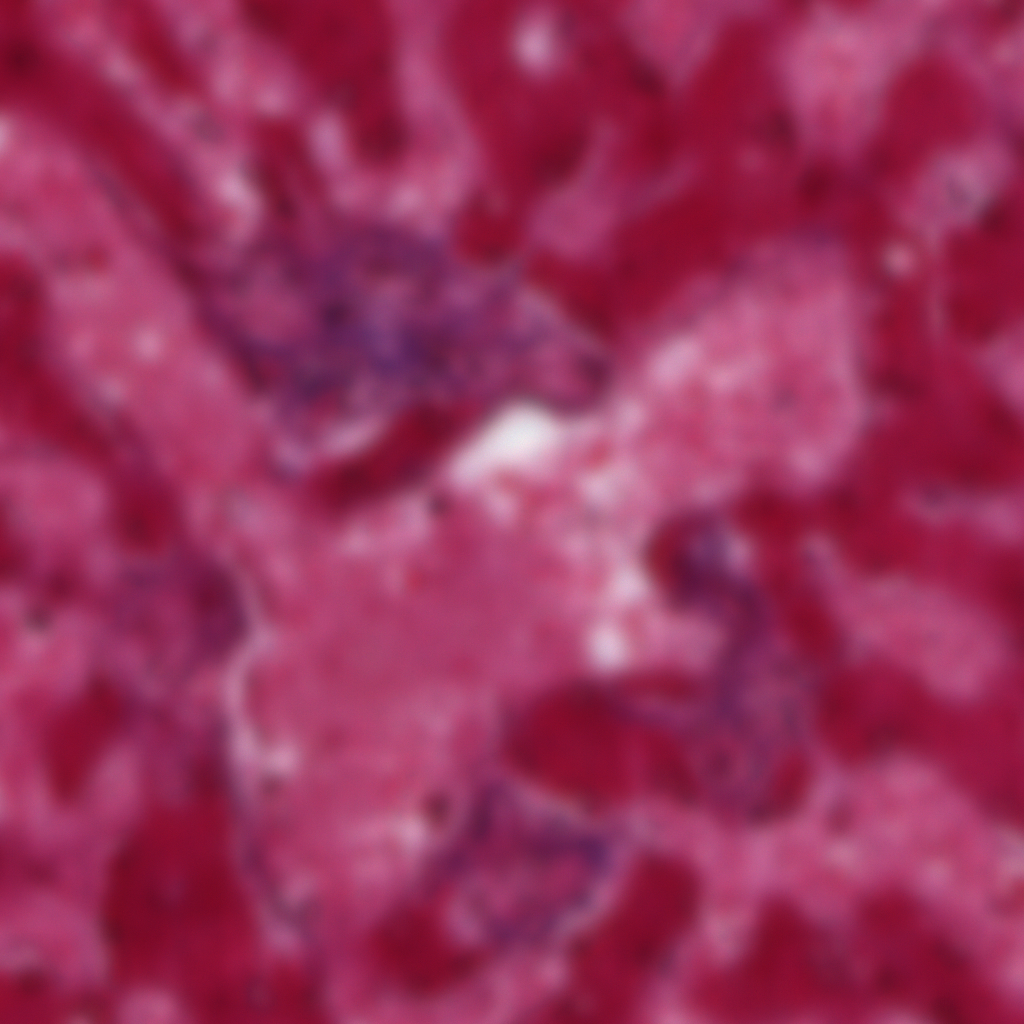}}\hspace{.01in}
}\vspace{-.05in}
\centerline{
\subfigure[LIVE]{\includegraphics[height=0.09\textwidth]{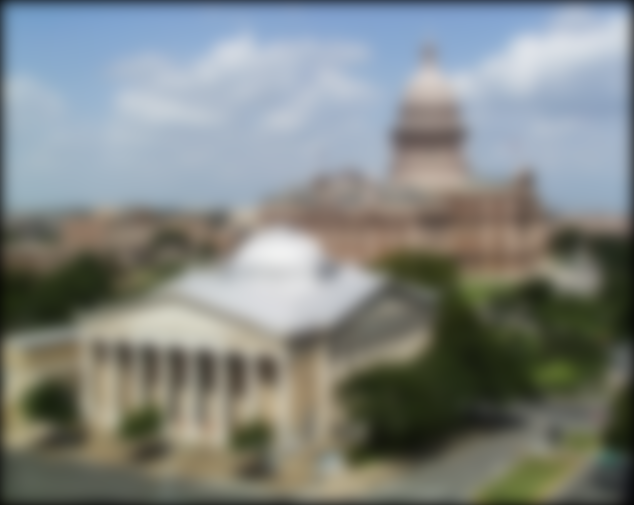}}\hspace{.01in}
\subfigure[CSIQ]{\includegraphics[height=0.09\textwidth]{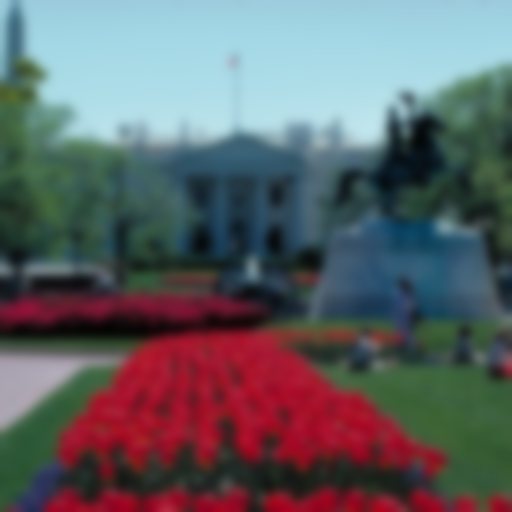}}\hspace{.01in}
\subfigure[TID2008]{\includegraphics[height=0.09\textwidth]{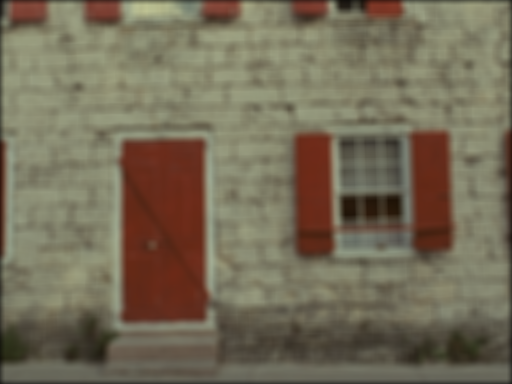}}\hspace{.01in}
\subfigure[TID2013]{\includegraphics[height=0.09\textwidth]{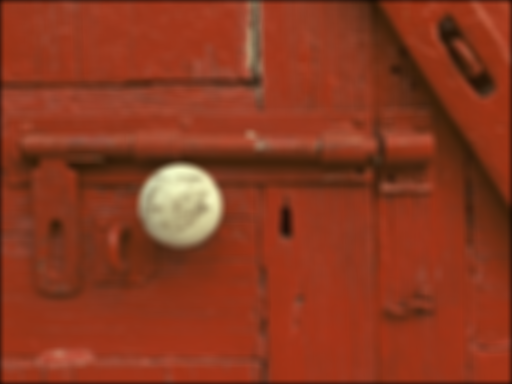}}\hspace{.01in}
}\vspace{-.05in}
\caption{One sample image from each natural (a)-(c) and synthetic (d)-(g) dataset.}
\label{sample image}
\end{figure}

Unlike the synthetic blur, the type of blur within natural dataset varies differently. BID contains out-of-focus blur, simple motion blur and complex motion blur, while CID 2013 contains blur caused by subject luminance and subject-camera distance. Due to the fact that human beings are not consistent with their subjective scoring, all the NR-ISA metrics cannot achieve a very high accuracy on such dataset. In addition, FocusPath mainly contains the out-of-focus blur caused by the focus misalignment between the tissue and objective lens that is more homogenous as opposed to BID and CID2013. Figure \ref{sample image} demonstrates one sample image from all seven dataset. 


\subsection{Performance Metric Selection}
The accuracy measurement of all NR-ISAs are reported using the commonly practiced accuracy measurements of Pearson Linear Correlation Coefficient (PLCC) and Spearman Rank Order Correlation (SRCC) indicating the linear correlation of strength and direction of monotonicity between objective and subjective scoring, respectively \cite{video2003final, sheikh2006statistical}. The aggregation of these metric performances across different synthetic blur dataset is usually achieved by weighted averaging across the number images per dataset. However, such aggregation on selected natural dataset is not a meaningful approach. This is because the type of the blur observed from all three natural dataset of BID, CID2013 and FocusPath are different, such that
\begin{itemize}
\item the ground-truth quality score is human-subjective based in BID and CID2013, whereas, this score is created by a statistical measure in FocusPath along the in-depth z-level of the lens camera,
\item the objective quality performances yield different range of values across different dataset.
\end{itemize}
To overcome the above-listed uncertainties, we deploy the statistical evaluations of the objective performance models introduced in \cite{krasula2016accuracy}. In particular, here we select three different statistical measures of (a) Area Under Curve (AUC) of the Receiver Operating Characteristic (ROC) in different vs. similar analysis ($\text{AUC}_{\text{DS}}$), (b) AUC of the ROC in better versus worse analysis ($\text{AUC}_{\text{BW}}$), and (c) percentage of correct classification ($\text{C}_{\text{0}}$). Here, $\text{AUC}_{\text{DS}}$ indicates the capability of a certain metric to distinguish between different and similar pairs, while $\text{AUC}_{\text{BW}}$ reveals the capability to recognize the stimulus of higher quality in the significantly different pair. In addition, $\text{C}_{\text{0}}$ shows the correct classification ability of a metric. To aggregate the results on different natural dataset, we merge the pair significance measures across each dataset and evaluate the latter statistical performance measures, accordingly. This is a meaningful approach since the subjective scores are transformed into binary sets of better/worse, and different/similar that is less invariant toward amplitude scoring. Please refer to \cite{krasula2016accuracy, krasula2017quality} for more information on the statistical performance measures.

\begin{table*}[htp]
\renewcommand{\arraystretch}{1.05}
\caption{Performance evaluation of 11 NR-ISA metrics over seven image blur dataset, including four synthetic and three natural blur. The performances are reported in terms of PLCC, SRCC, $\text{AUC}_\text{DS}$, $\text{AUC}_\text{BW}$, and $\text{C}_\text{0}$ scores. Furthermore, the average calculation time of each metric over different dataset are also reported in ``$\text{Avg. Process Time (sec)}$''. The overall performances are also computed separately using the statistical aggregation method for synthetic blur and natural blur dataset.}
\label{methods_table}
\centering
\scriptsize
\begin{tabular}{l|c|c|cccc|ccc|cc}
\hlinewd{1.5pt}
\mcrot{1}{l}{0}{\textcolor{mygray1}{NR-ISA Method}} &
\mcrot{1}{l}{0}{\textcolor{mygray1}{Year}} &
\mcrot{1}{l}{0}{\textcolor{mygray1}{Performance Metric}} &
\mcrot{1}{c}{45}{\textcolor{mygray1}{LIVE\cite{sheikh2006statistical}}} &
\mcrot{1}{c}{45}{\textcolor{mygray1}{CSIQ\cite{larson2010most}}} &
\mcrot{1}{c}{45}{\textcolor{mygray1}{TID2008\cite{ponomarenko2009tid2008}}} &
\mcrot{1}{c}{45}{\textcolor{mygray1}{TID2013\cite{ponomarenko2015image}}} &
\mcrot{1}{c}{45}{\textcolor{mygray1}{BID\cite{ciancio2011no}}} &
\mcrot{1}{c}{45}{\textcolor{mygray1}{CID2013\cite{virtanen2015cid2013}}} &
\mcrot{1}{c}{45}{\textcolor{mygray1}{FocusPath}} &
\mcrot{1}{c}{45}{\textcolor{mygray1}{Overall-Synthetic}} &
\mcrot{1}{c}{45}{\textcolor{mygray1}{Overall-Natural}}
\\ \hlinewd{1.5pt}
\multirow{6}{*}{$\text{S}_3$ \cite{vu2012bf}}   & \multirow{6}{*}{$2012$} & PLCC & $0.9771$ & $0.9175$ & $0.8555$ & $0.8816$ & $0.4234$ & $0.6573$ & $0.7921$ & $0.9136$ & $0.6466$ \\
                                                &                         & SRCC & $\textbf{\textcolor{mygreen2}{0.9917}}$ & $0.9058$ & $0.8643$ & $0.8609$ & $0.4097$ & $0.5944$ & $0.7925$ & $0.9110$ & $0.6271$ \\
                                                &                         & $\text{AUC}_\text{DS}$ & $0.7059$ & $0.7747$ & $0.4112$ & $0.3830$ & $0.5409$ & $0.6258$ & $0.7119$ & $0.4833$ & $\textbf{\textcolor{mygreen1}{0.7071}}$ \\
                                                &                         & $\text{AUC}_\text{BW}$ & $0.9721$ & $1.0000$ & $0.8354$ & $0.8371$ & $\textbf{\textcolor{mygreen1}{0.7374}}$ & $0.8655$ & $0.9794$ & $0.9327$ & $0.8281$ \\
                                                &                         & $\text{C}_\text{0}$ & $0.9339$ & $1.0000$ & $0.7343$ & $0.7272$ & $\textbf{\textcolor{mygreen1}{0.6813}}$ & $0.7768$ & $0.9263$ & $0.8588$ & $0.8177$ \\ \cline{3-12}
                                                &                         & $\text{Avg. Process Time (sec)}$ & $9.1456$ & $6.2996$ & $5.0259$ & $5.1421$ & $79.7839$ & $69.7957$ & $4.1336$ & $6.5700$ & $43.3513$ \\ \hlinewd{.5pt}																									
\multirow{6}{*}{MLV \cite{bahrami2014fast}}			& \multirow{6}{*}{$2014$} & PLCC & $0.9784$ & $0.9489$ & $0.8584$ & $0.8830$ & $0.3642$ & $0.6847$ & $\textbf{\textcolor{mygreen2}{0.8569}}$ & $0.9239$ & $0.6644$ \\
																								&                         & SRCC & $0.9749$ & $0.9246$ & $0.8546$ & $0.8785$ & $0.3256$ & $0.5369$ & $\textbf{\textcolor{mygreen2}{0.8636}}$ & $0.9141$ & $0.6193$ \\
                                                &                         & $\text{AUC}_\text{DS}$ & $0.7923$ & $0.8882$ & $0.4800$ & $0.4776$ & $0.5412$ & $0.6085$ & $\textbf{\textcolor{mygreen1}{0.7664}}$ & $0.5430$ & $\textbf{\textcolor{mygreen2}{0.7164}}$ \\
                                                &                         & $\text{AUC}_\text{BW}$ & $0.9798$ & $1.0000$ & $0.8414$ & $0.8366$ & $0.6926$ & $0.8458$ & $\textbf{\textcolor{mygreen1}{0.9929}}$ & $0.9360$ & $0.8503$ \\
                                                &                         & $\text{C}_\text{0}$ & $0.9219$ & $1.0000$ & $0.7400$ & $0.7305$ & $0.6416$ & $0.7525$ & $\textbf{\textcolor{mygreen2}{0.9662}}$ & $0.8545$ & $0.8188$ \\ \cline{3-12}
                                                &                         & $\text{Avg. Process Time (sec)}$ & $0.0655$ & $0.0493$ & $0.0320$ & $0.0342$ & $0.5175$ & $0.3836$ & $0.2151$ & $0.0469$ & $0.3487$ \\ \hlinewd{.5pt}	
\multirow{2}{*}{Kang's CNN \cite{kang2014convolutional}}   & \multirow{2}{*}{$2014$}   & PLCC & $0.9625$ & $0.7743$ & $0.8803$ & $\textbf{\textcolor{mygreen1}{0.9308}}$ & $-$ & $-$ & $-$ & $0.8848$ & $-$ \\
																								&                                      & SRCC & $0.9831$ & $0.7806$ & $0.8496$ & $0.9215$ & $-$ & $-$ & $-$ & $0.8842$ & $-$ \\ \hlinewd{.5pt}
\multirow{6}{*}{$\text{ARISM}_{\text{C}}$ \cite{gu2015no}}   & \multirow{6}{*}{$2015$} & PLCC & $0.9136$ & $0.9482$ & $0.8544$ & $0.8974$ & $0.1841$ & $0.5530$ & $0.4389$ & $0.9083$ & $0.3893$ \\
																								&                         & SRCC & $0.9704$ & $0.9322$ & $0.8679$ & $0.9012$ & $0.1841$ & $0.4440$ & $0.4382$ & $0.9230$ & $0.3622$ \\
                                                &                         & $\text{AUC}_\text{DS}$ & $0.7984$ & $0.9056$ & $0.4614$ & $0.4142$ & $0.5330$ & $0.5908$ & $0.5394$ & $0.4789$ & $0.5007$ \\
                                                &                         & $\text{AUC}_\text{BW}$ & $0.9871$ & $1.0000$ & $\textbf{\textcolor{mygreen1}{0.8584}}$ & $\textbf{\textcolor{mygreen2}{0.8470}}$ & $0.4813$ & $0.7494$ & $0.3144$ & $\textbf{\textcolor{mygreen2}{0.9465}}$ & $0.4402$ \\
                                                &                         & $\text{C}_\text{0}$ & $0.9446$ & $1.0000$ & $0.7542$ & $0.7353$ & $0.4931$ & $0.6934$ & $0.3355$ & $\textbf{\textcolor{mygreen1}{0.8694}}$ & $0.4642$ \\ \cline{3-12}
                                                &                         & $\text{Avg. Process Time (sec)}$ & $10.2492$ & $7.8289$ & $5.7262$ & $5.7574$ & $109.6214$ & $78.3470$ & $71.8967$ & $7.6015$ & $84.9758$ \\ \hlinewd{.5pt}
\multirow{6}{*}{GPC \cite{leclaire2015no}}      & \multirow{6}{*}{$2015$} & PLCC & $0.9243$ & $0.9018$ & $0.8690$ & $0.8671$ & $\textbf{\textcolor{mygreen1}{0.4409}}$ & $0.6520$ & $0.7499$ & $0.8934$ & $0.6317$ \\
																								&                         & SRCC & $0.8372$ & $0.8648$ & $0.8741$ & $0.8681$ & $\textbf{\textcolor{mygreen1}{0.4361}}$ & $0.6080$ & $0.7811$ & $0.8597$ & $0.6334$ \\
                                                &                         & $\text{AUC}_\text{DS}$ & $0.6375$ & $0.7036$ & $0.4258$ & $0.4054$ & $0.5119$ & $0.6269$ & $0.6310$ & $0.5022$ & $0.6797$ \\
                                                &                         & $\text{AUC}_\text{BW}$ & $0.9328$ & $1.0000$ & $\textbf{\textcolor{mygreen2}{0.8669}}$ & $\textbf{\textcolor{mygreen1}{0.8439}}$ & $0.6962$ & $0.8554$ & $0.9703$ & $0.9177$ & $0.7976$ \\
                                                &                         & $\text{C}_\text{0}$ & $0.8771$ & $0.9995$ & $\textbf{\textcolor{mygreen2}{0.7741}}$ & $\textbf{\textcolor{mygreen2}{0.7453}}$ & $\textbf{\textcolor{mygreen2}{0.6830}}$ & $\textbf{\textcolor{mygreen1}{0.7905}}$ & $0.9235$ & $0.8409$ & $0.8199$ \\ \cline{3-12}
                                                &                         & $\text{Avg. Process Time (sec)}$ & $0.0553$ & $0.0353$ & $0.0269$ & $0.0268$ & $0.6734$ & $0.4868$ & $0.2737$ & $0.0372$ & $0.4479$ \\ \hlinewd{.5pt}
\multirow{6}{*}{SPARISH \cite{li2016image}}   & \multirow{6}{*}{$2016$}   & PLCC & $0.9447$ & $0.9380$ & $0.8900$ & $0.9020$ & $0.3448$ & $0.5726$ & $0.3459$ & $0.9220$ & $0.4013$ \\
																								&                         & SRCC & $0.9856$ & $0.9139$ & $0.8836$ & $0.8940$ & $0.3408$ & $0.5581$ & $0.3566$ & $0.9233$ & $0.4014$ \\
                                                &                         & $\text{AUC}_\text{DS}$ & $\textbf{\textcolor{mygreen1}{0.8411}}$ & $0.8747$ & $0.4985$ & $0.5040$ & $0.5403$ & $\textbf{\textcolor{mygreen1}{0.6484}}$ & $0.5821$ & $0.5603$ & $0.6295$ \\
                                                &                         & $\text{AUC}_\text{BW}$ & $\textbf{\textcolor{mygreen1}{0.9901}}$ & $1.0000$ & $0.8471$ & $0.8290$ & $0.6813$ & $0.8589$ & $0.6620$ & $0.9364$ & $0.7179$ \\
                                                &                         & $\text{C}_\text{0}$ & $\textbf{\textcolor{mygreen1}{0.9510}}$ & $0.9989$ & $0.7424$ & $0.7283$ & $0.6386$ & $0.7869$ & $0.6377$ & $0.8689$ & $0.6715$ \\ \cline{3-12}
                                                &                         & $\text{Avg. Process Time (sec)}$ & $2.6738$ & $1.9937$ & $1.5043$ & $1.5386$ & $27.6209$ & $16.1626$ & $14.5033$ & $1.9798$ & $18.9073$ \\ \hlinewd{.5pt}
\multirow{6}{*}{RISE \cite{li2017no}}   				& \multirow{6}{*}{$2017$} & PLCC & $\textbf{\textcolor{mygreen2}{0.9897}}$ & $\textbf{\textcolor{mygreen1}{0.9572}}$ & $0.9171$ & $0.9279$ & $0.1927$ & $0.3839$ & $0.6509$ & $\textbf{\textcolor{mygreen1}{0.9515}}$ & $0.4456$ \\
																								&                         & SRCC & $\textbf{\textcolor{mygreen1}{0.9886}}$ & $\textbf{\textcolor{mygreen1}{0.9389}}$ & $\textbf{\textcolor{mygreen1}{0.9135}}$ & $\textbf{\textcolor{mygreen1}{0.9318}}$ & $0.2099$ & $0.2178$ & $0.6566$ & $\textbf{\textcolor{mygreen1}{0.9462}}$ & $0.4125$ \\
                                                &                         & $\text{AUC}_\text{DS}$ & $\textbf{\textcolor{mygreen2}{0.8877}}$ & $\textbf{\textcolor{mygreen1}{0.9522}}$ & $0.5109$ & $0.5511$ & $0.5227$ & $0.5383$ & $0.6543$ & $\textbf{\textcolor{mygreen1}{0.5869}}$ & $0.5723$ \\
                                                &                         & $\text{AUC}_\text{BW}$ & $\textbf{\textcolor{mygreen2}{0.9931}}$ & $1.0000$ & $0.8459$ & $0.8236$ & $0.4204$ & $0.3894$ & $0.9179$ & $\textbf{\textcolor{mygreen1}{0.9392}}$ & $0.6790$ \\
                                                &                         & $\text{C}_\text{0}$ & $\textbf{\textcolor{mygreen2}{0.9594}}$ & $1.0000$ & $0.7468$ & $0.7342$ & $0.4520$ & $0.4331$ & $0.8442$ & $\textbf{\textcolor{mygreen2}{0.8754}}$ & $0.6317$ \\ \cline{3-12}
                                                &                         & $\text{Avg. Process Time (sec)}$ & $0.7881$ & $0.6009$ & $0.4631$ & $0.4663$ & $10.6461$ & $7.2797$ & $3.8973$ & $0.5942$ & $6.7861$ \\ \hlinewd{.5pt}
\multirow{2}{*}{Yu's CNN \cite{yu2017shallow}}   & \multirow{2}{*}{$2017$}& PLCC & $0.9735$ & $0.9416$ & $\textbf{\textcolor{mygreen2}{0.9374}}$ & $0.9221$ & $-$ & $-$ & $-$ & $0.9449$ & $-$ \\
																								&                         & SRCC & $0.9646$ & $0.9253$ & $0.9189$ & $0.9135$ & $-$ & $-$ & $-$ & $0.9322$ & $-$ \\ \hlinewd{.5pt}
\multirow{6}{*}{Synthetic-MaxPol \cite{mahdi2018image}}  & \multirow{6}{*}{$2018$} 	& PLCC & $0.9773$ & $\textbf{\textcolor{mygreen2}{0.9655}}$ & $\textbf{\textcolor{mygreen1}{0.9331}}$ & $\textbf{\textcolor{mygreen2}{0.9414}}$ & $0.3237$ & $0.6582$ & $0.8063$ & $\textbf{\textcolor{mygreen2}{0.9567}}$ & $0.6228$ \\
																								&                         & SRCC & $0.9706$ & $\textbf{\textcolor{mygreen2}{0.9478}}$ & $\textbf{\textcolor{mygreen2}{0.9397}}$ & $\textbf{\textcolor{mygreen2}{0.9454}}$ & $0.2717$ & $0.5310$ & $0.8153$ & $\textbf{\textcolor{mygreen2}{0.9520}}$ & $0.5797$ \\
                                                &                         & $\text{AUC}_\text{DS}$ & $0.8211$ & $\textbf{\textcolor{mygreen2}{0.9611}}$ & $0.5536$ & $0.5529$ & $0.5478$ & $0.5969$ & $0.7553$ & $0.5579$ & $0.6434$ \\
                                                &                         & $\text{AUC}_\text{BW}$ & $0.9810$ & $1.0000$ & $0.8532$ & $0.8269$ & $0.6445$ & $0.8382$ & $0.9841$ & $0.9327$ & $0.8719$ \\
                                                &                         & $\text{C}_\text{0}$ & $0.9308$ & $1.0000$ & $\textbf{\textcolor{mygreen1}{0.7610}}$ & $0.7347$ & $0.5967$ & $0.7436$ & $0.9380$ & $0.8633$ & $0.7897$ \\ \cline{3-12}
                                                &                         & $\text{Avg. Process Time (sec)}$ & $0.0819$ & $0.0571$ & $0.0443$ & $0.0442$ & $0.8030$ & $0.6132$ & $0.3059$ & $0.0584$ & $0.5330$ \\ \hlinewd{.5pt}
\multirow{6}{*}{HVS-MaxPol-1}                   & \multirow{6}{*}{$2019$} & PLCC & $0.9783$ & $0.9511$ & $0.8618$ & $0.8891$ & $0.3601$ & $\textbf{\textcolor{mygreen2}{0.7620}}$ & $0.8216$ & $0.9266$ & $\textbf{\textcolor{mygreen1}{0.6663}}$ \\
																								&                         & SRCC & $0.9850$ & $0.9240$ & $0.8602$ & $0.8802$ & $0.3625$ & $\textbf{\textcolor{mygreen2}{0.7081}}$ & $0.8146$ & $0.9182$ & $\textbf{\textcolor{mygreen1}{0.6506}}$ \\
                                                &                         & $\text{AUC}_\text{DS}$ & $0.8178$ & $0.9499$ & $\textbf{\textcolor{mygreen2}{0.5749}}$ & $\textbf{\textcolor{mygreen2}{0.5917}}$ & $\textbf{\textcolor{mygreen1}{0.5649}}$ & $\textbf{\textcolor{mygreen2}{0.6635}}$ & $0.7561$ & $0.5140$ & $0.6838$ \\
                                                &                         & $\text{AUC}_\text{BW}$ & $0.9802$ & $1.0000$ & $0.8398$ & $0.8226$ & $0.7207$ & $\textbf{\textcolor{mygreen2}{0.9185}}$ & $0.9841$ & $0.9181$ & $\textbf{\textcolor{mygreen1}{0.8954}}$ \\
                                                &                         & $\text{C}_\text{0}$ & $0.9244$ & $1.0000$ & $0.7529$ & $\textbf{\textcolor{mygreen1}{0.7380}}$ & $0.6591$ & $\textbf{\textcolor{mygreen2}{0.8336}}$ & $0.9343$ & $0.8596$ & $\textbf{\textcolor{mygreen1}{0.8274}}$ \\ \cline{3-12}
                                                &                         & $\text{Avg. Process Time (sec)}$ & $\textbf{\textcolor{mygreen2}{0.0176}}$ & $\textbf{\textcolor{mygreen2}{0.0128}}$ & $\textbf{\textcolor{mygreen2}{0.0100}}$ & $\textbf{\textcolor{mygreen2}{0.0099}}$ & $\textbf{\textcolor{mygreen2}{0.1752}}$ & $\textbf{\textcolor{mygreen2}{0.1342}}$ & $\textbf{\textcolor{mygreen2}{0.0560}}$ & $\textbf{\textcolor{mygreen2}{0.0129}}$ & $\textbf{\textcolor{mygreen2}{0.1116}}$ \\ \hlinewd{.5pt}
\multirow{6}{*}{HVS-MaxPol-2}                   & \multirow{6}{*}{$2019$}& PLCC & $\textbf{\textcolor{mygreen1}{0.9868}}$ & $0.9450$ & $0.8549$ & $0.8867$ & $\textbf{\textcolor{mygreen2}{0.4659}}$ & $\textbf{\textcolor{mygreen1}{0.7330}}$ & $\textbf{\textcolor{mygreen1}{0.8538}}$ & $0.9253$ & $\textbf{\textcolor{mygreen2}{0.7059}}$ \\
																								&                         & SRCC & $0.9485$ & $0.9208$ & $0.8507$ & $0.8746$ & $\textbf{\textcolor{mygreen2}{0.4475}}$ & $\textbf{\textcolor{mygreen1}{0.6107}}$ & $\textbf{\textcolor{mygreen1}{0.8574}}$ & $0.9039$ & $\textbf{\textcolor{mygreen2}{0.6718}}$ \\
                                                &                         & $\text{AUC}_\text{DS}$ & $0.7915$ & $0.9424$ & $\textbf{\textcolor{mygreen1}{0.5715}}$ & $\textbf{\textcolor{mygreen1}{0.5825}}$ & $\textbf{\textcolor{mygreen2}{0.5859}}$ & $0.6290$ & $\textbf{\textcolor{mygreen2}{0.8101}}$ & $\textbf{\textcolor{mygreen2}{0.6807}}$ & $0.6760$ \\
                                                &                         & $\text{AUC}_\text{BW}$ & $0.9707$ & $1.0000$ & $0.8374$ & $0.8218$ & $\textbf{\textcolor{mygreen2}{0.7533}}$ & $\textbf{\textcolor{mygreen1}{0.8815}}$ & $\textbf{\textcolor{mygreen2}{0.9948}}$ & $0.9368$ & $\textbf{\textcolor{mygreen2}{0.9144}}$ \\
                                                &                         & $\text{C}_\text{0}$ & $0.9057$ & $0.9995$ & $0.7475$ & $0.7309$ & $0.6800$ & $0.7865$ & $\textbf{\textcolor{mygreen1}{0.9621}}$ & $0.8476$ & $\textbf{\textcolor{mygreen2}{0.8362}}$ \\ \cline{3-12}
                                                &                         & $\text{Avg. Process Time (sec)}$ & $\textbf{\textcolor{mygreen1}{0.0338}}$ & $\textbf{\textcolor{mygreen1}{0.0250}}$ & $\textbf{\textcolor{mygreen1}{0.0193}}$ & $\textbf{\textcolor{mygreen1}{0.0196}}$ & $\textbf{\textcolor{mygreen1}{0.3493}}$ & $\textbf{\textcolor{mygreen1}{0.2597}}$ & $\textbf{\textcolor{mygreen1}{0.1095}}$ & $\textbf{\textcolor{mygreen1}{0.0251}}$ & $\textbf{\textcolor{mygreen1}{0.2195}}$ \\ \hlinewd{.5pt}
\end{tabular}
\end{table*}

\subsection{Performance Evaluation}
Table \ref{methods_table} demonstrates the performance scores for all NR-ISAs using five different dataset. In natural imaging, HVS-MaxPol-2 achieves the highest overall scores in four categories of PLCC, SRCC, $\text{AUC}_\text{BW}$, and $\text{C}_\text{0}$, leading the second best metric HVS-MaxPol-1 for about $3.96\%$, $2.12\%$, $1.9\%$, and $0.88\%$, respectively. It worth noting that both metrics are the fastest compared to the other competing NR-ISAs. Furthermore, both MLV and $\text{S}_{3}$ provide better separability between different and similar blurred image pair i.e. $\text{AUC}_\text{DS}$ ($0.7164$ and $0.7071$, respectively) compared to the rest of the NR-ISAs. However, $\text{S}_{3}$ takes $43.3513$ second in average to process one natural image which is $198$ times slower than the best overall method HVS-MaxPol-2 (MLV is slightly slower for about 1.58 in ratio). With respect to individual natural dataset, HVS-MaxPol-2 shows its robustness toward scoring diverse homogenous and non-homogenous blur types across three dataset. Furthermore, HVS-MaxPol-1 provides superior performance on CID2013 related to luminance and distance blur. MLV also achieves three best performance on FocusPath (PLCC, SRCC, and $\text{C}_0$), leading the second best HVS-MaxPol-2. GPC and $\text{S}_{3}$ also lead the best blur classification rate in BID showing their performance scoring across different blur types.

For synthetic blur scoring, Synthetic-MaxPol outperforms the best overall performances on PLCC and SRCC measures. In addition, HVS-MaxPol-2, $\text{ARISM}_{\text{C}}$, and RISE achieve the best performance on $\text{AUC}_\text{DS}$, $\text{AUC}_\text{BW}$, and $\text{C}_\text{0}$, respectively. It also worth noting that we have reported RISE metric performance using all data. Notices the high performance in LIVE which is trained over synthetic blur dataset.

\begin{table*}[htp]
\renewcommand{\arraystretch}{1.3}
\caption{Statistical significance performances on Natural image blur dataset of BID, CID2013, and FocusPath using nine different NR-ISAs. The green-color-squares indicate the selected metric in the row performs significantly better (`+1') than the metric in the column, the red-color-squares indicate the opposite, and black-color squares indicate that there is no statistically significant difference between performances.}
\label{statistical_performance_overall_natural_databases}\vspace{-.1in}
\begin{center}
\begin{tabular}{|c|c|c|c|c|}
\cline{2-5}
\multicolumn{1}{c|}{} &  BID & CID2013 & FocusPath & All Natural Datasets \\
\cline{1-5} 
{\hspace{-.05in}\begin{sideways} \hspace{.7in}AUC-DS \end{sideways}\hspace{-.05in}} &
{\hspace{-.05in}\includegraphics[width=0.225\textwidth]{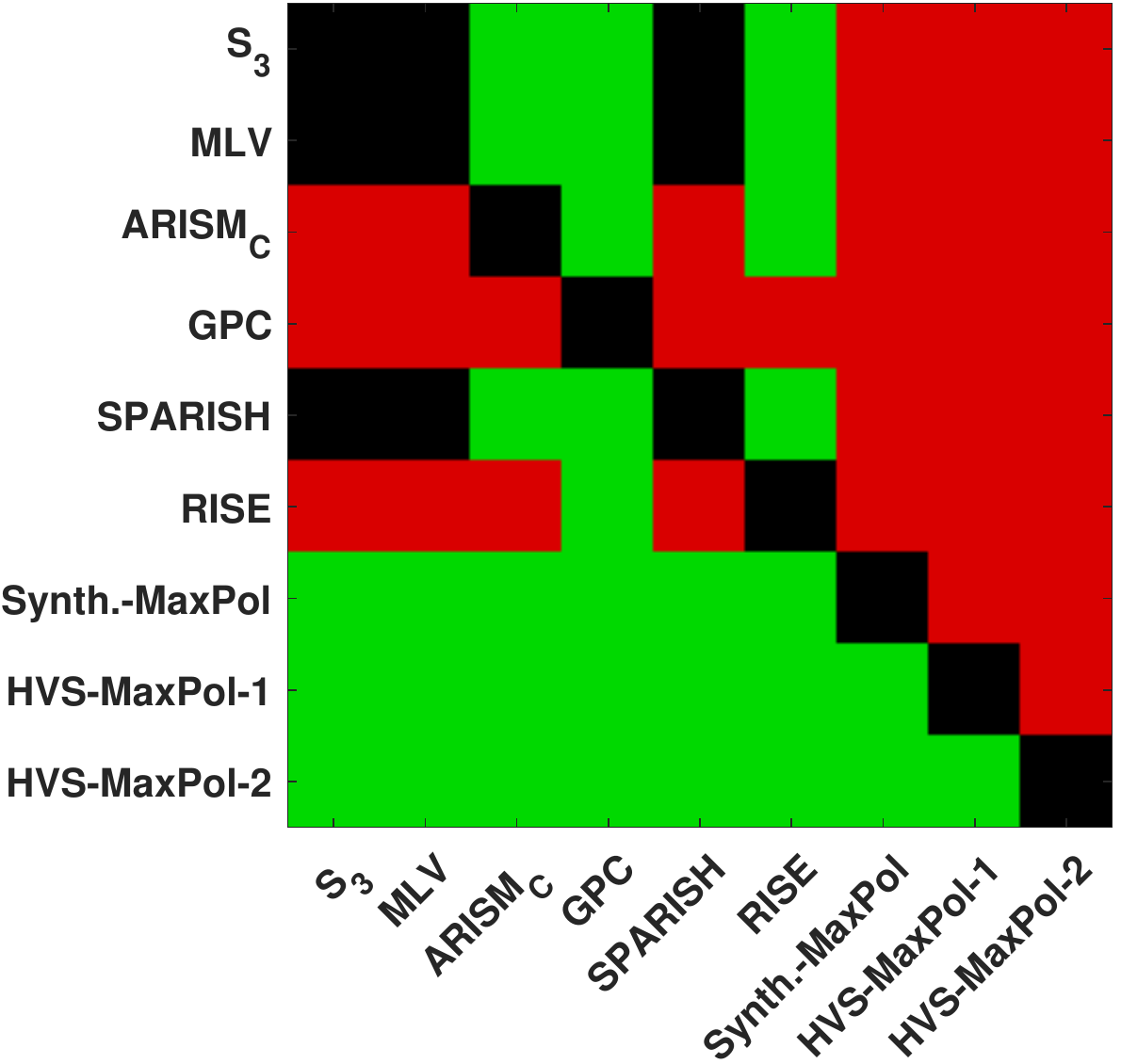}\hspace{-.05in}} &
{\hspace{-.05in}\includegraphics[width=0.225\textwidth]{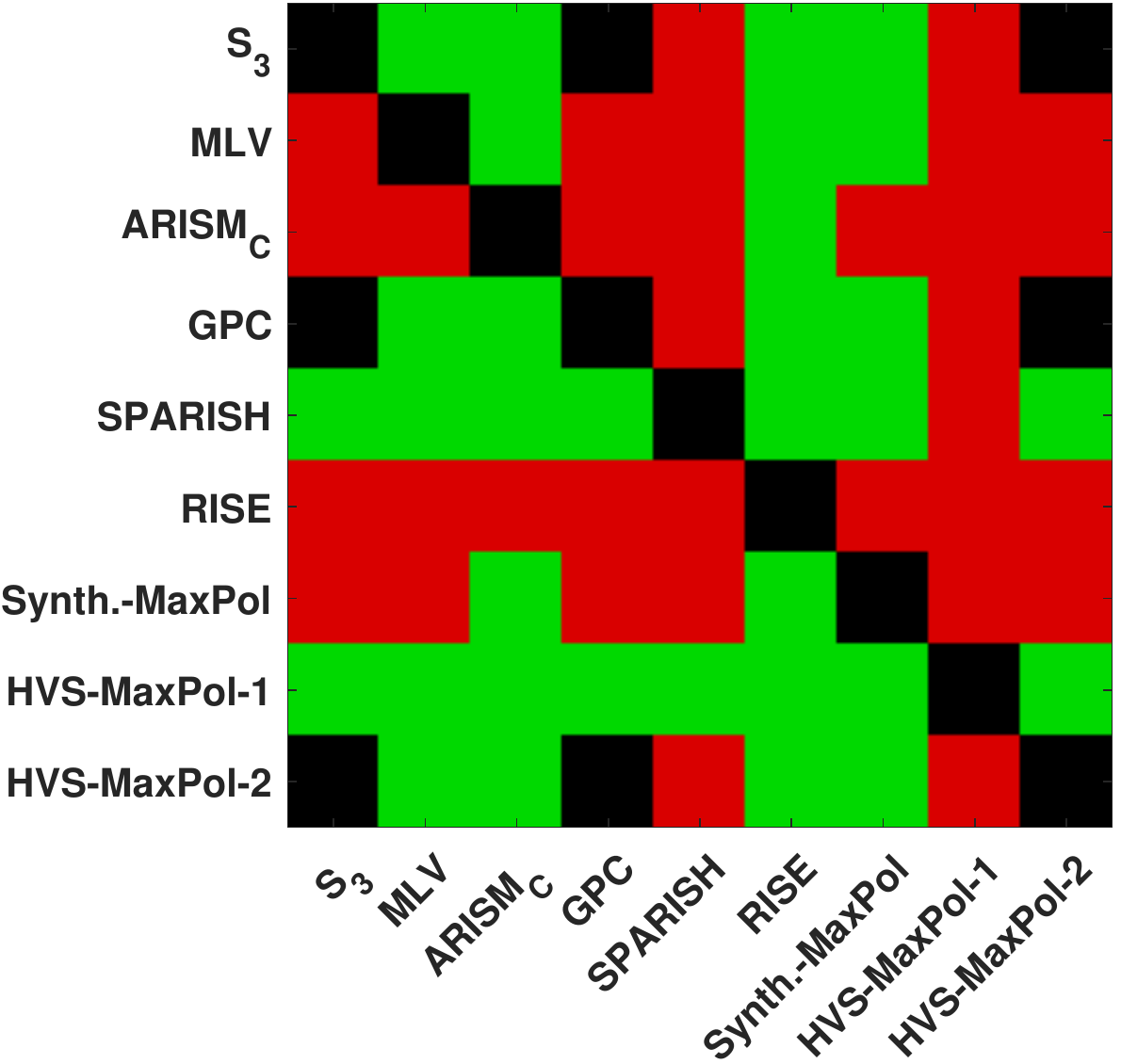}\hspace{-.05in}} &
{\hspace{-.05in}\includegraphics[width=0.225\textwidth]{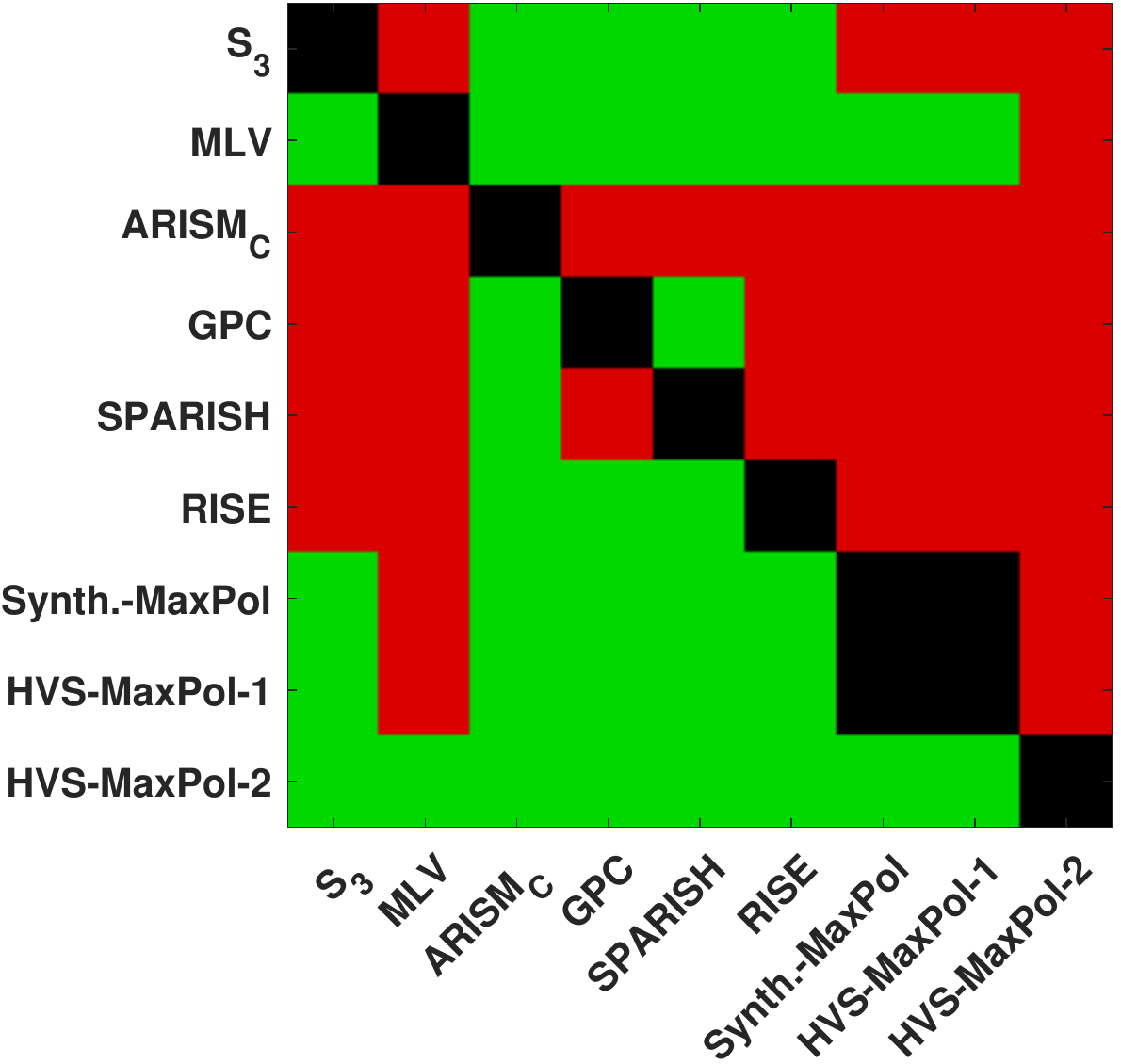}\hspace{-.05in}} &
{\hspace{-.05in}\includegraphics[width=0.225\textwidth]{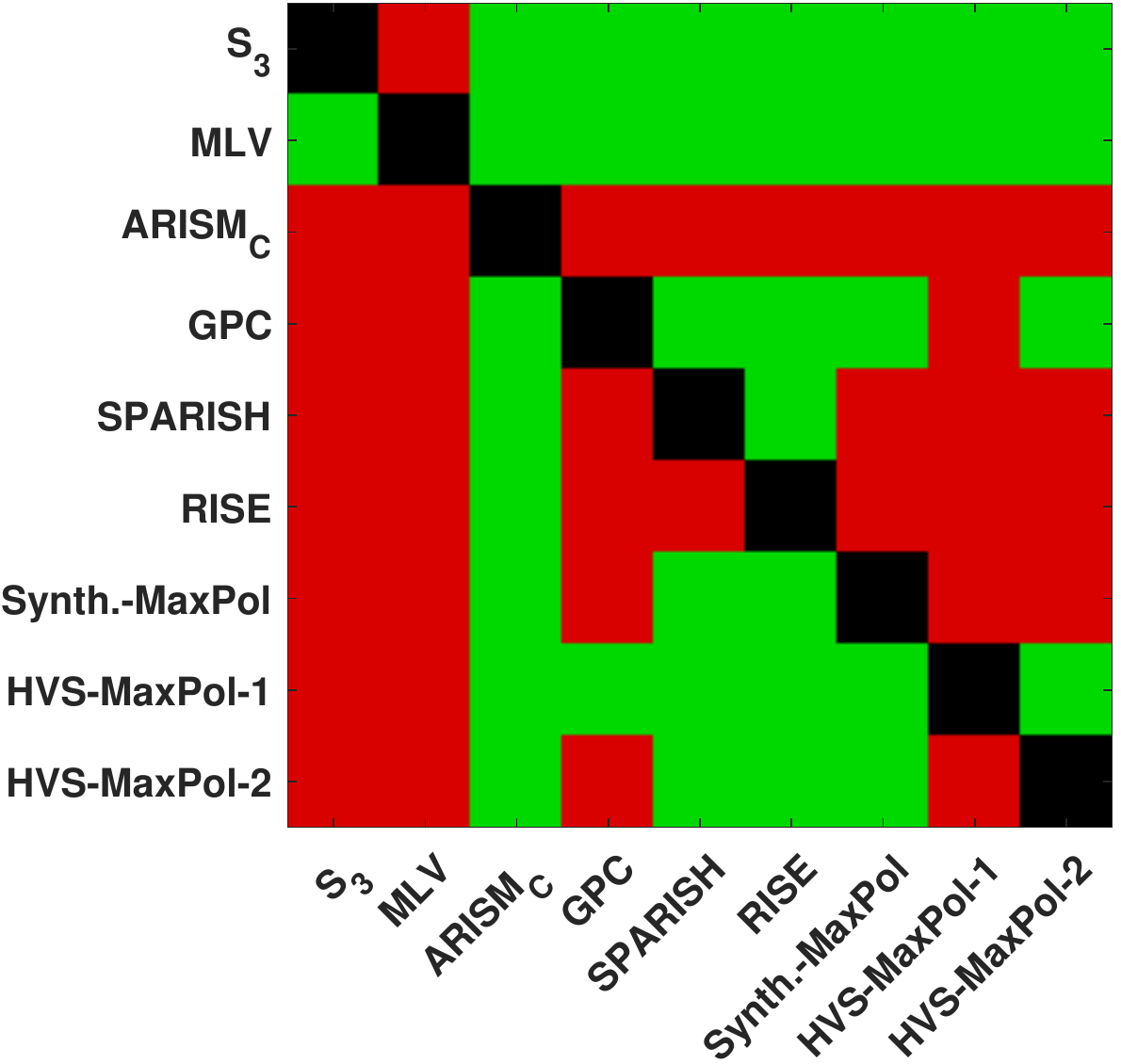}\hspace{-.05in}} \\
\hline
\cline{1-5}
{\hspace{-.05in}\begin{sideways} \hspace{.7in}C0-BW \end{sideways}\hspace{-.05in}} &
{\hspace{-.05in}\includegraphics[width=0.225\textwidth]{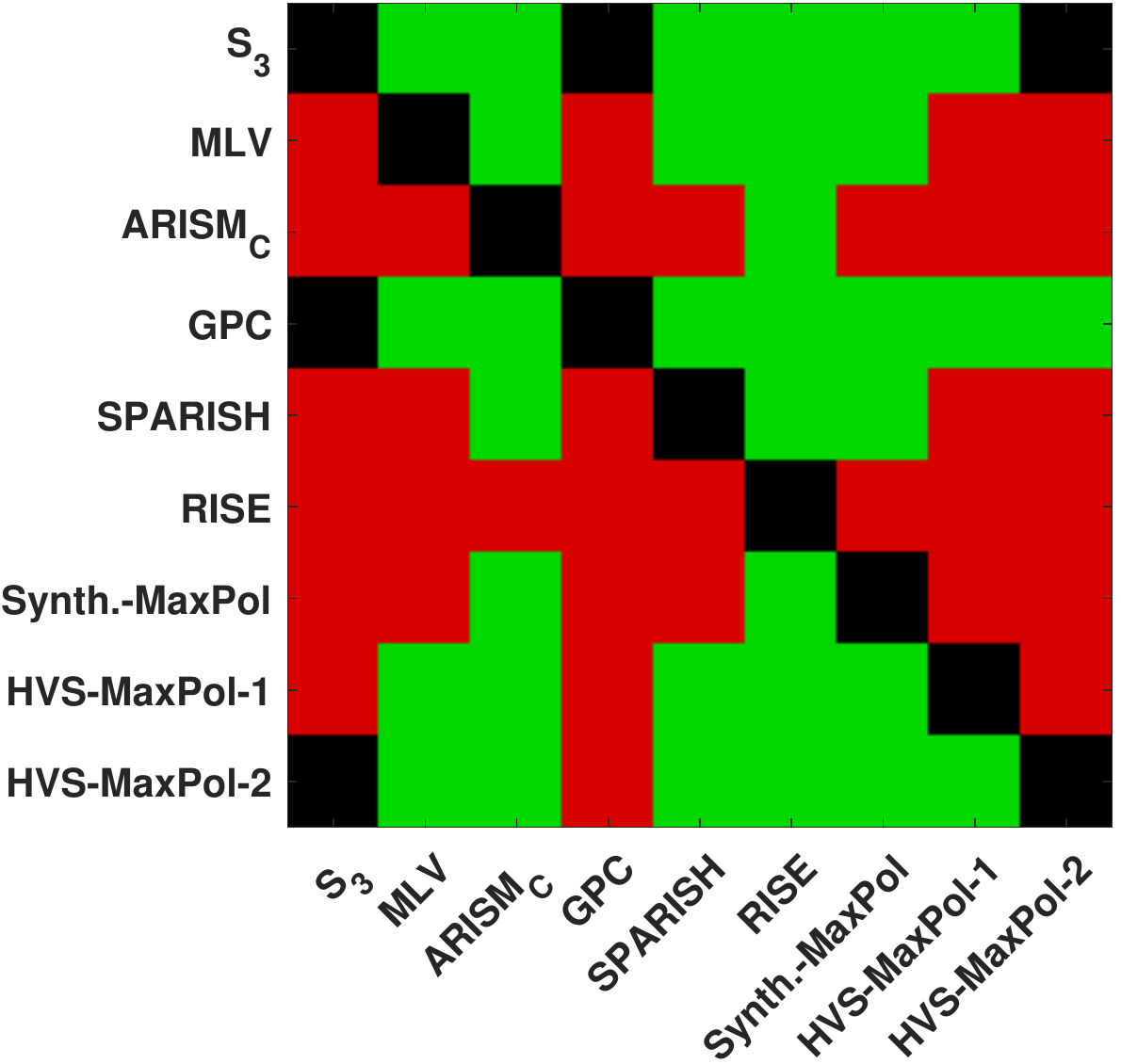}\hspace{-.05in}} &
{\hspace{-.05in}\includegraphics[width=0.225\textwidth]{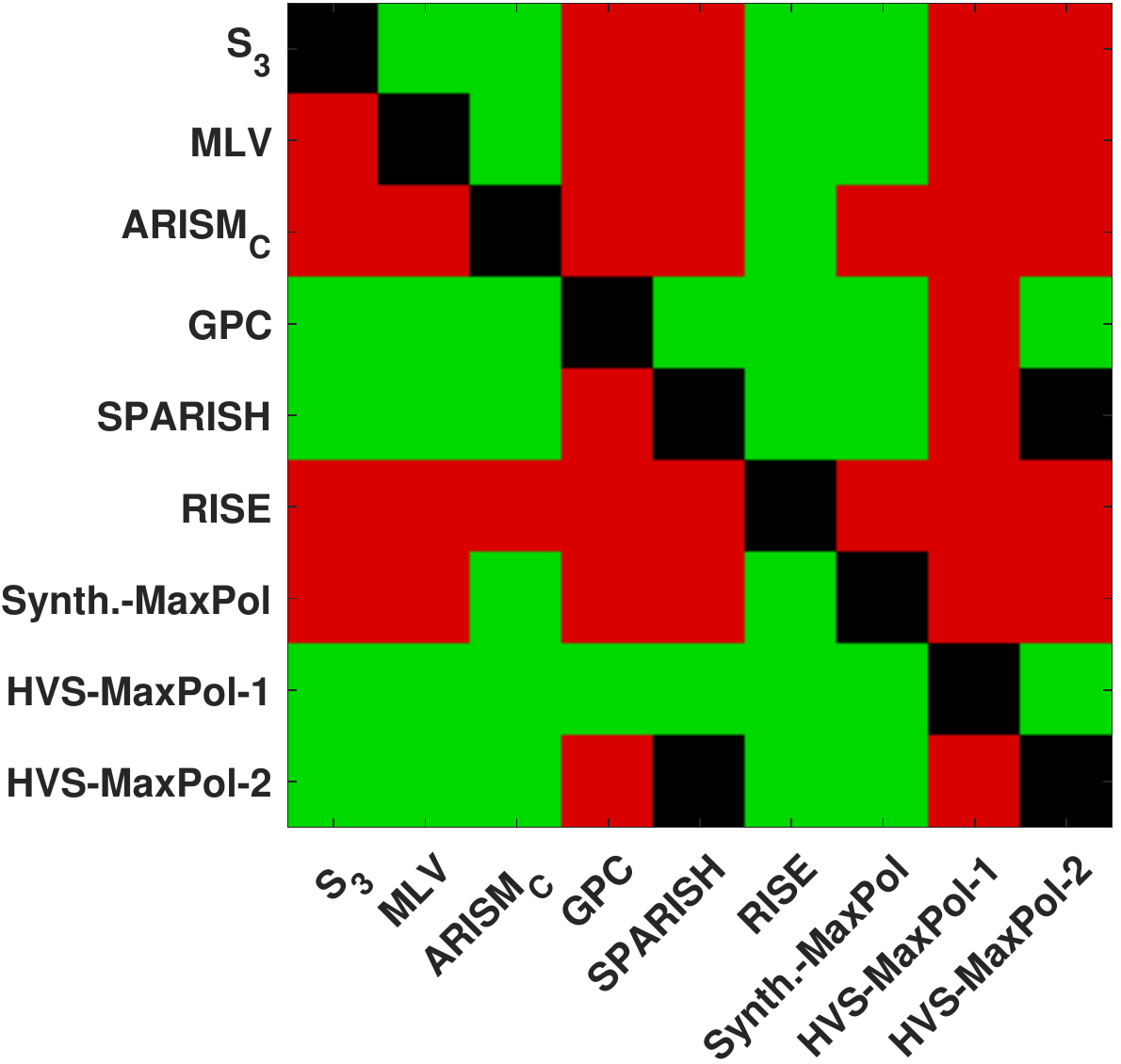}\hspace{-.05in}} &
{\hspace{-.05in}\includegraphics[width=0.225\textwidth]{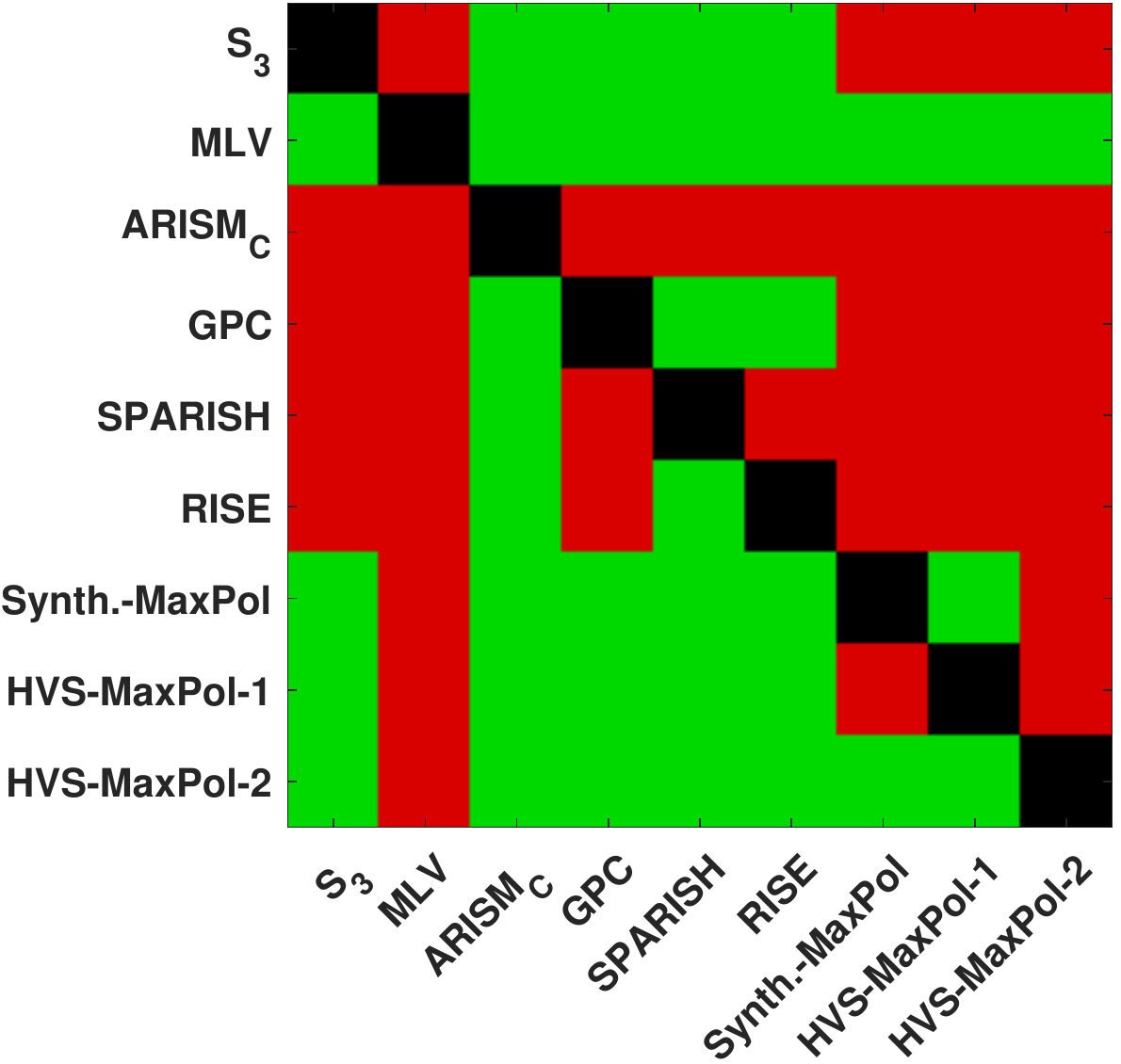}\hspace{-.05in}} &
{\hspace{-.05in}\includegraphics[width=0.225\textwidth]{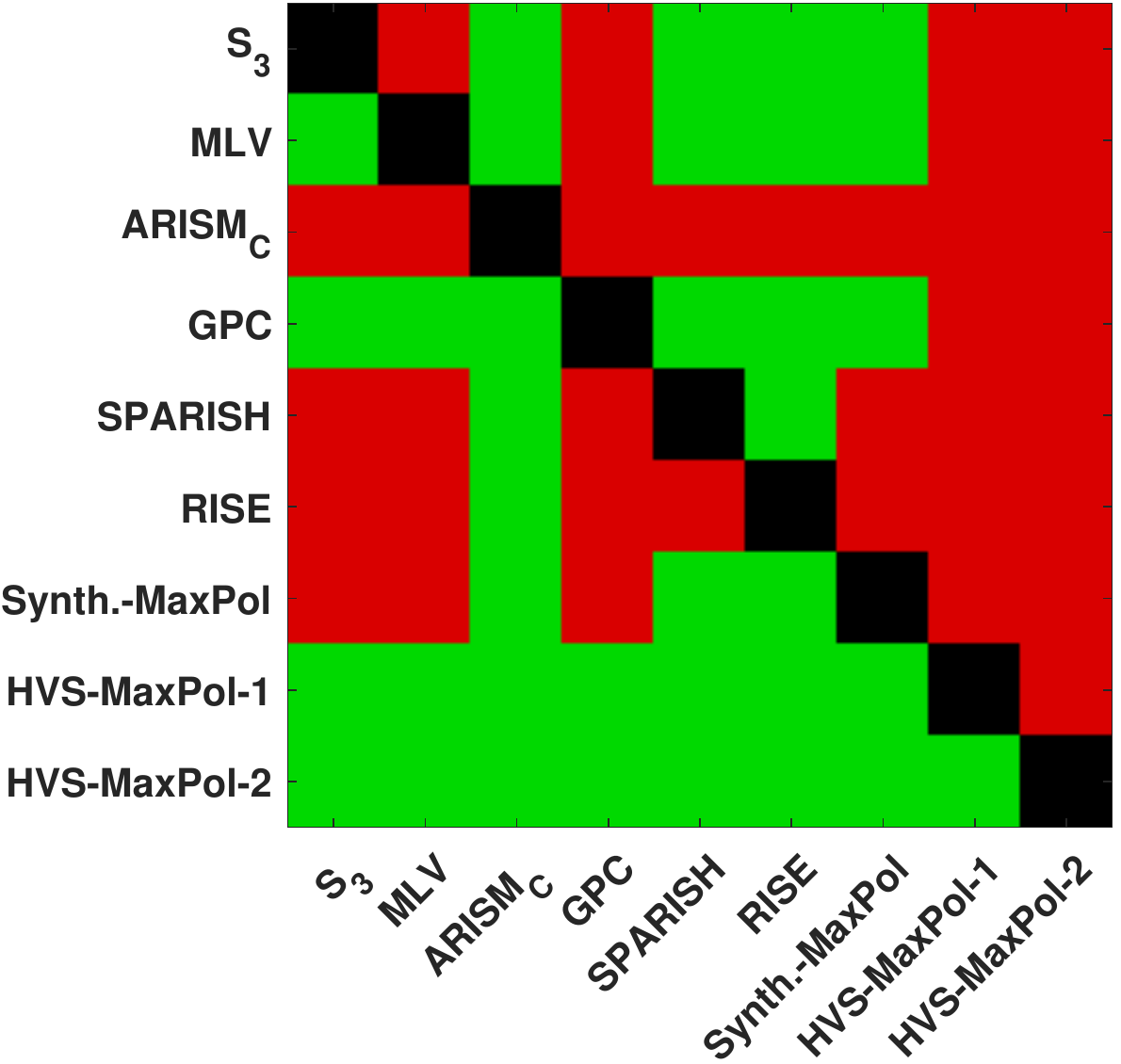}\hspace{-.05in}} \\
\hline
\cline{1-5} 
{\hspace{-.05in}\begin{sideways} \hspace{.7in}AUC-BW \end{sideways}\hspace{-.05in}} &
{\hspace{-.05in}\includegraphics[width=0.225\textwidth]{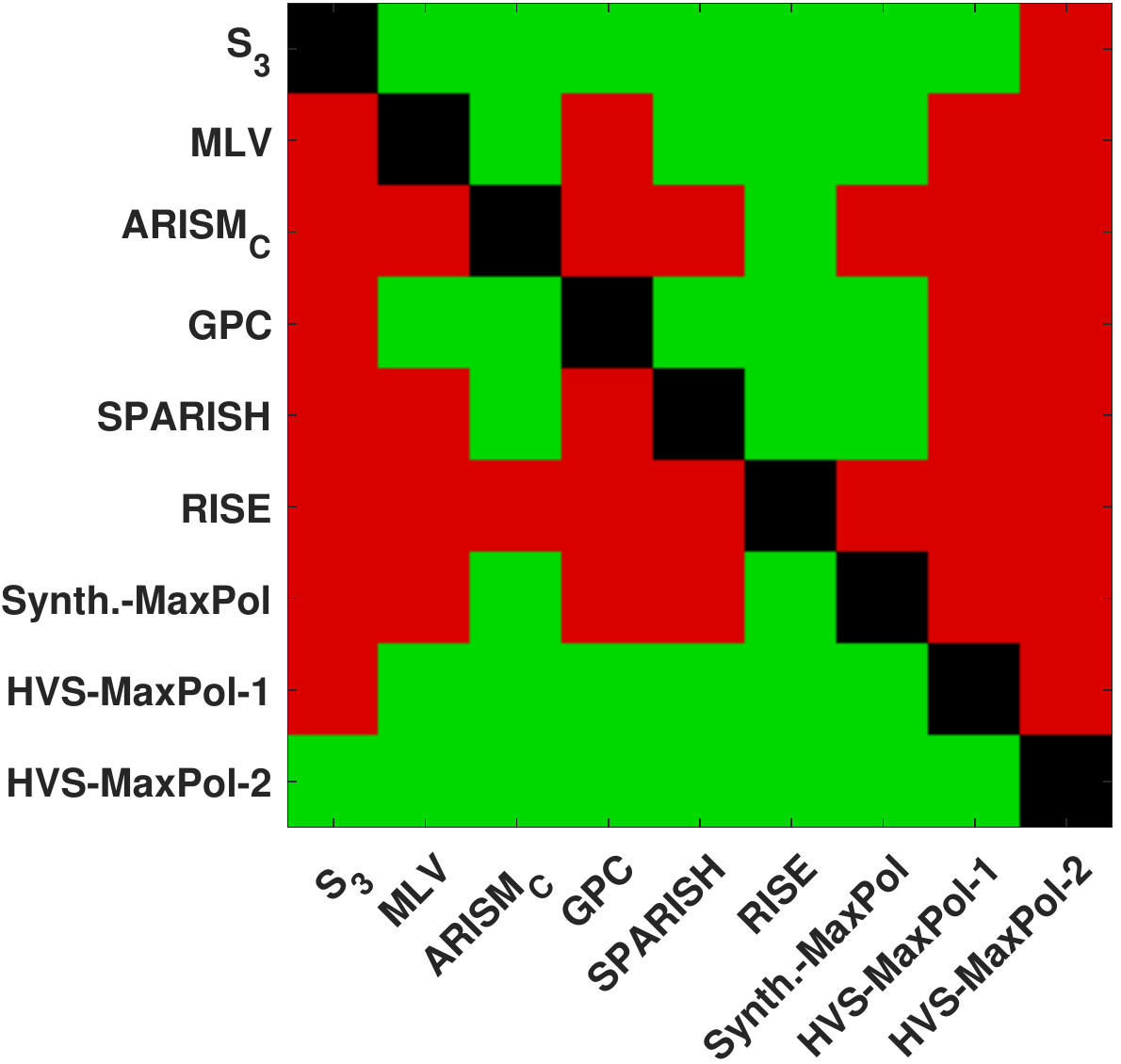}\hspace{-.05in}} &
{\hspace{-.05in}\includegraphics[width=0.225\textwidth]{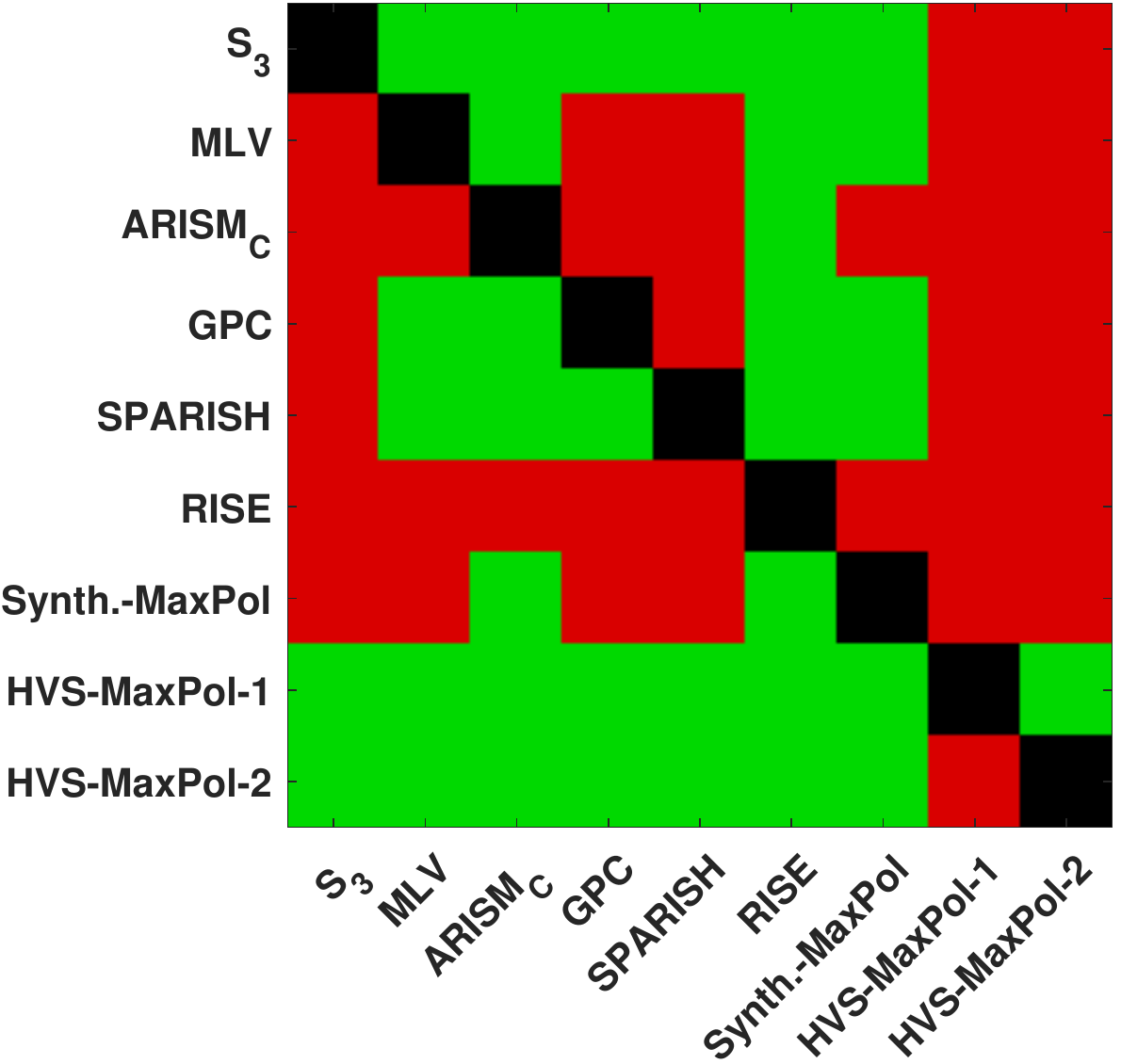}\hspace{-.05in}} &
{\hspace{-.05in}\includegraphics[width=0.225\textwidth]{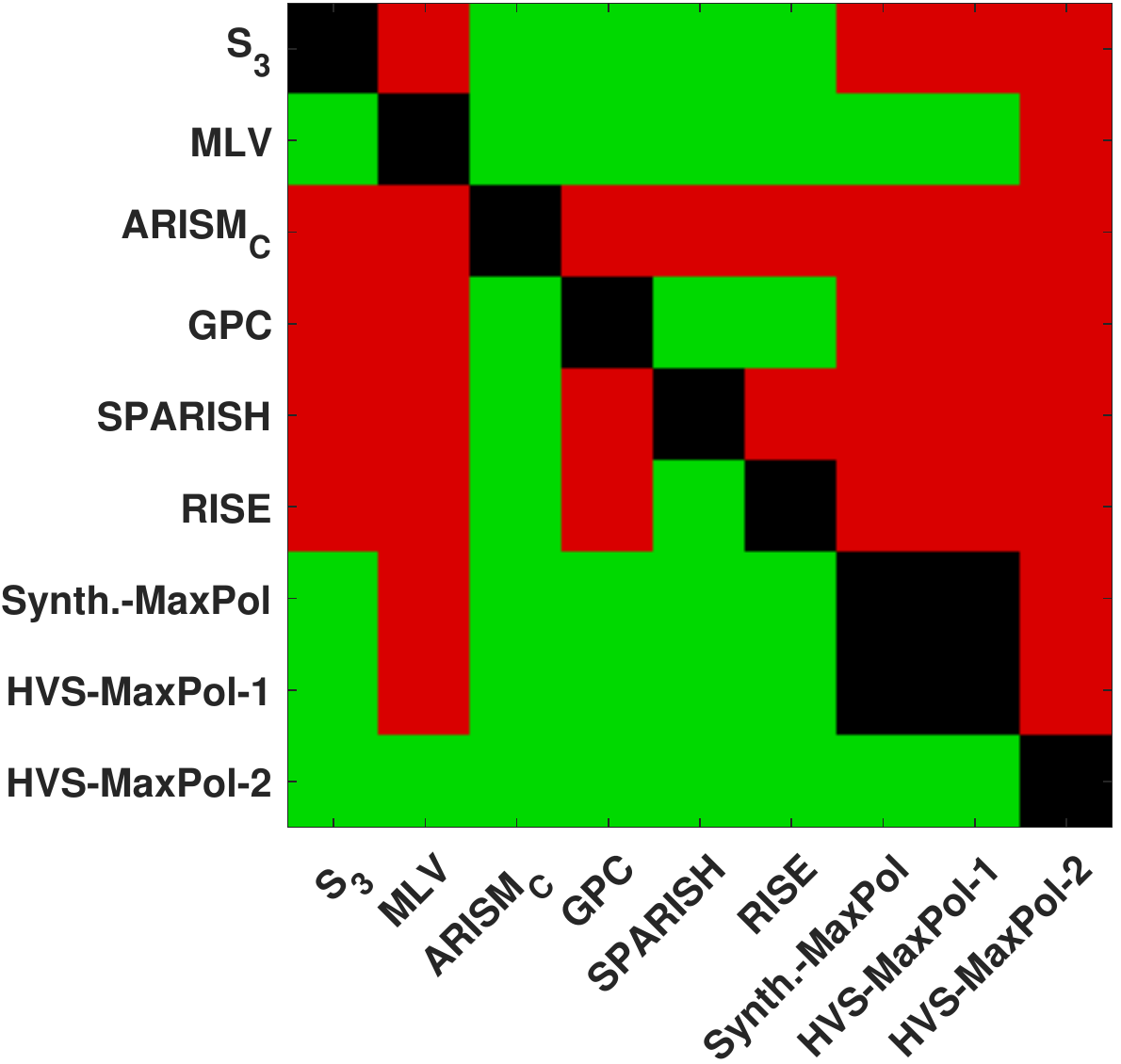}\hspace{-.05in}} &
{\hspace{-.05in}\includegraphics[width=0.225\textwidth]{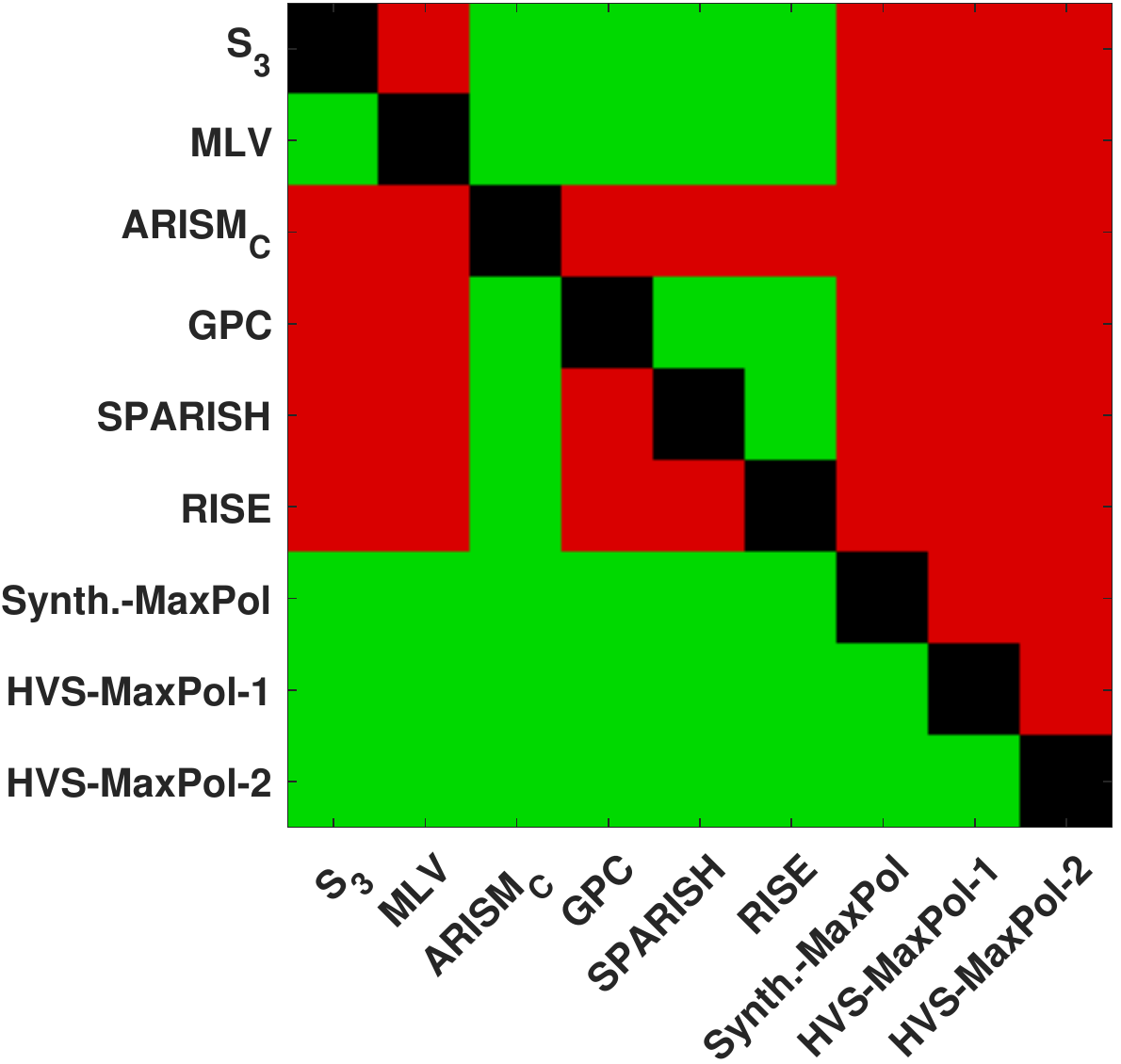}\hspace{-.05in}} \\
\hline
\end{tabular}
\end{center}
\end{table*}

We conduct the statistical significance analysis across all NR-ISAs over three natural dataset by calculating the probability of the critical ratio between AUCs defined in \cite{krasula2016accuracy}. $95\%$ confidence interval is used to define the significance over the statistical distribution. The results are visualized in Figure \ref{statistical_performance_overall_natural_databases}. Green square between metrics indicate that the corresponding metric across the row is significantly better than the metric across its column, while the red square indicates the opposite. The black square also is an indication of no statistical significance is observed between crossing metrics. Both HVS-MaxPol-1 and HVS-MaxPol-2 outperform the other methods in overall natural dataset under $\text{C}_\text{0}$ and $\text{AUC}_\text{BW}$ performance metrics, indicating that both measures provide better correct classification rate and recognizing capability in significantly different pair of blured images. Whereas, MLV outperforms the other methods under $\text{AUC}_\text{DS}$ implying better separation between different/similar blurred image pair.

\begin{figure*}[htp]
\scriptsize
\centerline{
\subfigure[LIVE]{\includegraphics[height=0.2\textwidth]{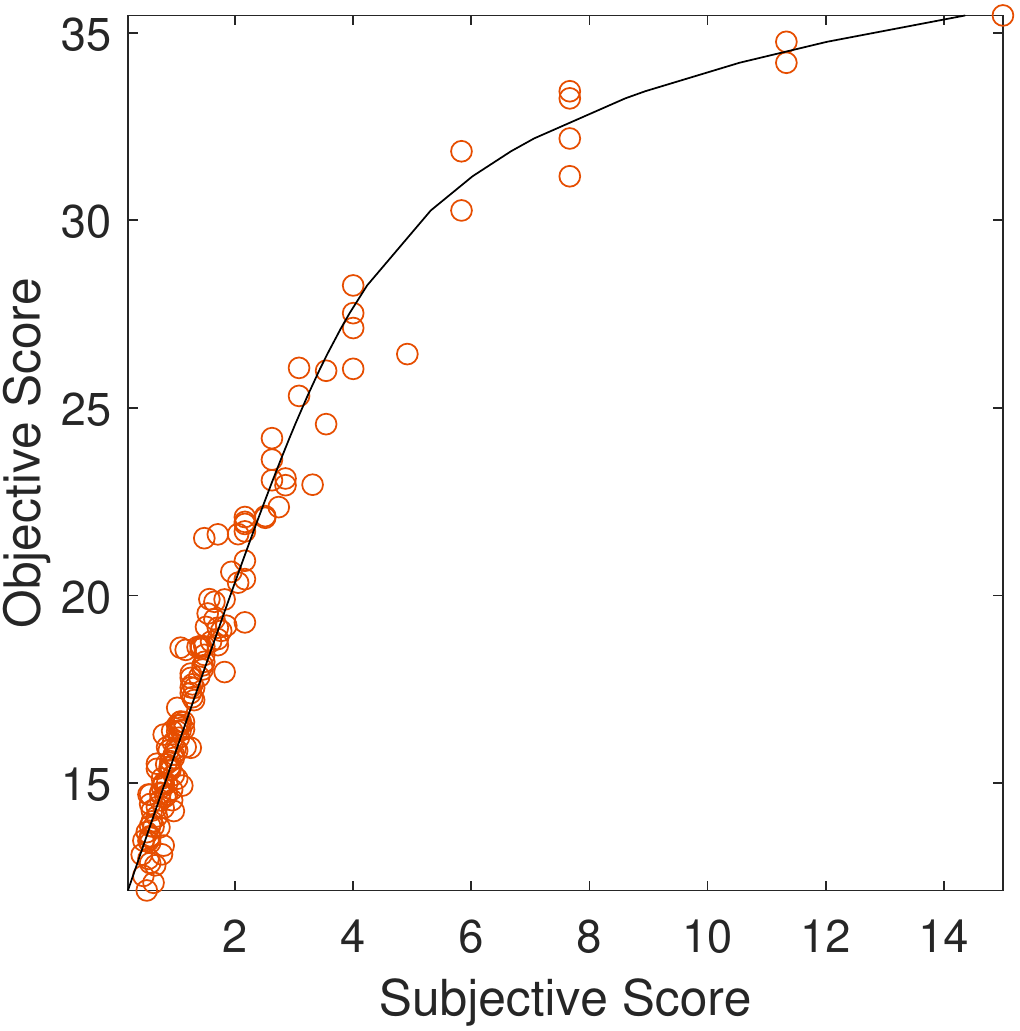}}\hspace{.12in}
\subfigure[CSIQ]{\includegraphics[height=0.2\textwidth]{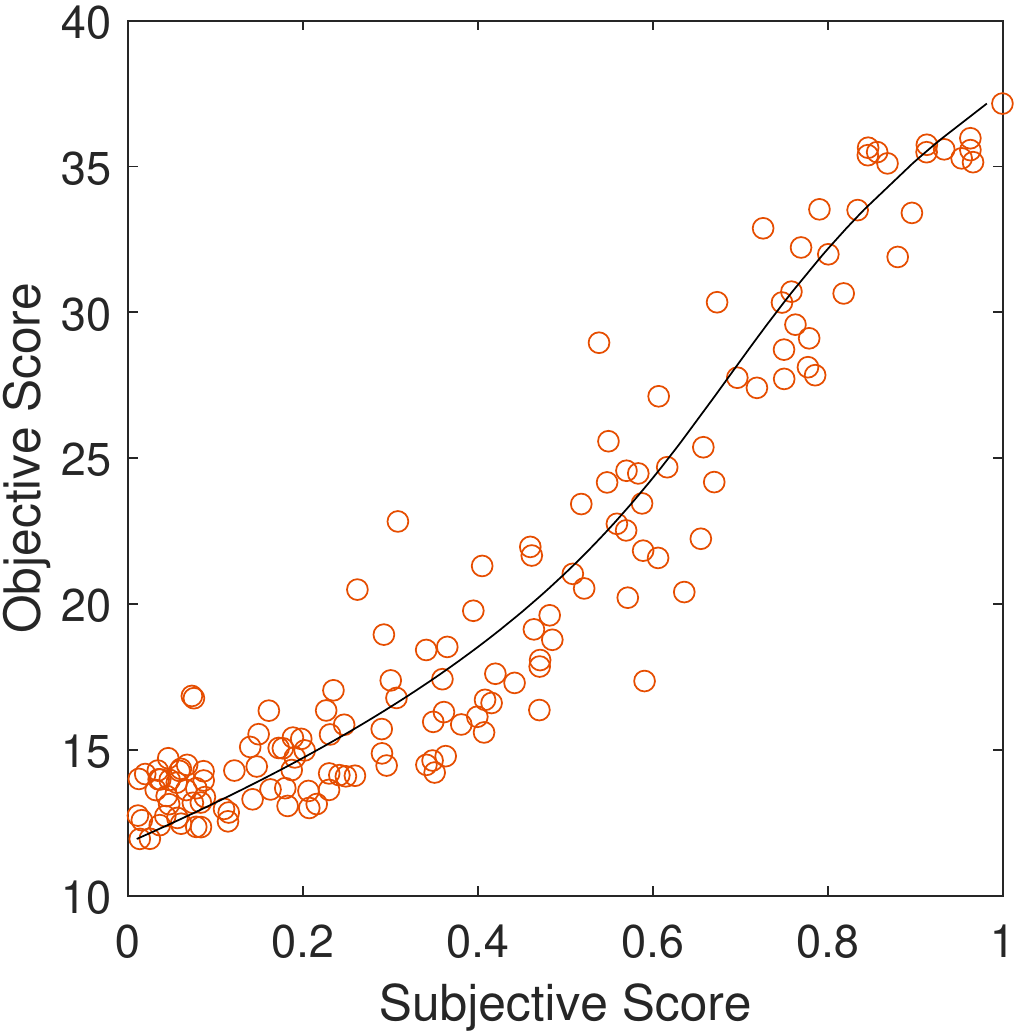}}\hspace{.12in}
\subfigure[TID2008]{\includegraphics[height=0.2\textwidth]{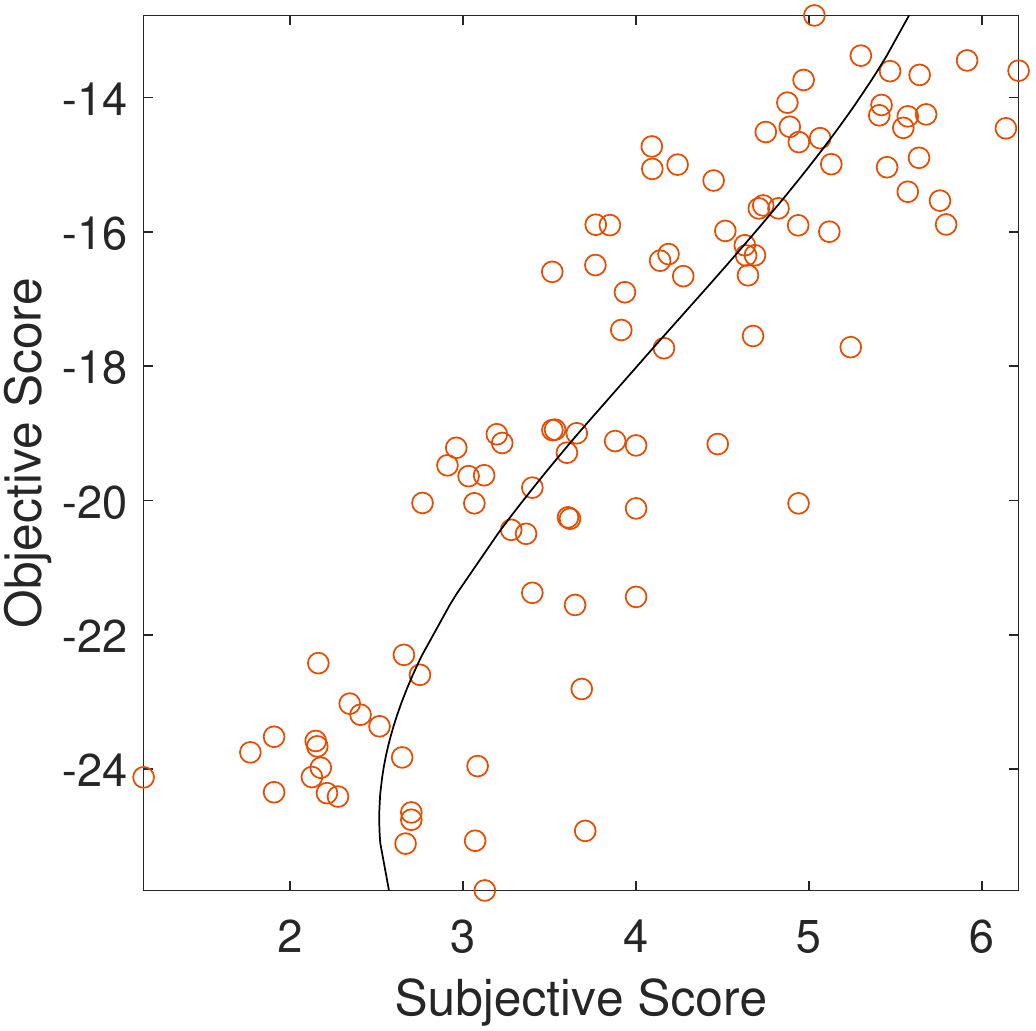}}\hspace{.12in}
\subfigure[TID2013]{\includegraphics[height=0.2\textwidth]{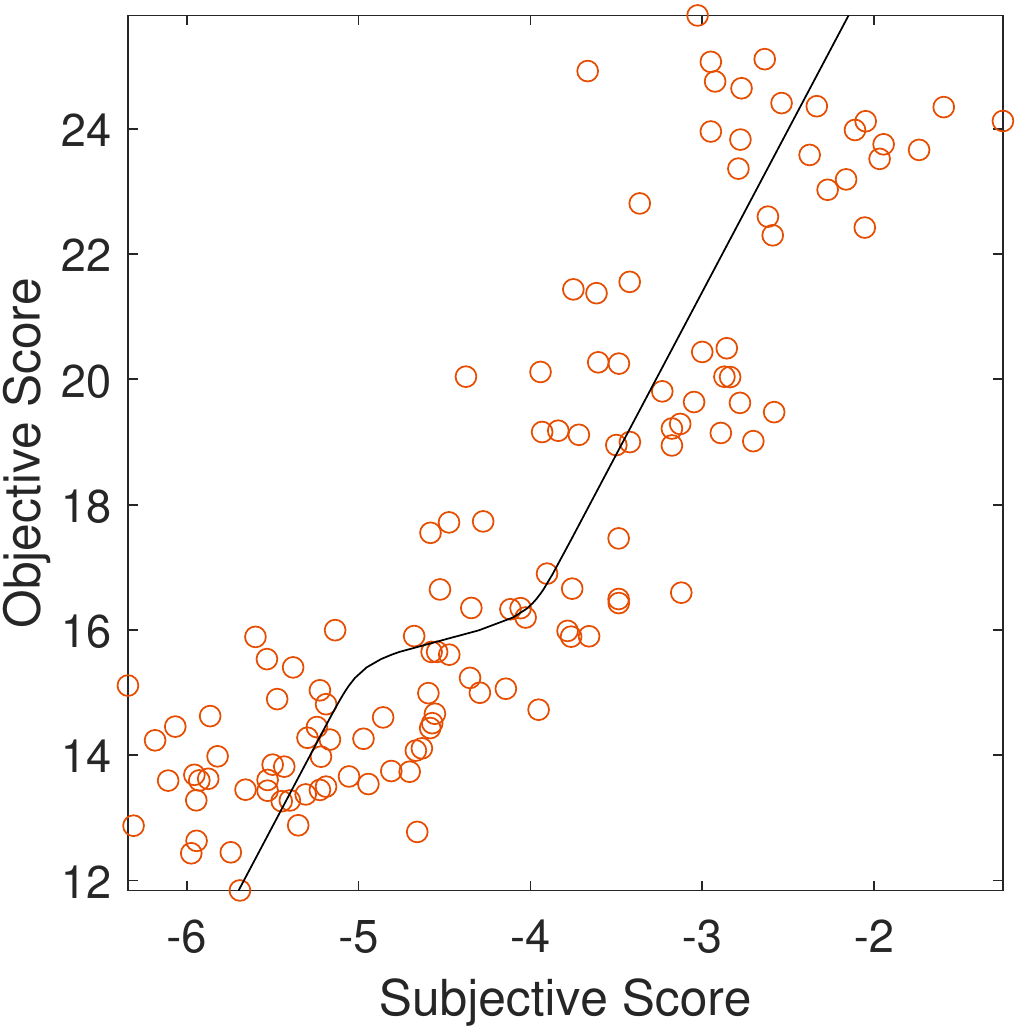}}\hspace{.12in}
}\vspace{-.05in}
\centerline{
\subfigure[BID]{\includegraphics[height=0.2\textwidth]{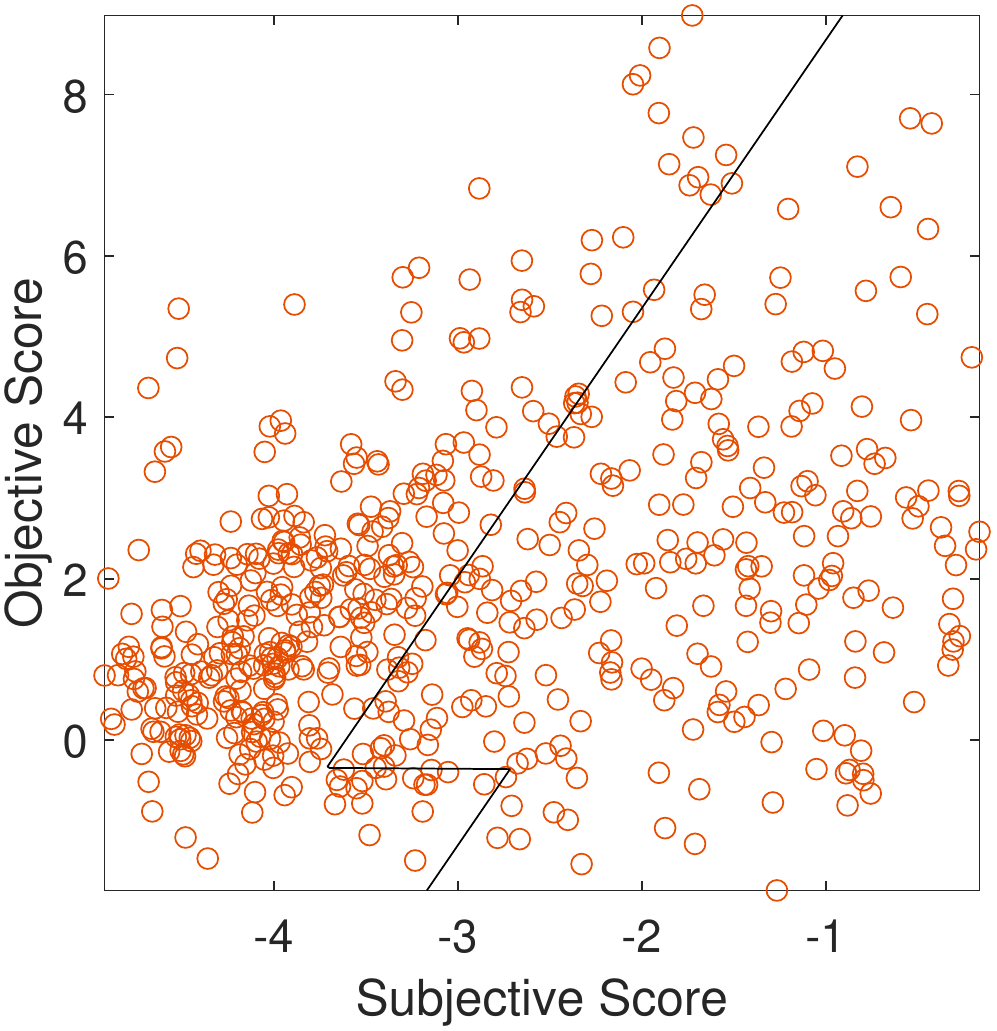}}\hspace{.12in}
\subfigure[CID2013]{\includegraphics[height=0.2\textwidth]{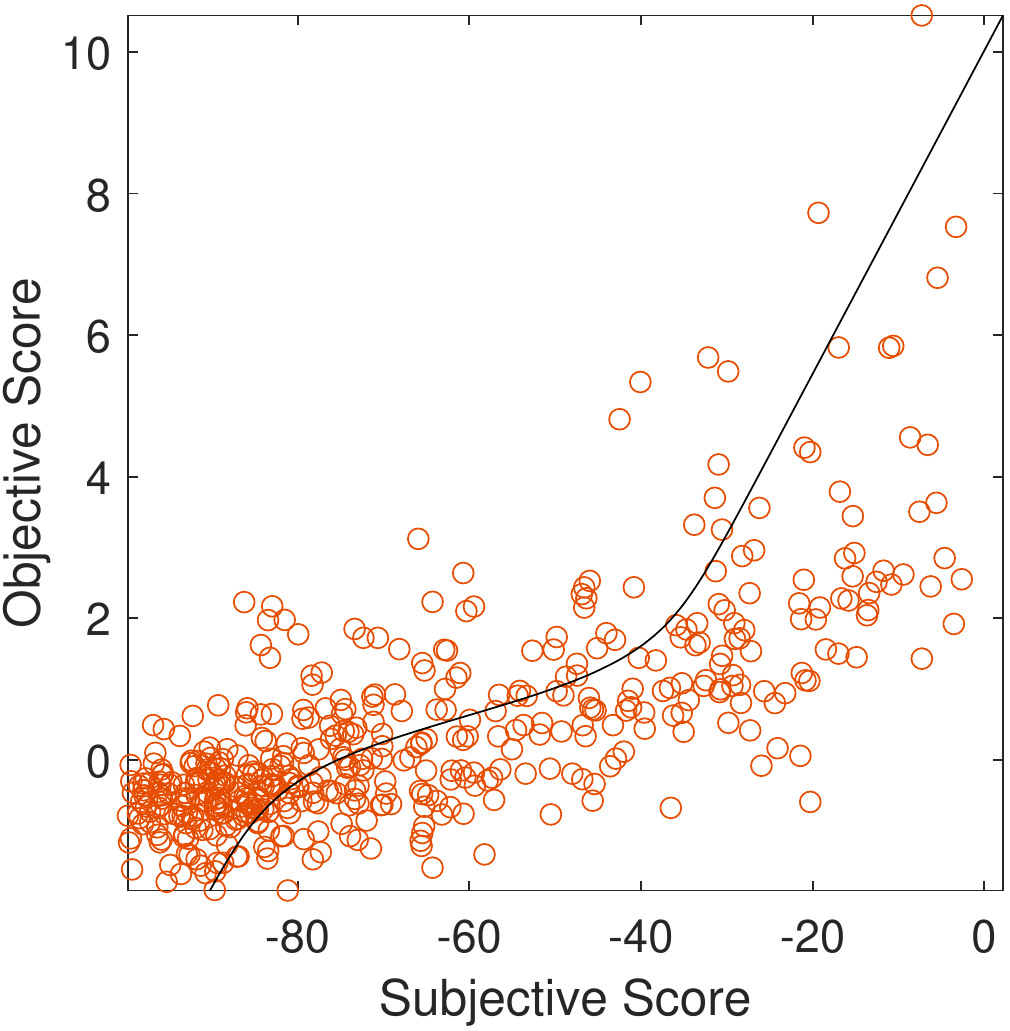}}\hspace{.12in}
\subfigure[FocusPath]{\includegraphics[height=0.2\textwidth]{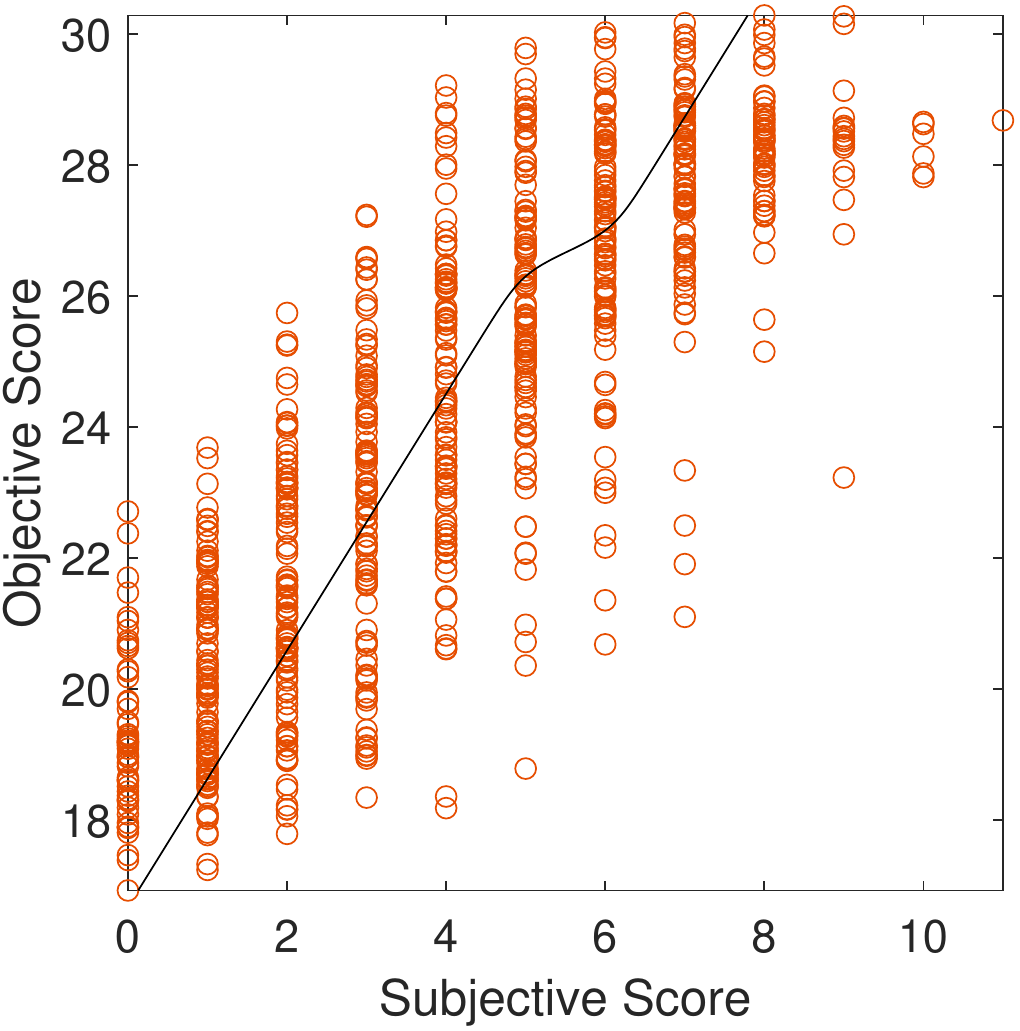}}\hspace{.12in}
}\vspace{-.05in}
\centerline{
\subfigure[LIVE]{\includegraphics[height=0.2\textwidth]{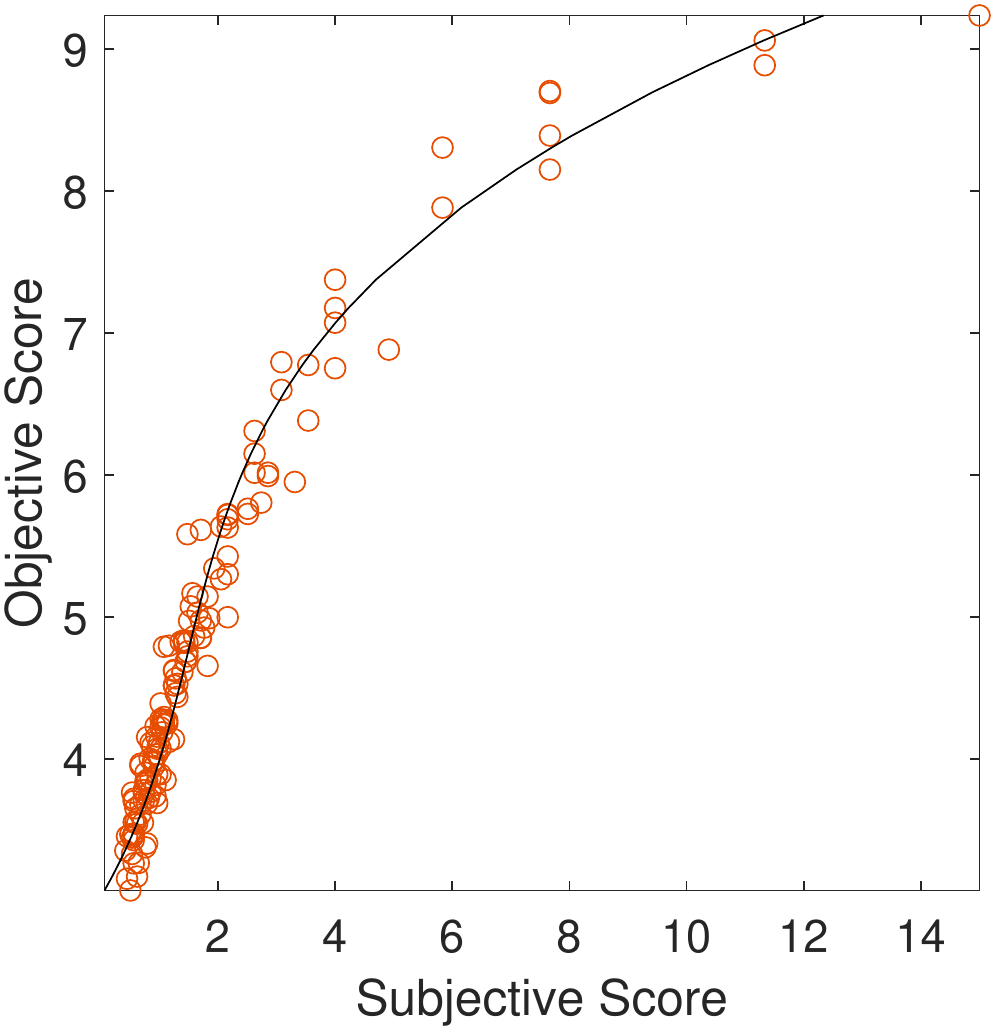}}\hspace{.12in}
\subfigure[CSIQ]{\includegraphics[height=0.2\textwidth]{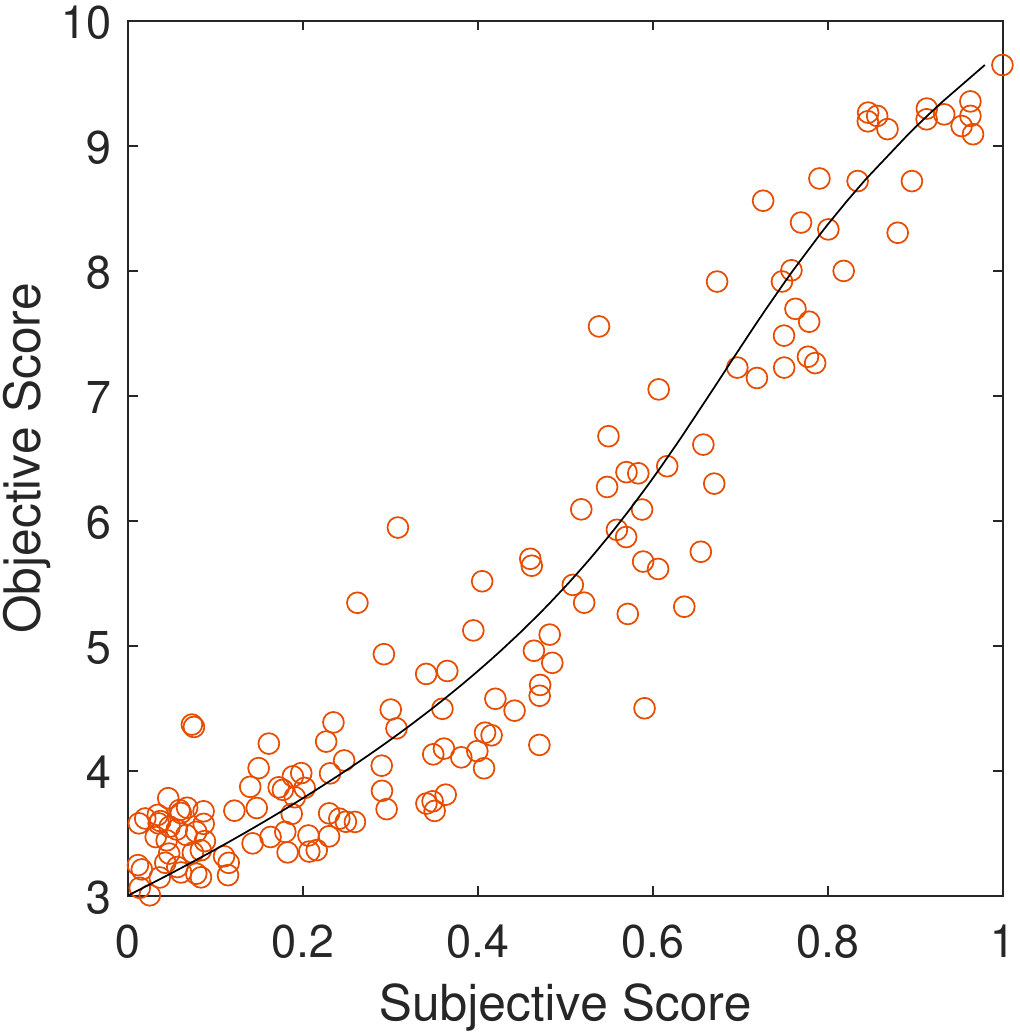}}\hspace{.12in}
\subfigure[TID2008]{\includegraphics[height=0.2\textwidth]{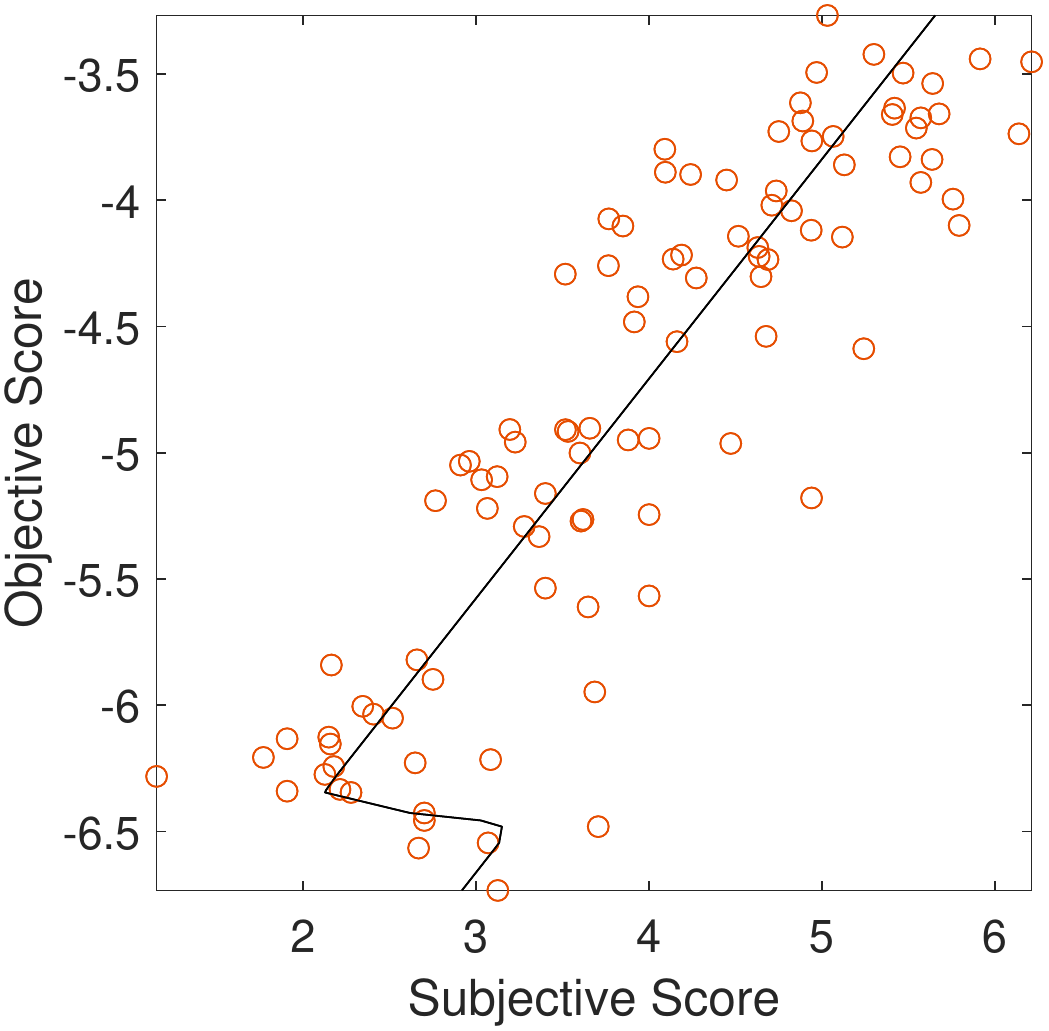}}\hspace{.12in}
\subfigure[TID2013]{\includegraphics[height=0.2\textwidth]{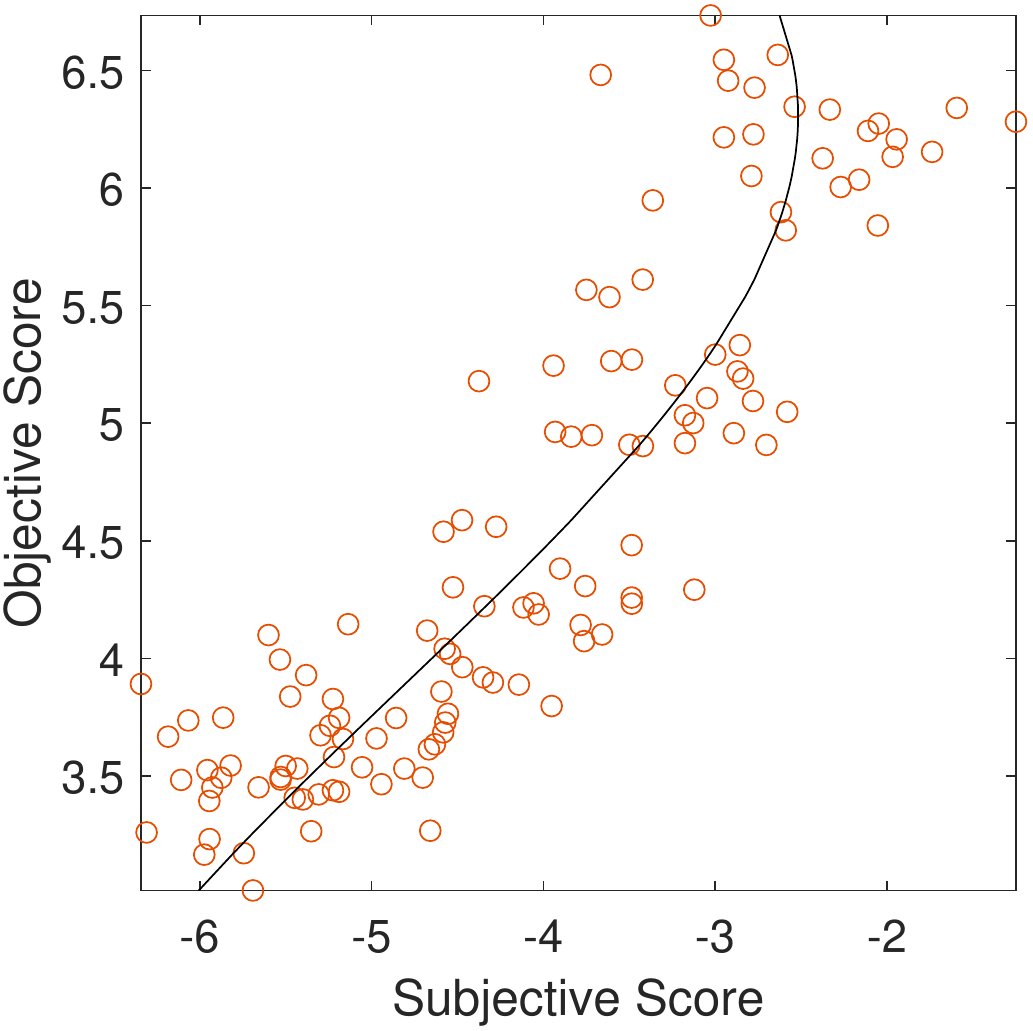}}\hspace{.12in}
}\vspace{-.05in}
\centerline{
\subfigure[BID]{\includegraphics[height=0.2\textwidth]{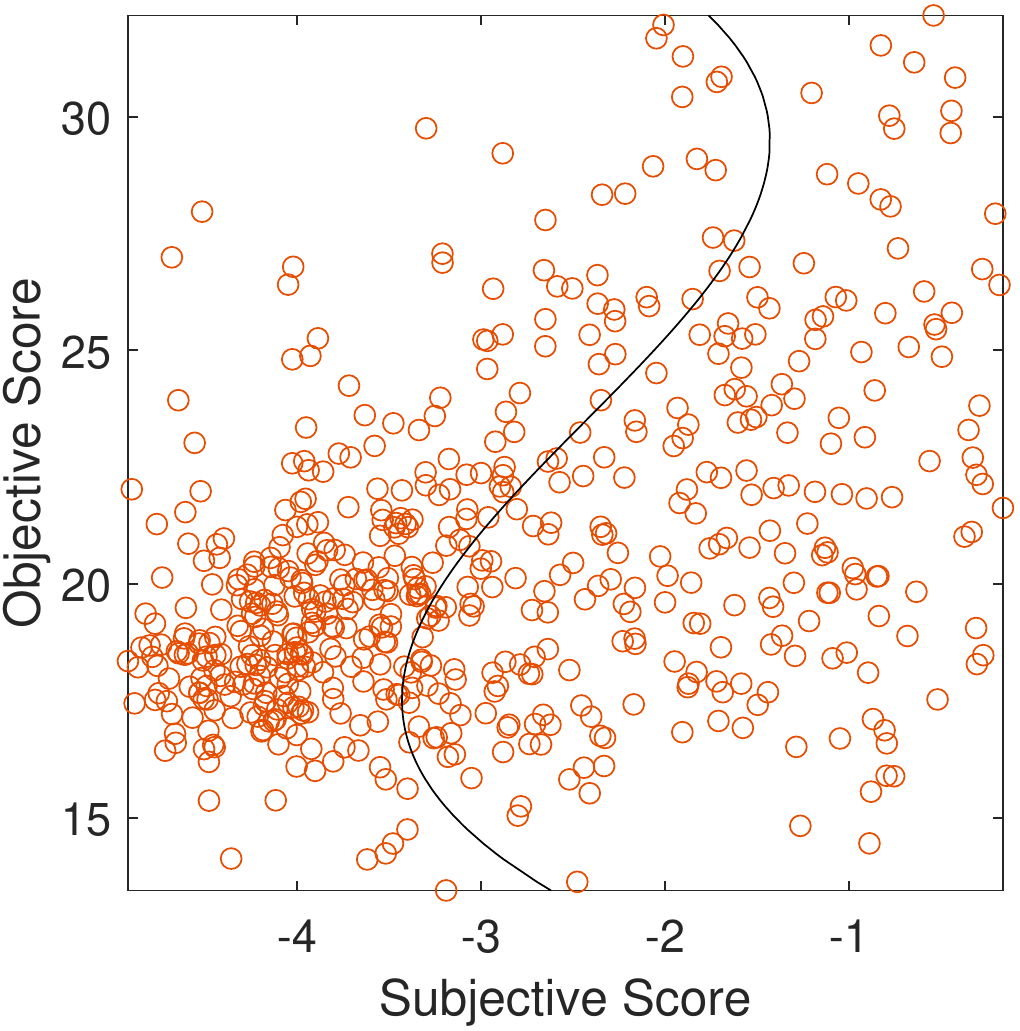}}\hspace{.12in}
\subfigure[CID2013]{\includegraphics[height=0.2\textwidth]{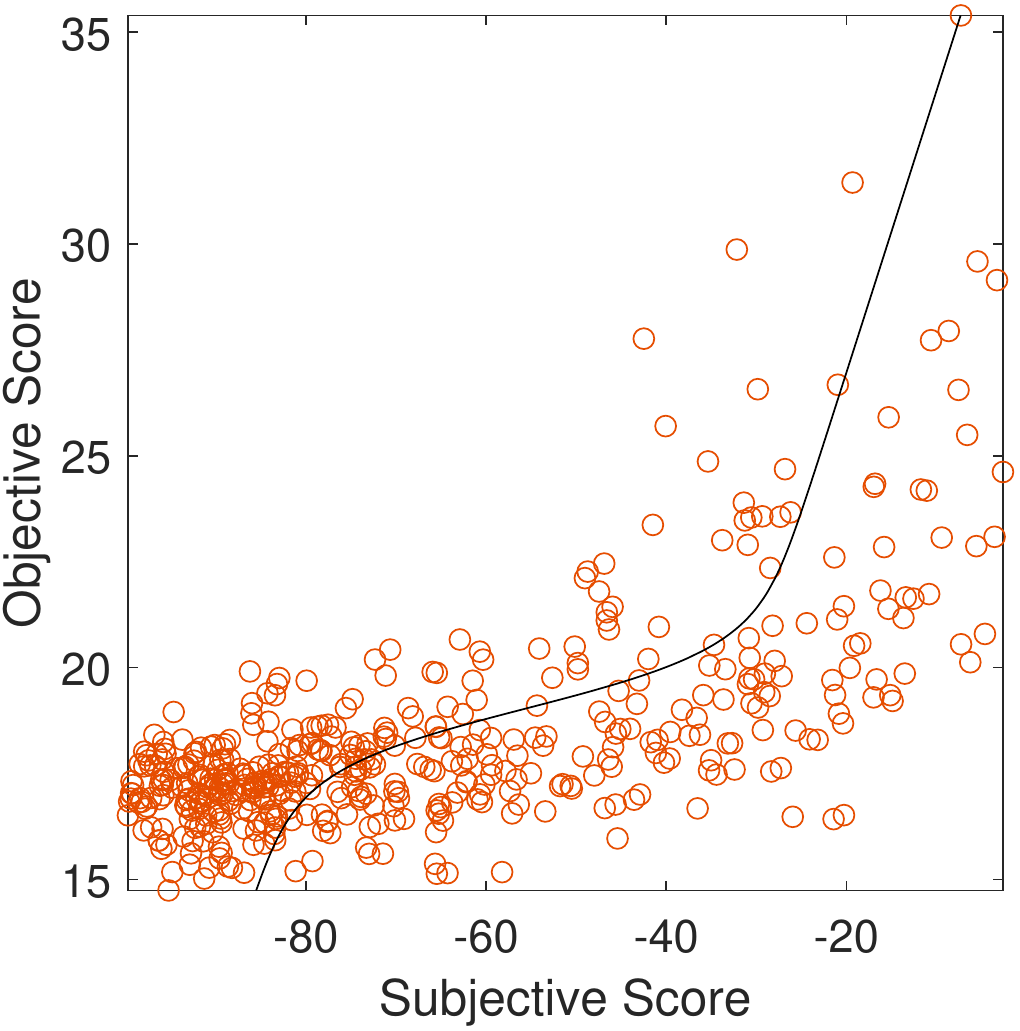}}\hspace{.12in}
\subfigure[FocusPath]{\includegraphics[height=0.2\textwidth]{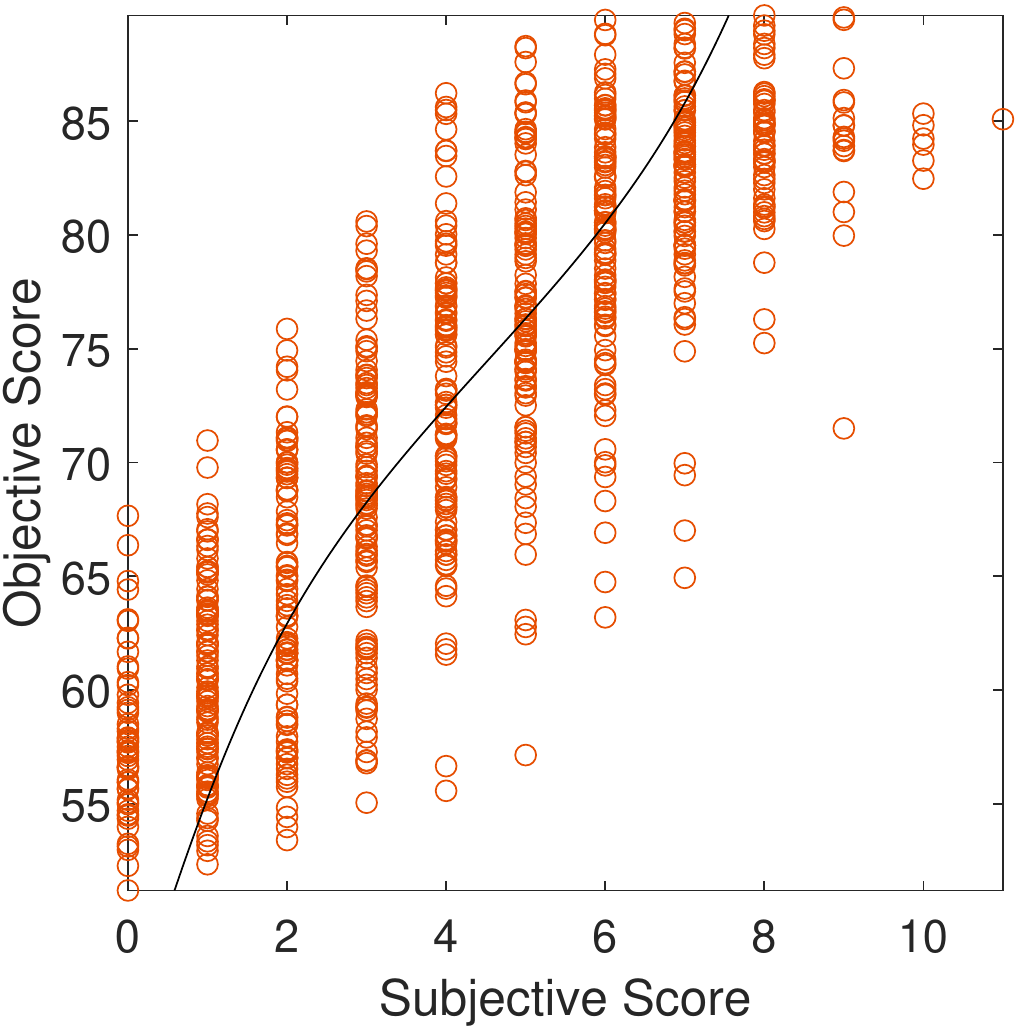}}\hspace{.12in}
}\vspace{-.05in}
\centering
\caption{Scatter plots of subjective MOS versus objective scoring using HVS MaxPol-1 (first row) and HVS MaxPol-2 (second row). The experiments are done for synthetic and natural blur dataset as shown in the labels. The fitted regression curve is overlaid by a black solid line on all plots. Note that the BID dataset has the lowest performance.}
\label{score scatter}
\end{figure*}

To better visualize the performance of HVS-MaxPol-1 and HVS-MaxPol-2, we include the scatter plots of generated objective scores versus the corresponding subjective scores on every dataset in Figure \ref{score scatter}. The plots also include the regression curves that are overlaid on the figures. The concentration of the scatter points around this curve implies the closeness of the metric to the subjective ratings. The deviation of points from the regression curve in natural images is higher compared to synthetic images, yielding inferior performance in general as expected. Furthermore, the objective scores are widely spread along the vertical axes with respect to their subjective scores, indicating that the measure has direct relationship (approximately linear) with subjective scores. Such relationship implies a better interpretation of the objective scores in the context the of HVS \cite{seshadrinathan2010study}.

Overall, the proposed HVS-MaxPol-1 and HVS-MaxPol-2 perform well on scoring both synthetic and natural blur images. Our proposed metric does not require any training and can be reliably and effectively employed in practice regardless of the source and type of image blur. In addition, Synthetic-MaxPol is especially excellent in scoring synthetic blur images, and thus it can be another optimal option for synthetic blur assessment. It worth noting that we did not evaluate Yu's CNN and Kang's CNN on natural dataset due to the lack of an existing pre-trained model. We also did not train the networks on these dataset from scratch due to the complexity of their training procedural guidelines.

\subsection{Scalability Analysis}	
\begin{figure*}[htp]
\scriptsize
\centerline{
\subfigure[HVS MaxPol-1]{\includegraphics[height=0.25\textwidth]{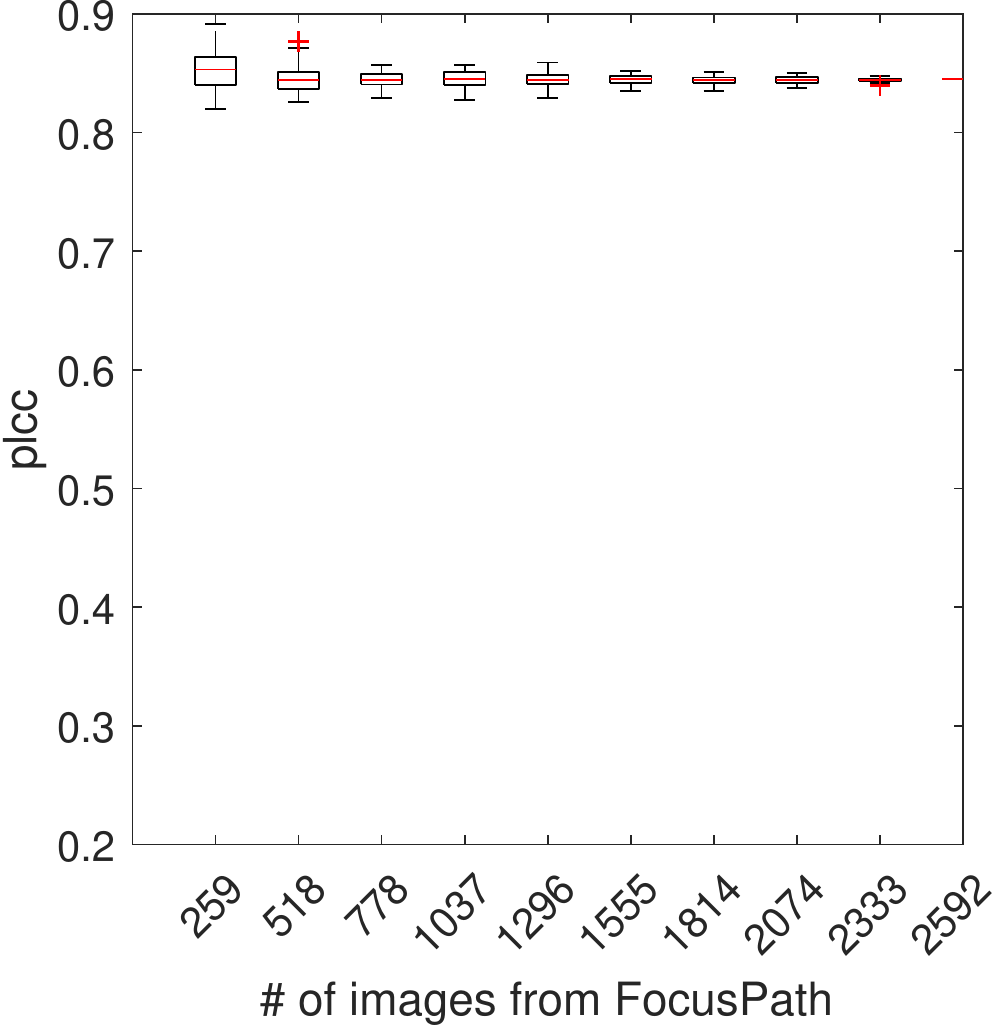}}\hspace{.15in}
\subfigure[HVS MaxPol-2]{\includegraphics[height=0.25\textwidth]{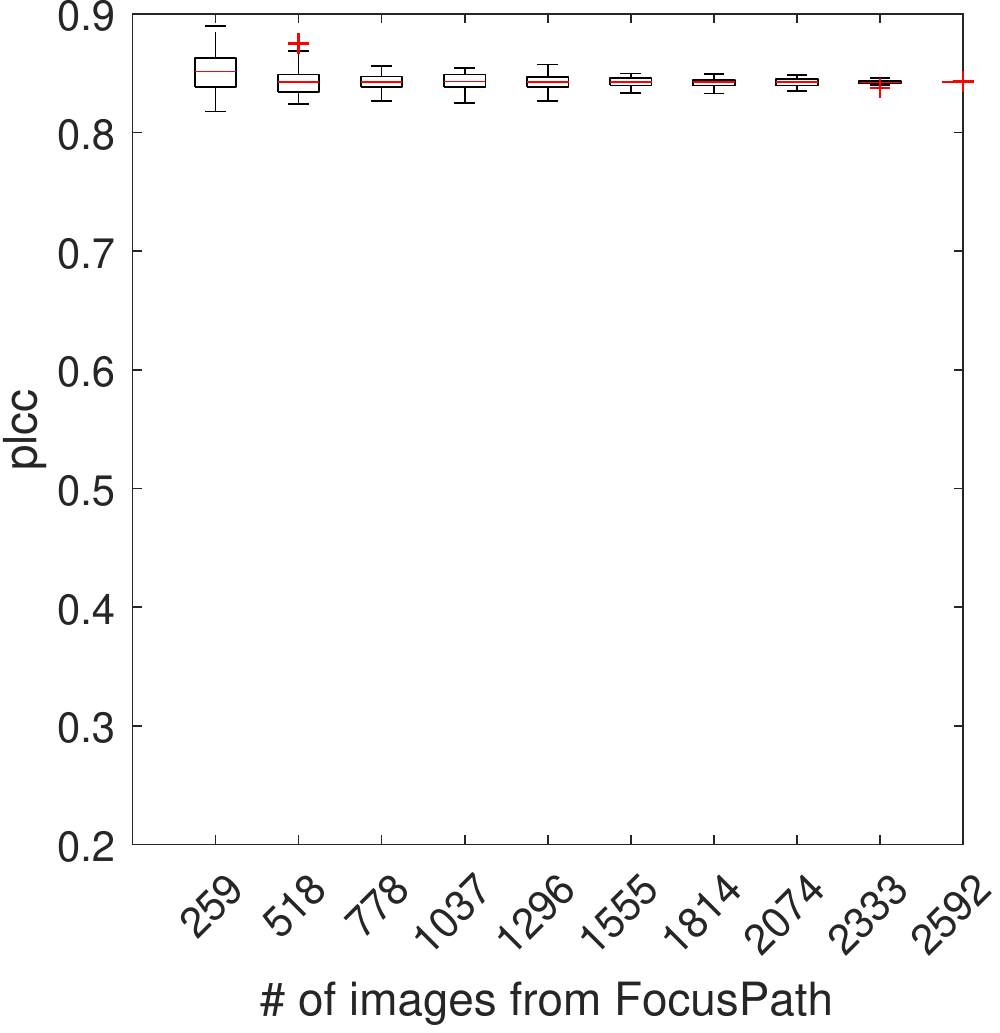}}\hspace{.15in}
\subfigure[Synthetic-MaxPol \cite{mahdi2018image}]{\includegraphics[height=0.25\textwidth]{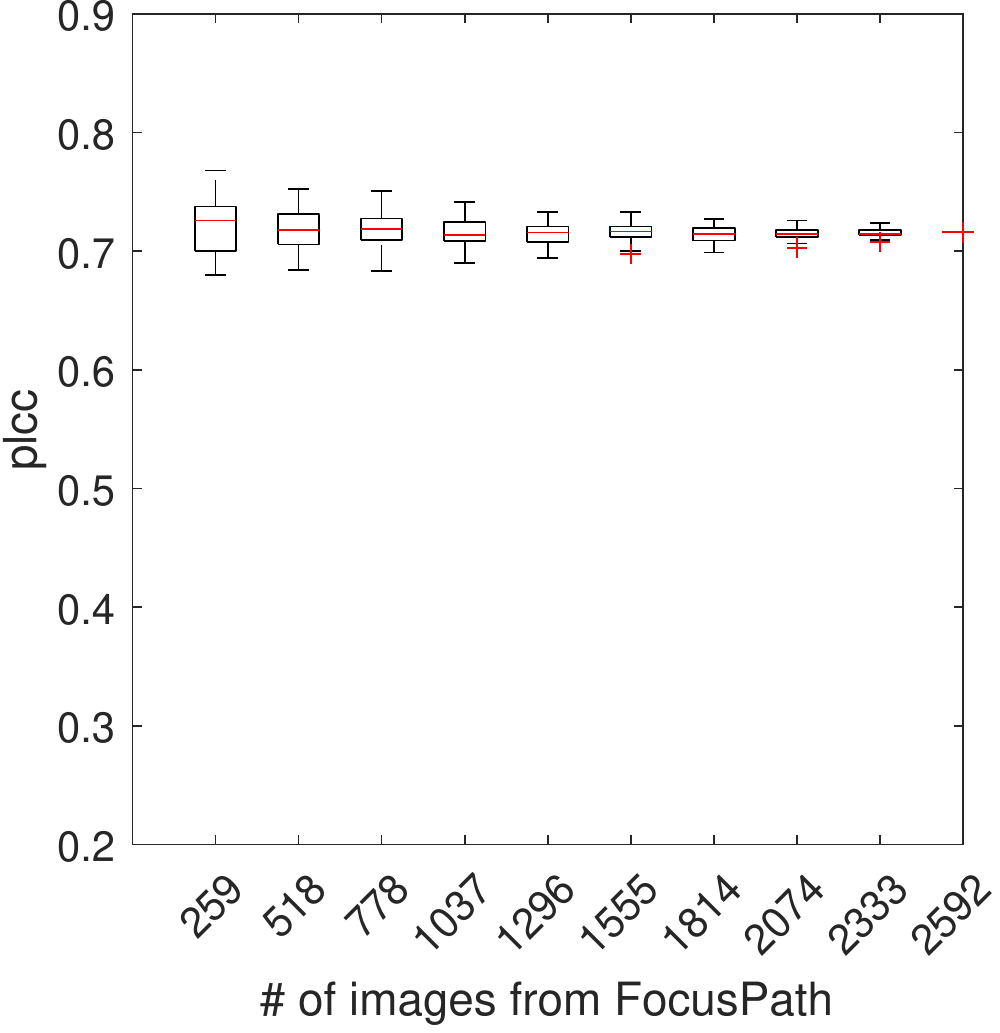}}\hspace{.15in}
}\vspace{-.05in}
\centerline{
\subfigure[MLV \cite{bahrami2014fast}]{\includegraphics[height=0.25\textwidth]{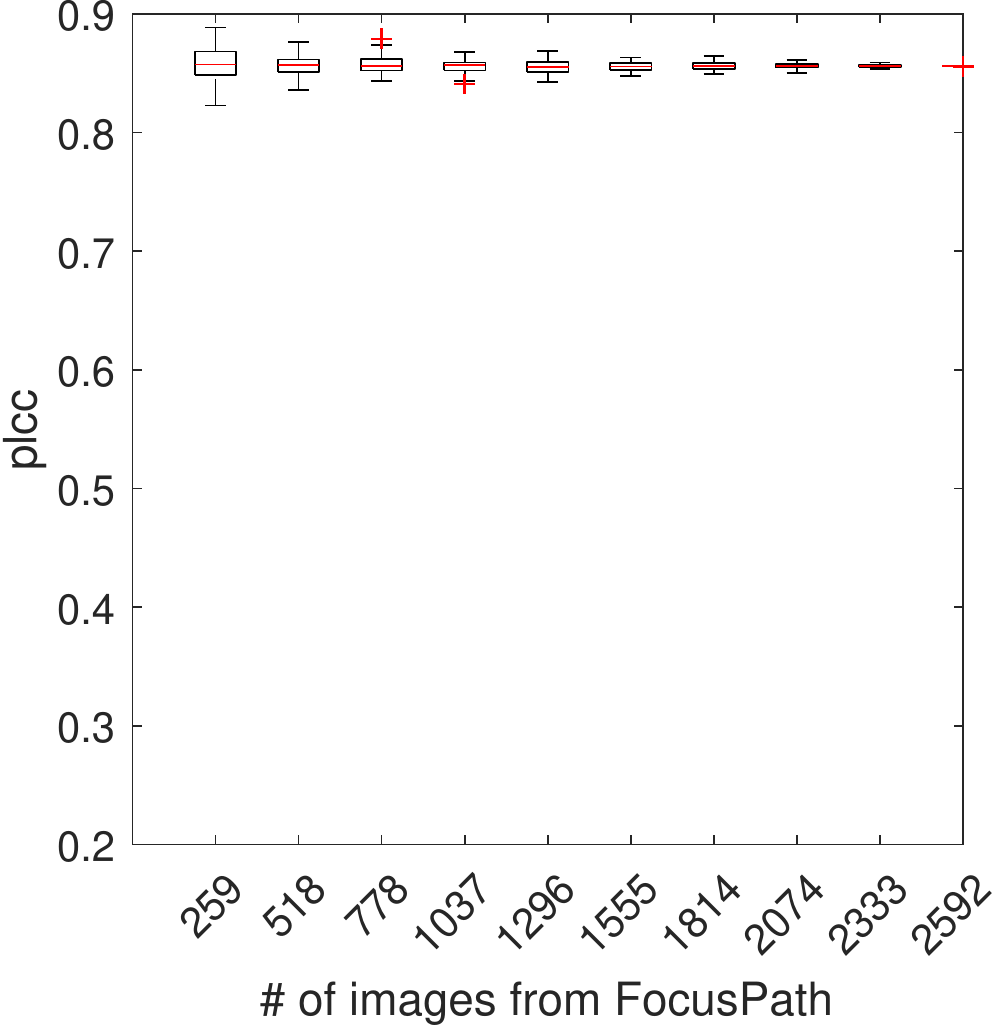}}\hspace{.15in}
\subfigure[RISE \cite{li2017no}]{\includegraphics[height=0.25\textwidth]{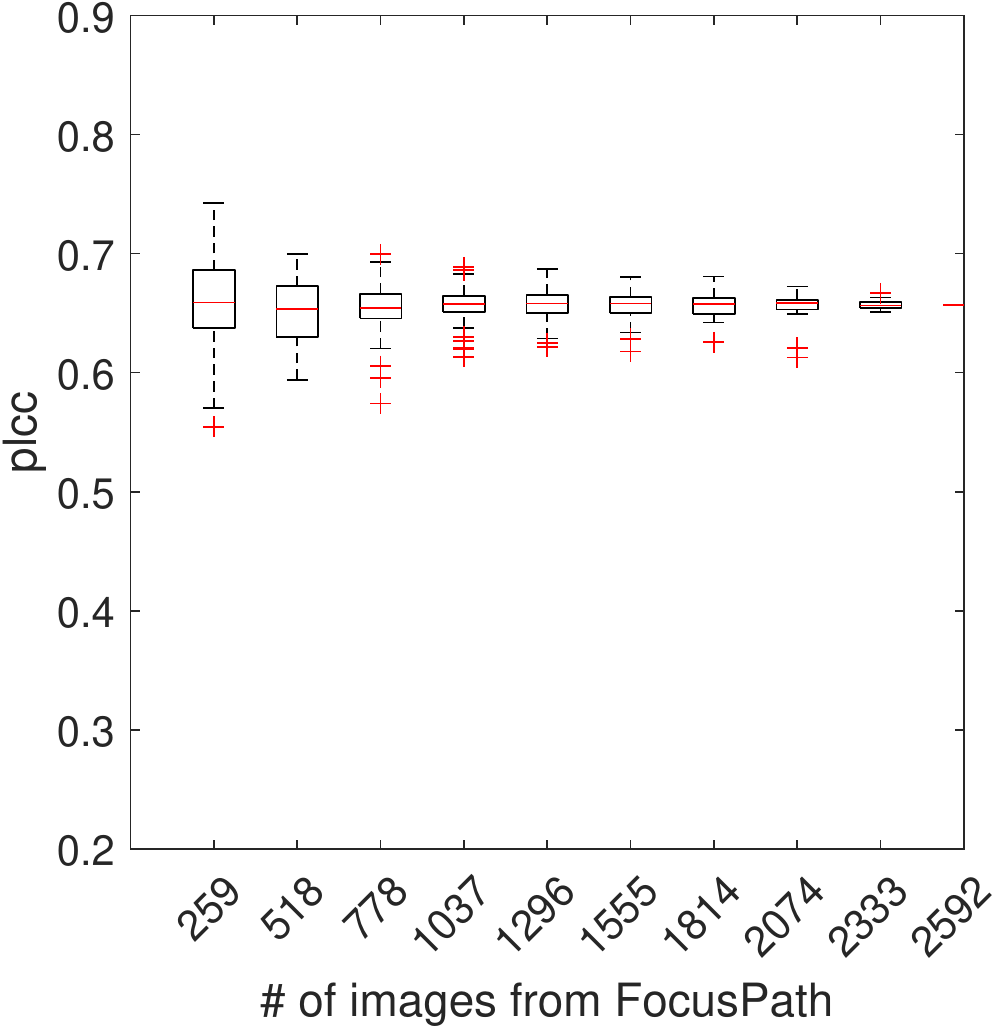}}\hspace{.15in}
\subfigure[$\text{ARISM}_{\text{C}}$ \cite{gu2015no}]{\includegraphics[height=0.25\textwidth]{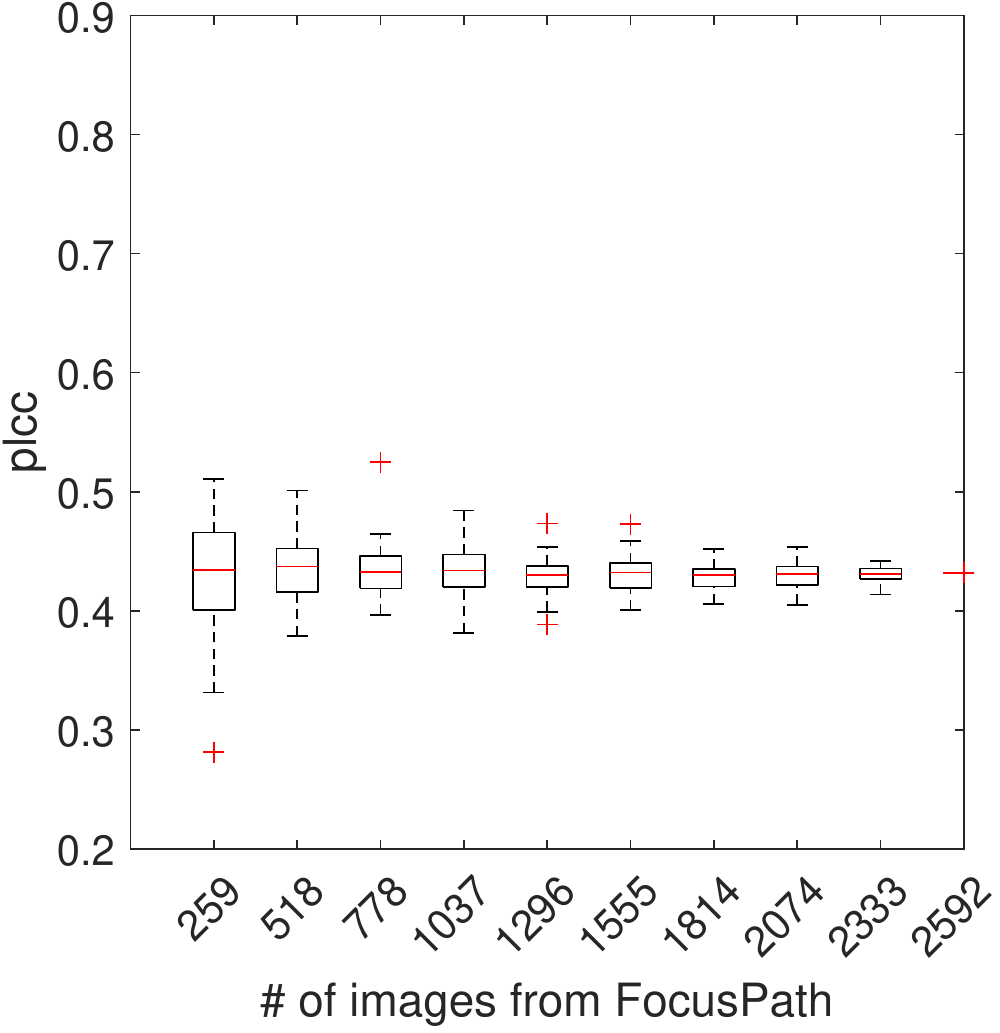}}\hspace{.15in}
}\vspace{-.05in}
\centerline{
\subfigure[GPC \cite{leclaire2015no}]{\includegraphics[height=0.25\textwidth]{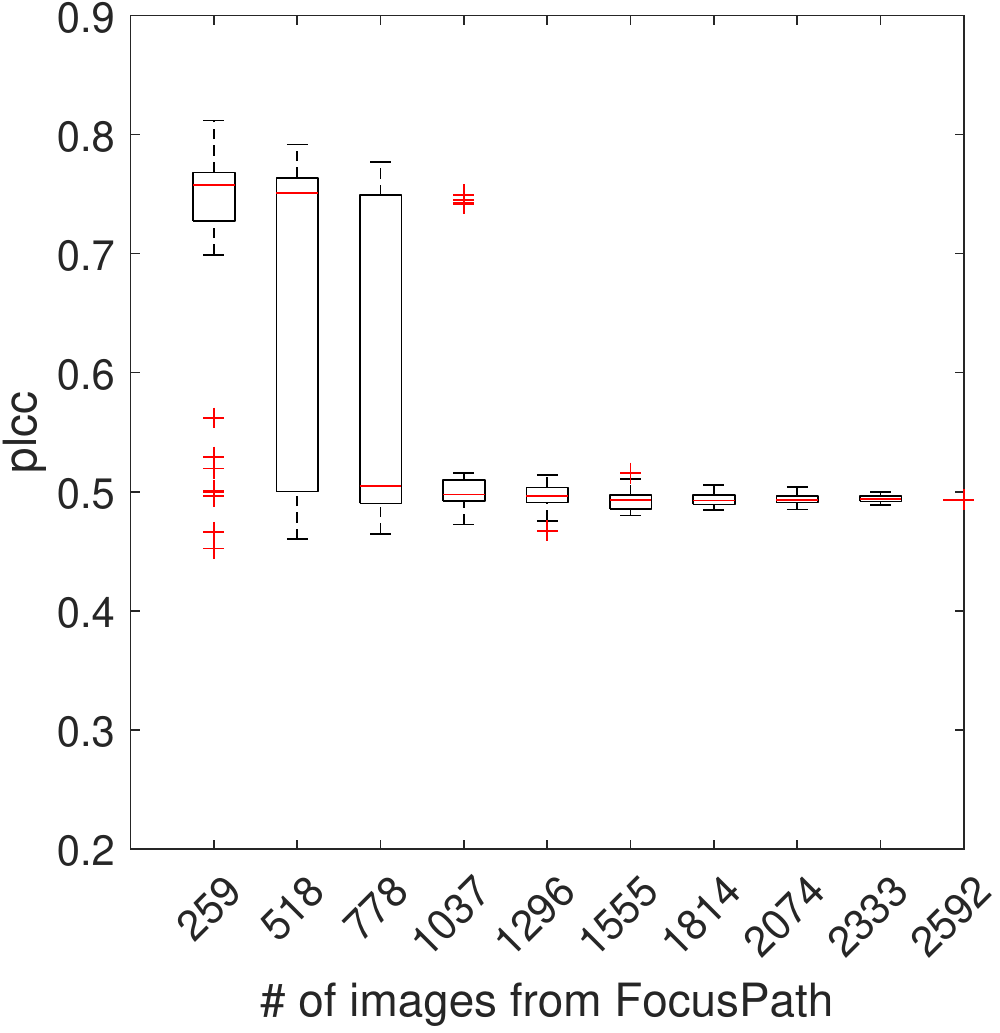}}\hspace{.15in}
\subfigure[$\text{S}_3$ \cite{vu2012bf}]{\includegraphics[height=0.25\textwidth]{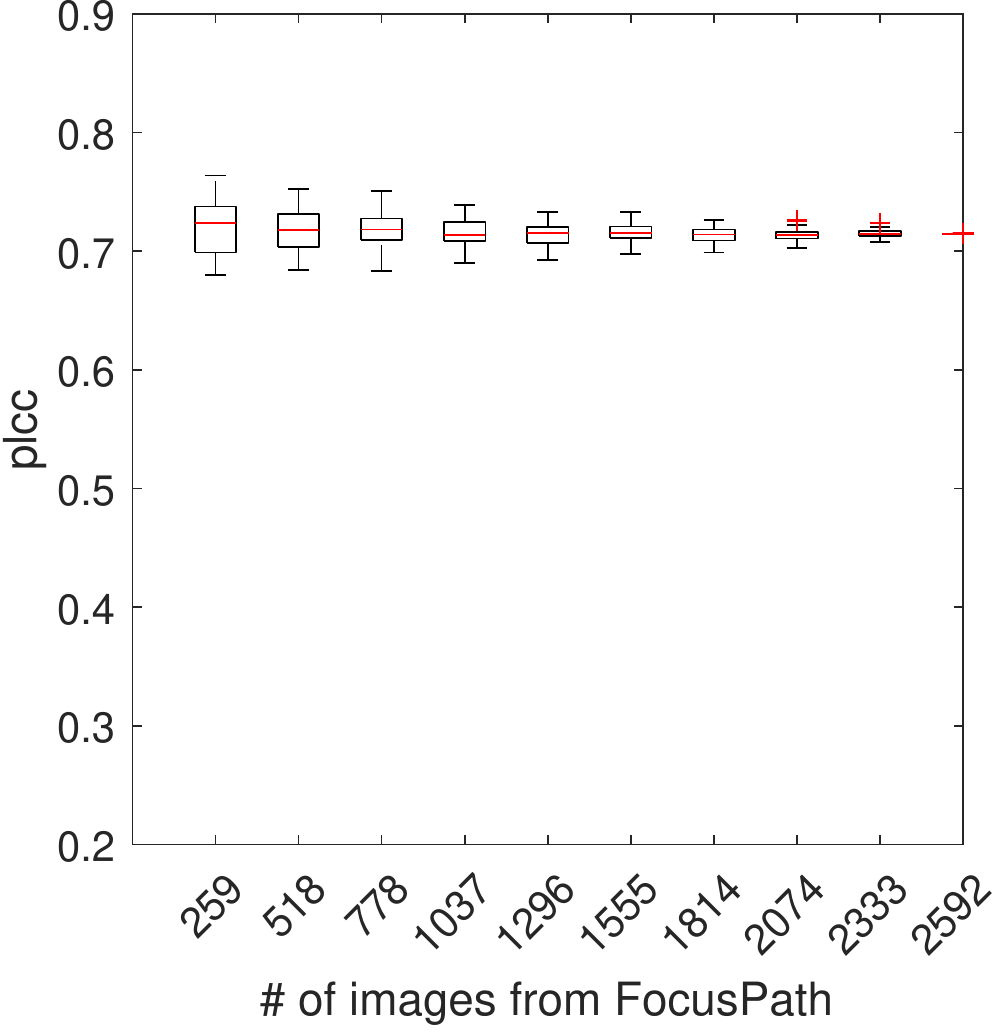}}\hspace{.15in}
\subfigure[SPARISH \cite{li2016image}]{\includegraphics[height=0.25\textwidth]{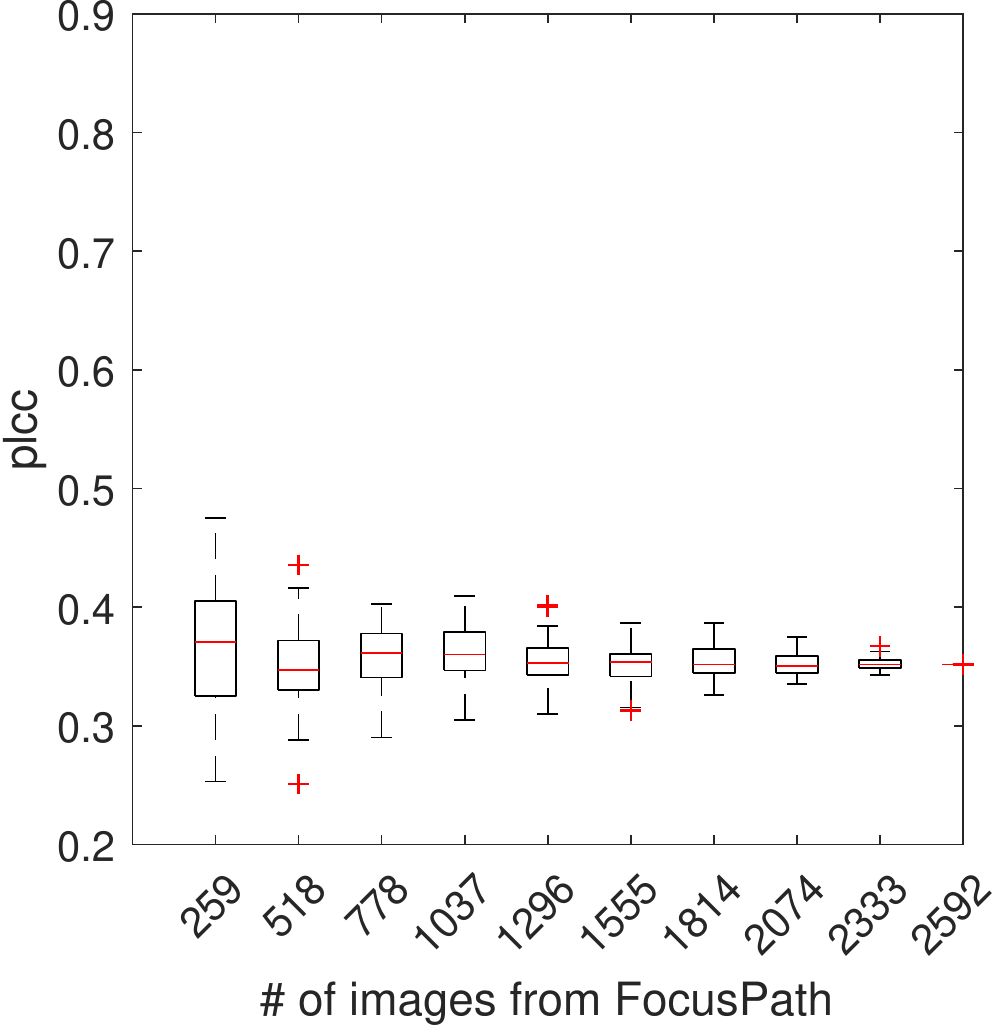}}\hspace{.15in}
}\vspace{-.05in}

\centering
\caption{Box plots for each method on different subsets of FocusPath. The size of box shows the variance of plcc values with respect to $50$ trials under the same percentage of FocusPath. The smaller the size of the box and the more even the height of the box, the better scalability of the method. Here, HVS-MaxPol-1, HVS-MaxPol-2, and MLV have the least fluctuations and variance with respect to different sizes, showing they have the best scalability among all methods.}
\label{box plot}
\end{figure*}

In this section, we evaluate the scalability of all NR-ISA metrics. We define the scalability here as the consistency of performance over different image dataset volume. For the sake of experiment, we have selected the extended version of FocusPath which contains $8640$ blur images in total for ISA analysis. We randomly select different dataset sizes of $\{3\%, 6\%, 9\%, \hdots, 30\%\}$ from the whole $8640$ images. Monte-Carlo simulation is done over $50$ different random selections from each dataset size. For each shuffle, we evaluate the correlation accuracy of different NR-ISA metrics and show them with a box-plot in the Figure \ref{box plot}. The bottom line and top line of the box represent 25th and 75th percentile of all data and the central red line stands for the median value. A good metric should (a) yield high median value with minimal box size related to the standard deviation of the data, and (b) provide consistent performance over different dataset size. Overall, HVS-MaxPol-1 and HVS-MaxPol-2, and MLV provide the best scalability with respect to the criteria defined in (a) and (b), where all three methods methods generate the smallest box size and remain on the same level along different dataset size. We can also observe that S3 and Synthetic-MaxPol are slightly worse than the top three methods. Moreover, the performance of GPC drops dramatically as the size of dataset increases, showing its poor scalability.

\subsection{Computational Complexity Analysis}
\begin{figure}[htp]
\centerline{
\subfigure[Synthetic Dataset]{\includegraphics[height=0.4\textwidth]{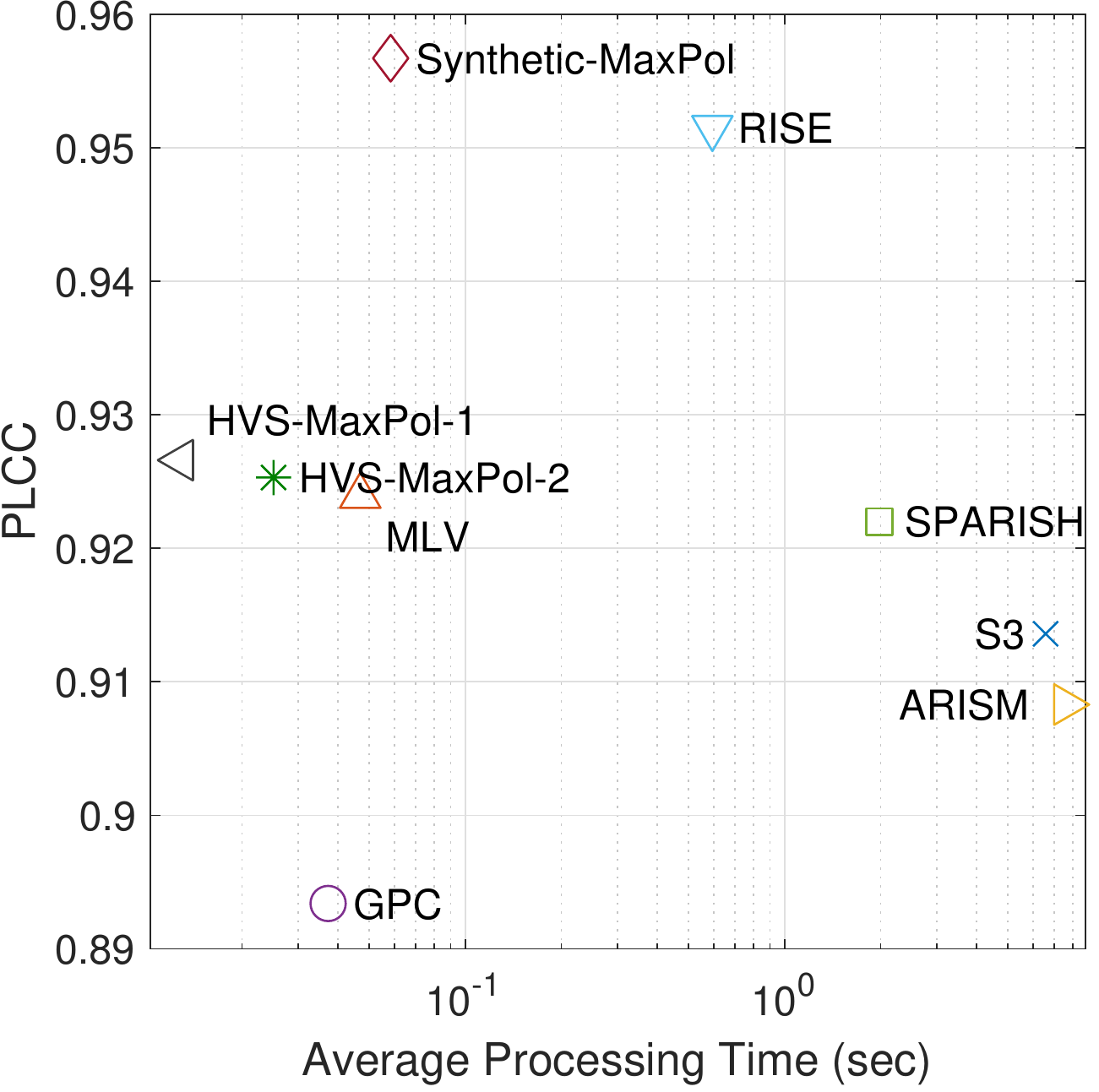}\label{time_plot_synthetic}}\vspace{.05in}
}\vspace{-.05in}
\centerline{
\subfigure[Natural Dataset]{\includegraphics[height=0.4\textwidth]{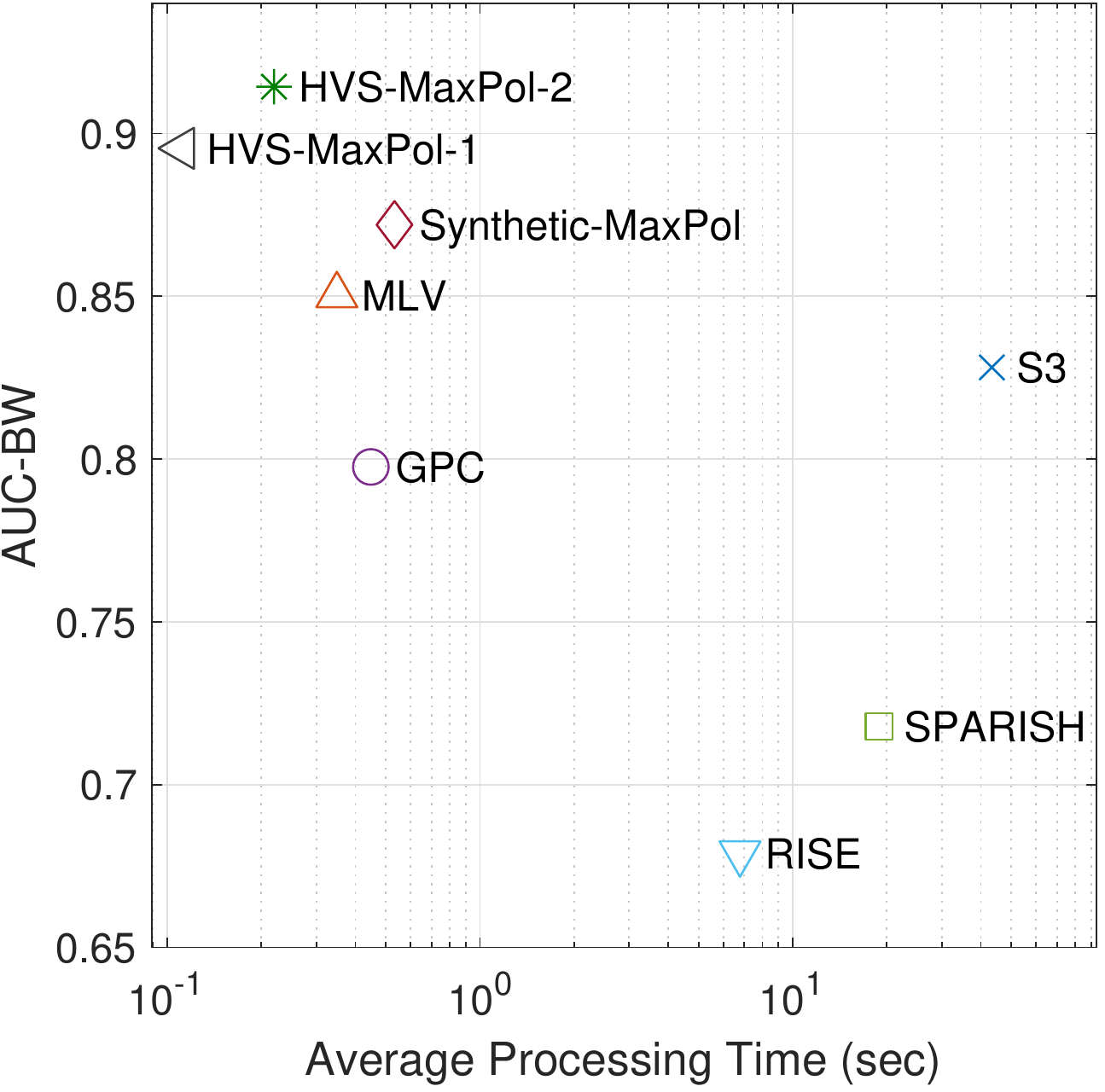}\label{time_plot_natural}}
}\vspace{-.05in}
\centering
\caption{PLCC values vs. CPU time among all methods on synthetic and natural dataset. Markers located at the top-left corner are desirable, meaning the method has high accuracy and speed.}
\label{time plot}
\end{figure}

Our final analysis is to study the computational complexity of all NR-ISA metrics. The speed of metric calculation in digital computing is highly critical for practical considerations. In particular, high-speed NR-ISA metric is a must in high-throughput digital image archiving platforms such as whole slide imaging in digital pathology and planetary observations in reconnaissance satellite orbiters such as MRO \cite{mcewen2007mars} and LRO \cite{robinson2010lunar}. In this section, we design two sets of experiments. The first one is the assessment of correlation accuracy of NR-ISA metrics versus CPU time over both natural and synthetic dataset. We study the second experiment by analyzing the complexity of CPU time versus different image sizes. For the CPU time measure, all the experiments are done on a Windows station with an AMD FX-8370E 8-Core CPU 3.30 GHz. The relationship between PLCC and $\text{AUC}_{BW}$ performances and the average CPU time of each method is shown in Figure \ref{time plot}. Here, the overall weighted PLCC is used for synthetic, while for the natural we demonstrate the overall aggregated performances. A large y-axis value indicates a high accuracy and a small x-axis value indicates higher speed for processing. Accordingly, an excellent method should locate at the top-left corner in each plot. As it shown in Figure \ref{time_plot_synthetic}, Synthetic-MaxPol, HVS-MaxPol-1 and HVS-MaxPol-2 are located at the top-left corner, indicating excellent accuracy performance for synthetic blur images. In particular, Synthetic-MaxPol has the highest accuracy and HVS MaxPol-1 consumes the least computational time. Additionally, although RISE has a higher accuracy compared to HVS MaxPol-1 and HVS MaxPol-2 over synthetic dataset, its time consumption are approximately 100 times higher than the two HVS-MaxPol metrics. With respect to the natural image dataset shown in the second plot of Figure \ref{time plot}, the top three metrics located on the top-left corners are HVS-MaxPol-1, HVS-MaxPol-2, and MLV. Among these three, HVS-MaxPol-1 is the most time efficient and HVS-MaxPol-2 is the most accurate.
	
\begin{figure}[htp]
\centerline{
\includegraphics[width=0.4\textwidth]{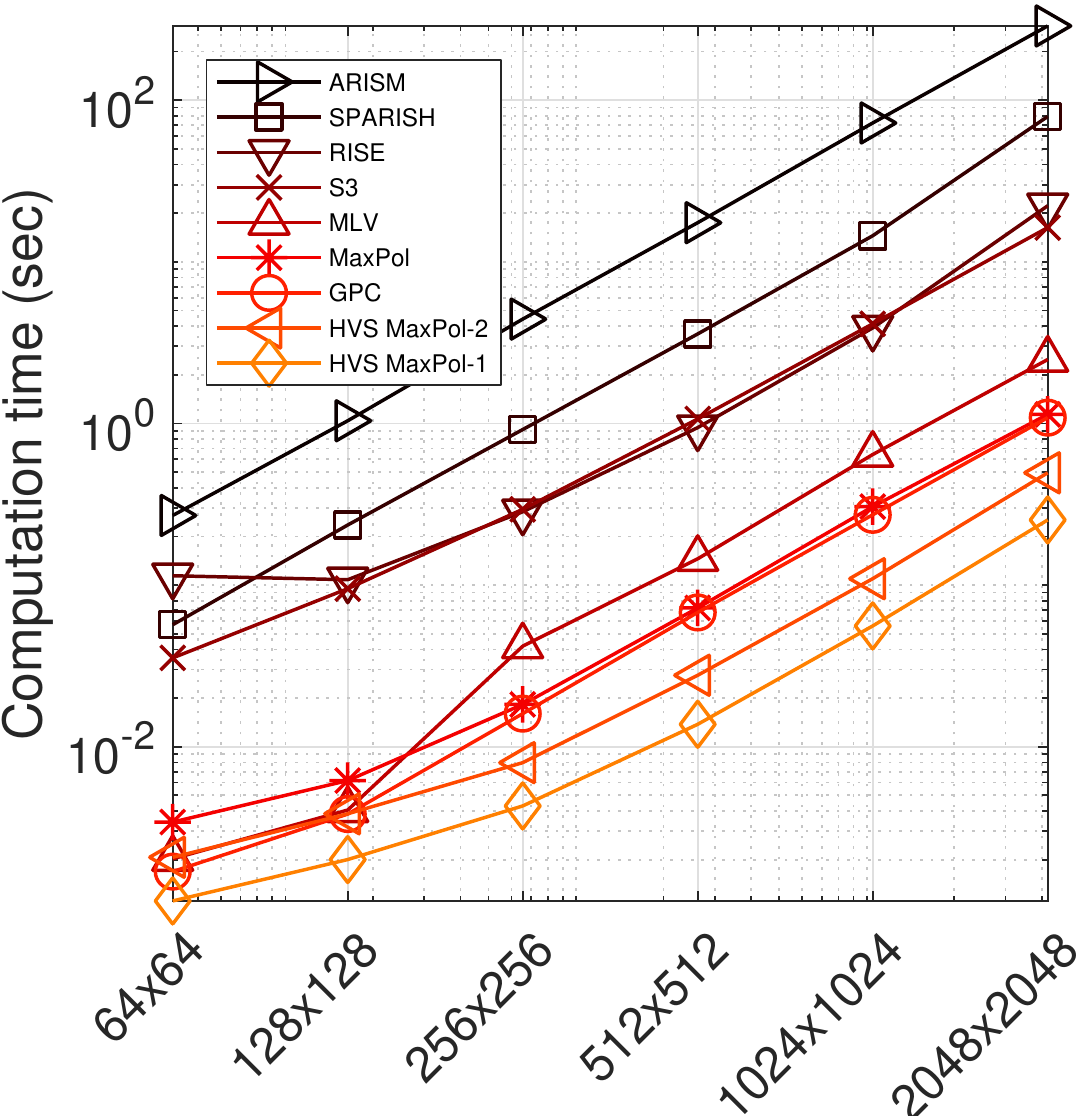}
}\vspace{-.05in}
\caption{The computational time vs the pixel number of an image using different methods. Lower curve shows computational efficiency.}
\label{computation plot}
\end{figure}

We conduct further experiment to analyze the computational complexity (in the CPU time) over different image sizes. For this experiment, we select $20$ samples of pathological images in square sizes $\{64, 128, 256, 512, 1024, 2048\}$. The CPU time is averaged over all $20$ trials and the performances are shown in Figure \ref{computation plot}. Each curve in the figure indicates the computational time complexity of the corresponding NR-ISA metric. According to the results, the rank observation of all metrics are consistent over different image size. HVS-MaxPol-1 is the most time efficient metric among all the evaluated metrics. It is almost $1000$ times faster than ARISM, $100$ times faster than RISE and S3, and almost $10$ times faster than MLV. HVS-MaxPol-2 ranks second, two times slower than HVS-MaxPol-1, followed by GPC and MaxPol Note that since the authors of Yu'CNN \cite{yu2017shallow} and Kang's CNN \cite{kang2014convolutional} did not provide a pre-trained model, the computational complexity analysis is not performed on these two metric.

\section{Conclusion}\label{conclusion}
In this paper, we introduced a novel no-reference image sharpness assessment (NR-ISA) metric, which can process the input image efficiently and accurately over diverse image blur types. The foundation of this metric is based on the human vision system (HVS) response. We have simulated the HVS response using a symmetric FIR kernel as a superposition of multiple even-order derivative kernels which boosts high frequency domain magnitudes in a balanced way similar to the HVS response. The numerical approximation of the proposed HVS filter is made feasible by MaxPol filters for image feature extraction that are closely related to focus blur. Thorough experiments are conducted on four synthetic and three natural blur image dataset for the comparison of eight state-of-the-art NR-ISA. The results demonstrate that our metric significantly outperforms other metrics over both synthetic and natural blur databases. We also conduct time complexity experiment and scalability experiments, both of which validate the superior performance of our metric. The qualification overview of these metrics are demonstrated in Table \ref{summary table}. 

\begin{table}[htp]
\scriptsize
\centering
\caption{The qualification different NR-ISA metrics over the conducted analysis. $\textcolor{mygreen1}{\checkmark}$ indicates that the corresponding metric ranks the top four under a certain analysis and $\textcolor{myred1}{\times}$ indicates the opposite.}
\label{summary table}
\begin{tabular}{l|c|cc|cc|c}
\hlinewd{1.5pt}
\multirow{2}{*}{Metric} & \multirow{2}{*}{\begin{tabular}[c]{@{}c@{}}Speed\end{tabular}} & \multicolumn{2}{c|}{PLCC} & \multicolumn{2}{c|}{SRCC} & \multirow{2}{*}{Scalability} \\ 
                        &                                                                                     & Synth.           & Natr.           & Synth.           & Natr.           &                              \\ \hlinewd{1.5pt}
$\text{S}_3$ \cite{vu2012bf}                   & $\textcolor{myred1}{\times}$       & $\textcolor{myred1}{\times}$   & $\textcolor{myred1}{\times}$  & $\textcolor{myred1}{\times}$ & $\textcolor{myred1}{\times}$  & $\textcolor{myred1}{\times}$        \\ 
MLV \cite{bahrami2014fast}                     & $\textcolor{mygreen1}{\checkmark}$       & $\textcolor{myred1}{\times}$ & $\textcolor{mygreen1}{\checkmark}$  & $\textcolor{myred1}{\times}$ & $\textcolor{mygreen1}{\checkmark}$ & $\textcolor{mygreen1}{\checkmark}$    \\ 
Kang's CNN \cite{kang2014convolutional}        & -                                  & $\textcolor{myred1}{\times}$ & -                 & $\textcolor{myred1}{\times}$ & -                 & -                            \\ 
$\text{ARISM}_{\text{C}}$ \cite{gu2015no}      & $\textcolor{myred1}{\times}$       & $\textcolor{myred1}{\times}$  & $\textcolor{myred1}{\times}$  & $\textcolor{myred1}{\times}$ & $\textcolor{myred1}{\times}$ & $\textcolor{myred1}{\times}$  \\ 
GPC \cite{leclaire2015no}                      & $\textcolor{myred1}{\times}$ & $\textcolor{myred1}{\times}$   & $\textcolor{myred1}{\times}$ & $\textcolor{myred1}{\times}$ & $\textcolor{myred1}{\times}$  & $\textcolor{myred1}{\times}$    \\ 
SPARISH \cite{li2016image}                     & $\textcolor{myred1}{\times}$       & $\textcolor{myred1}{\times}$ & $\textcolor{myred1}{\times}$ & $\textcolor{myred1}{\times}$ & $\textcolor{myred1}{\times}$  & $\textcolor{myred1}{\times}$    \\ 
RISE \cite{li2017no}                           & $\textcolor{myred1}{\times}$       & $\textcolor{mygreen1}{\checkmark}$ & $\textcolor{mygreen1}{\checkmark}$  & $\textcolor{mygreen1}{\checkmark}$  & $\textcolor{mygreen1}{\checkmark}$ & $\textcolor{myred1}{\times}$       \\ 
Yu's CNN \cite{yu2017shallow}                  & -                                  & $\textcolor{mygreen1}{\checkmark}$ & -                 & $\textcolor{mygreen1}{\checkmark}$  & -                 & -                            \\ 
MaxPol \cite{mahdi2018image}                   & $\textcolor{mygreen1}{\checkmark}$ & $\textcolor{mygreen1}{\checkmark}$ & $\textcolor{myred1}{\times}$   & $\textcolor{mygreen1}{\checkmark}$   & $\textcolor{myred1}{\times}$ & $\textcolor{mygreen1}{\checkmark}$  \\ 
HVS-MaxPol-1                                   & $\textcolor{mygreen1}{\checkmark}$ & $\textcolor{myred1}{\times}$ & $\textcolor{mygreen1}{\checkmark}$   & $\textcolor{myred1}{\times}$   & $\textcolor{mygreen1}{\checkmark}$ & $\textcolor{mygreen1}{\checkmark}$          \\ 
HVS-MaxPol-2                                   & $\textcolor{mygreen1}{\checkmark}$ & $\textcolor{mygreen1}{\checkmark}$ & $\textcolor{mygreen1}{\checkmark}$  & $\textcolor{mygreen1}{\checkmark}$  & $\textcolor{mygreen1}{\checkmark}$  & $\textcolor{mygreen1}{\checkmark}$         \\ \hlinewd{1pt}
\end{tabular}
\end{table}

In this paper, we also introduce a new benchmark dataset for natural blur assessment in digital pathology named ``FocusPath''. Such a database can help the validation of NR-ISA for medical imaging platforms. Unlike the previous metrics, our metric has the unique advantage of optimizing both speed and accuracy. The metric can be used for both synthetic and natural image assessment with a common parameter setting, allowing assessment of blur in broad imaging applications. Our future development will concern utilizing this metric in big image archiving database such as whole slide imaging systems and reconnaissance orbital imaging in satellites for development of quality check control applications.

\section*{Acknowledgment}
The authors would like to thank Huron Digital Pathology Inc. for providing digital pathology scan image database and their fruitful discussion during the development of the NR-ISA metric. The first and second authors research was partially supported by an NSERC Collaborative Research and Development Grant (contract CRDPJ-515553-17)

\ifCLASSOPTIONcaptionsoff
  \newpage
\fi

\bibliographystyle{IEEEbib}

\end{document}